\documentclass{iopjournal}

\usepackage{fancyhdr}


\makeatletter
\AtBeginDocument{%
	\fancyhead[R]{Sarno A. \textit{et al}}%
}
\makeatother

\usepackage{amsmath}
\usepackage{graphicx}
\usepackage{subcaption}
\usepackage{url}
\newcommand{\ignore}[1]{}

\begin{document}
	
	\articletype{Topical Review} 
	
	\title{Quantum Gate Engineering in Superconducting NISQ Platforms: Experimental Methods and Emerging Directions}
	
	\author{Alessandro Sarno$^{1,*}$\orcid{0009-0005-1912-7540}, Viviana Stasino$^{1}$\orcid{0009-0009-4898-5834}, Antonio Pio Italiano$^{1}$\orcid{0009-0003-8324-996X}, Alessandra Pizza$^{1}$\orcid{0009-0007-8771-2789}, Carlo Cosenza$^{1}$\orcid{0009-0004-2426-4008}, Wladislav Guarino$^{1}$\orcid{0009-0005-5335-8676}, Domenico Montemurro$^{1}$\orcid{0000-0001-8944-0640}, Francesco Tafuri$^{1}$\orcid{0000-0003-0784-1454}, Davide Massarotti$^{2}$\orcid{0000-0001-7495-362X}, Halima Giovanna Ahmad$^{1,*}$\orcid{0000-0003-2627-2496}} 
	
	\affil{$^1$Dipartimento di Fisica ``Ettore Pancini", Università degli Studi di Napoli ``Federico II", Via Cinthia, 80126, Napoli, IT.}
	
	\affil{$^2$Dipartimento di Ingegneria Elettrica e delle Tecnologie dell’Informazione, Università degli Studi di Napoli Federico II, Via Claudio 21, 80125, Napoli, IT. }
	
	\affil{$^*$Author to whom any correspondence should be addressed.}
	
	\email{alessandro.sarno@unina.it; halimagiovanna.ahmad@unina.it}
	
	
	\begin{abstract}
		This review aims to provide researchers, experimentalists, and quantum engineers with an updated and technically grounded perspective on the implementation of quantum gates in superconducting NISQ devices. The goal is to establish a comprehensive bridge between theoretical quantum circuit synthesis and its practical, experimental implementation, focusing on superconducting quantum technologies. We systematically examine the diverse landscape of superconducting qubits and processor architectures, ranging from conventional transmons to alternative and unconventional designs, evaluating their unique design trade-offs, connectivity constraints, device physics, and materials impact. Moving from theory to actual implementation, we detail how single- and two-qubit operations are synthesized via microwave-driven and flux-activated pulses. We here present our pulse engineering, wave shaping, calibration workflows, demonstrating how to bridge the gap between abstract gate models and real-world hardware limitations in the direction of a hardware-software co-design framework. To validate these operations, we choose  benchmarking methodologies with the purpose to get the more efficient metric for the specific evaluation aspect, including truth tables and quantum process tomography. Therefore, we both discuss the theoretical concepts behind the gate synthesis and show experimental implementation of most fundamental gates by proposing a practical workflow.
		
	\end{abstract}
	
	\section{Introduction}
	\label{Introduction}

	Since Deutsch established the framework for a universal quantum computer by extending the Turing machine model to the quantum domain \cite{Deutsch1985}, it became conceivable that the laws of quantum mechanics could be harnessed to perform computations beyond the practical reach of classical machines. The subsequent development of Shor's \cite{Shor94} and Grover's \cite{Grover96} algorithms provided the first concrete evidence of this potential, demonstrating quantum speedups for problems of direct technological importance. These advances laid the foundations for the modern vision of quantum computing as a tool for addressing societal challenges ranging from secure communications to big data processing \cite{Bennett2014, Sanders2023, Zheng2026}. 
	
	Another compelling motivation for quantum computation is the efficient simulation of nature itself. Following Feynman’s original insight \cite{Feynman82}, quantum computers also provide a natural platform for the simulation of quantum many-body systems \cite{Fauseweh2024}, strongly correlated matter \cite{HeadMarsden2021}, complex quantum dynamics \cite{Ollitrault2021}, and molecular electronic structure \cite{NAEIJ2024}, whose complexity scales exponentially on classical hardware. This is fundamental in particularly interesting use-case applications in the framework of drug and materials discovery \cite{Zhou2026}, solutions for energy efficiency \cite{Kim2026}, in economics and finance \cite{Ahmad26, Woerner2019}, as well as in logistic \cite{Weinberg2023}. Finally, extending Feynman’s vision of quantum devices as tools for simulating complex quantum phenomena, the integration of quantum computing with machine learning \cite{Biamonte2017} and artificial intelligence \cite{Alexeev2025, Sivak2026} may further broaden the range of accessible applications, spanning many-body physics, quantum chemistry, optimization, and data-driven scientific discovery, marking a significant interdisciplinary use of highly strategic technologies.

	Present-day quantum processors operate in a regime where errors are not merely technological imperfections, but are deeply rooted in the intrinsic fragility of qubits, the fundamental logical element in quantum computing, and their unavoidable coupling to external environment \cite{caldeira1983}, which render them highly susceptible to decoherence and imperfect control. Consequently, large-scale quantum algorithms remain constrained by the limitations of current quantum hardware technologies \cite{Preskill2018}. Most importantly, research is conducted across various hardware platforms (e.g., trapped ions \cite{Main2025}, superconducting circuits \cite{Acharya2025}, and spin qubits \cite{Takeda2022}), each exhibiting different physical characteristics, operational modalities, noise sensitivities, and control architectures.  Therefore, the full potential of quantum computing ultimately relies on the removal or mitigation of noise in any type of hardware. They are known as \emph{fault-tolerant protocols}.
	
	To overcome the intrinsic noise and imperfections of quantum hardware, quantum error correction (QEC) is currently the most promising solution to address the fragility of quantum information \cite{gottesman2010,Preskill2018}. Unlike classical error correction, QEC must cope with the no-cloning theorem \cite{Cerf97}, which does not allow for copying quantum information as in classical computing. To bypass the no-cloning theorem, QEC encodes the quantum information non-locally across several physical qubits and provides codes to extract error syndromes without collapsing the logical state and actively suppressing the effect of decoherence, control imperfections, and operational noise. Fault-tolerant quantum computing aims at providing an abstraction layer between algorithms and physical devices by suppressing noise and errors in a quantum computer.
	
	Several families of QEC codes have been proposed to enable fault-tolerant computation, among the others, repetition, surface and color codes \cite{Guillaud19, Acharya2025, Takada24}. A complete overview can be found in reference \cite{Fowler2021}. However, QEC codes' logical performance can differ substantially depending not only on the algorithm, but also, and most importantly, on the noise properties of the physical implementation on the hardware, which reflects the absence of a universally optimal strategy for protecting quantum information. Hardware platforms support different error models, connectivity graphs, control mechanisms, and fault-tolerant protocols, leading to markedly different logical error rates and resource requirements \cite{Acharya2025, Egan2021, Schneider23}. Moreover, one must also be aware that the realization of fault-tolerant quantum computation extends beyond logical encoding, and the implementation of protected logical operations is also critical. Protecting quantum information through encoding alone is, in fact, insufficient for scalable quantum computation \cite{campbell2017}.
	
	Consequently, the development of practically useful quantum applications increasingly requires \emph{hardware-aware} approaches \cite{Ahmad26}, in which algorithm design and compilation strategies are tailored to the capabilities and limitations of the underlying quantum processor.   
	
	\subsection{Quantum utility in the NISQ era}
	\label{QuantumUtility}
	
	To respond to the need for negligible logical error rates \cite{riverlane2025qec}, a growing scientific, industrial, and strategic interest has emerged around the question of how current quantum processors can deliver tangible computational value before fully fault-tolerant machines become available. In this context, the demonstration of logical qubits, error-corrected memories, and fault-tolerant operations, while essential for validating the long-term roadmap, does not alone answer the broader question of quantum utility \cite{Field2024,Chakraborty2024, Herrmann23}.   
	This demand has stimulated a complementary research direction focused on extracting useful computational capabilities from noisy hardware through hardware-aware algorithms that can operate within the capabilities of current NISQ hardware, even just in proof-of-concept settings. 
	
	On one hand, algorithmic developments are often pursued under idealized assumptions that neglect architectural constraints, control limitations, and realistic noise mechanisms \cite{Montanaro2016}. On the other hand, hardware efforts frequently emphasize device-level metrics, such as coherence times and gate fidelities, without fully addressing how logical operations, gate synthesis, compilation overhead, and algorithmic execution ultimately determine computational performance \cite{Lao22}. Quantum Benchmarking Initiatives have increasingly emphasized system-level evaluation criteria extending beyond isolated hardware metrics toward application- and scalability-oriented assessment frameworks \cite{darpaqbisolicitation} and the importance of scalable fabrication and standardization procedures for next-generation quantum hardware and integration with High-Performance Computing (HPC) \cite{eurohpc}. Nevertheless, quantum computing research remains strongly compartmentalized between algorithmic and hardware-oriented communities \cite{Karuppasamy2025}.
	This calls for a genuine hardware-software co-design paradigm, in which device physics, control engineering, compilation, and algorithm development are treated as inseparable components of the same computational stack \cite{Chong2017, Haener_2018, Fellous23}.
	
	The review first outlines the current roadmap towards quantum utility from the hardware perspective, especially focusing on superconducting technologies. The goal is to stress the challenge of a fully agnostic quantum computing paradigm, the strong correlation between algorithm development strategies and the hardware capabilities, and the need to bridge the gap between quantum hardware engineering and quantum computing specialists \cite{Haener_2018}. We outline both the theoretical and experimental background of elementary quantum gates for superconducting digital (or gate-based) quantum computing, highlighting current approaches to achieve high-fidelity quantum operations, e.g. single- and two-qubit native (or atomic) gates calibration workflows, and a discussion on benchmarking techniques that effectively allow both to run quantum algorithms in NISQ superconducting platforms, and to scale the number of physical qubits. Thus, the review aims at giving a practical example of experimental gate synthesis. 
	
	\section{From quantum computing theory to superconducting quantum hardware}
	\label{QuantumHardware}
	
	The roadmap toward scalable quantum computing and quantum utility is commonly articulated along two parallel axes: a hardware-driven approach aimed at engineering increasingly coherent, low-noise, and intrinsically protected qubits, and a software-oriented approach aimed at developing the software stack required for fault-tolerant or NISQ computation \cite{Acharya2023, Bravyi2024,  Acharya2025, Paetznick2026}. 
	
	Any discussion of scalable digital quantum processors inevitably begins with the requirements commonly referred to as Di Vincenzo's criteria \cite{DiVincenzo}, which also serve well as a conceptual framework for discussing the hardware challenges associated with scalable digital quantum processors. The key ingredients for a quantum computer are:
	\begin{enumerate}
		\item A scalable physical system with well-characterized qubits;
		\item The ability to initialize the state of the qubits to a simple fiducial state;
		\item Long relevant quantum coherence times;
		\item A ``universal" set of quantum gates;
		\item A qubit-specific measurement capability.
	\end{enumerate}
	
	Especially relevant for solid state systems, e.g., superconducting hardware, efforts are focused on optimizing circuit design \cite{wong2026, Siddiqi2021} and tailoring fabrication techniques that allow for substantial scaling of the number of qubits while keeping coherence performances stable \cite{MURRAY2021, Bylander2017}, eventually answering to Di Vincenzo's criteria 1. and 3.. At the same time, circuit and materials engineering must not compromise the capability to interface with the quantum processor, and perform the most fundamental operations for computation: control, through the application of quantum logic gates, and readout (e.g., Di Vincenzo's criteria 2., 4. and 5.). Despite the longstanding interest in developing the ``perfect" superconducting qubit in the direction of a more software-oriented approach are the development of novel error correction techniques (e.g., surface codes and dynamical decoupling \cite{lidar2008}), but also error mitigation techniques \cite{Cai2023}, to fill the gap to the fault-tolerant regime and use current state-of-the-art processors in the finite coherence time limit.  
	
	Among the various hardware platforms available, superconducting technologies have made significant breakthroughs, establishing themselves as a valid choice for their predisposition to scalability requirements, industrial fabrication, and the operational speed they offer \cite{Vool2017, Arute2019, Gold2021, Walter26}. Superconducting qubits are often referred to as \emph{artificial atoms} \cite{devoret2004}. They use Josephson junctions (JJs) \cite{JOSEPHSON1962, tafuri2019}, a circuital non-linear element used in various electronics applications: from energy-efficient digital circuits \cite{Tolpygo2016} to sensors \cite{Degen2017} and metrology applications \cite{Ghosh2026}. 
	
	A JJ is a trilayer made of two superconducting electrodes, separated by a non-superconducting barrier, in which the current flow and the voltage are regulated by the Josephson laws, and directly depend on the quantum phase-difference of the macroscopic wavefunctions of the superconducting electrodes \cite{Barone1982}. The dependence of macroscopic physical quantities, current and voltage, on a quantum variable is the most fundamental feature exploited in quantum technologies. As a matter of fact, the cosinusoidal potential energy of a JJ can be quantized if the device is operated below the effective temperature related to its plasma frequency. Low critical temperature JJs, e.g. most standard JJs based on aluminum technology, have typical plasma frequencies of the order of few $GHz$ \cite{devoret2004,Vool2017, Oliver_2013}, which corresponds to temperatures of the order of hundreds of $mK$. This is mainly the reason why the development of cryogenic infrastructure able to allocate as many qubits as possible, i.e., with sufficient cooling power, is also critical \cite{Hornibrook2015}. The integration of non-linear superconducting circuital elements, as the JJs, into an electrical circuit comprising capacitors, inductors and microwave elements, e.g. transmission lines and resonators, and the smart engineering of the potential energy of the circuit (in the framework of the more general circuit quantum electrodynamics - cQED) \cite{Vool2017} then allows to eventually isolate the first levels of the artificial atom, defining a \emph{quantum bit}, or qubit.  
	
	The term superconducting qubit then refers not to a unique physical implementation but to a broad class of superconducting artificial atoms, characterized by different circuit topologies and operating regimes, whose properties can be engineered by tailoring the relative contributions of three main energy scales: the charging energy $E_c$, which weights the qubit anharmonicity; the Josephson energy $E_J$; and the inductive energy $E_L$. An outstanding summary, a ``periodic table of superconducting qubits", has been reported in references \cite{Vool2017, kjaergaard2020superconducting, Jiang25n, Clarke2008}. As follows, we summarize the main qubit archetypes to highlight the fundamental differences that can occur from the operational point of view, and their advantages and disadvantages for quantum gates implementation. 
	
	\subsection{Circuits engineering towards scalable coherent superconducting processors}
	\label{circuits_engineering}
	
	Early implementations, such as the Cooper pair box (CPB) \cite{Nakamura99}, encoded quantum information in the charge degree of freedom but suffered from strong sensitivity to charge noise. Quantitatively, charge noise arises when the ratio between the Josephson and the charging energy is small: $E_J/E_C\ll 1$.
	The primary motivation for this architecture was the conceptual simplicity of electrostatic manipulation. Operationally, however, the CPB required precise static tuning via a voltage gate line. Because charge dispersion increases exponentially as $E_J/E_C$ decreases, the CPB proved drastically vulnerable to substrate charge noise. This vulnerability capped coherence times to a few nanoseconds, rendering the baseline configuration obsolete for scalable quantum processors.
	This limitation led to the development of the \emph{transmon} (transmission-line shunted plasma oscillation qubit) by Koch et al. \cite{Koch07}, which operates in a regime of reduced charge dispersion. The trick was to decrease the charging energy by shunting the Cooper pair box with a large capacitance, hence achieving a ratio $E_J/E_c \sim 50 -100$ \cite{Krantz19, Koch07}. 
	
	The transmon has become the dominant architecture in contemporary superconducting quantum processing units (sQPUs). Within the transmon family, both \emph{fixed-frequency} and \emph{flux-tunable} variants have been extensively explored, the latter enabling in-situ frequency control through SQUID-based designs that can be implemented using either symmetric or asymmetric junction configurations \cite{Hutchings2017}. 
	Besides the investigation on structural design to face different intrinsic noise sources, integration or packaging designs also had a big relevance in the roadmap of scaling. Starting from the transmon design \cite{Koch07}, the \emph{Xmon} \cite{Barends13} and the \emph{coaxmon}  \cite{Rahamim17} have been developed. The Xmon rearranges the transmon's shunt capacitor into a cross-shaped layout. This structural shift leaves the underlying Hamiltonian unchanged but spatially decouples four distinct terminal access points per qubit: one for readout, one for control, one for qubit frequency tuning, and one for inter-qubit coupling. This geometric decoupling minimized microwave crosstalk and laid the physical groundwork for executing surface codes on processors like Google Sycamore \cite{Arute2019}.
	The coaxmon adopts a concentric topology featuring a central capacitive disk surrounded by a grounded outer ring. This structure allows the system to break free from planar (2D) routing limits, enabling vertical three-dimensional wiring. Control and readout signals are delivered from the rear or the top of the packaging via integrated coaxial pins, substantially shrinking the surface footprint and eliminating planar microwave crosstalk.
	
	Beyond transmons, alternative superconducting artificial atoms, such as the \emph{fluxonium}, exploit large inductive shunts to achieve enhanced anharmonicity and potentially improved coherence. First synthesized by Manucharyan et al. in 2009 \cite{Manucharyan2009}, the \textit{fluxonium} emerged as a structural countermeasure to the transmon's weak anharmonicity. It replaces the massive shunt capacitor with a ``superinductor", i.e. a linear array of dozens of JJs connected in series, providing a high kinetic inductance while remaining completely non-dissipative. This array shunts a single JJ, maintaining exceptionally large intrinsic anharmonicity, frequently exceeding $1 \, GHz$ and eliminating spectral crowding and the associated risks of leakage into higher non-computational states. This feature makes the fluxonium one of the most promising architectures to achieve record coherence times of milliseconds \cite{Somoroff2023}.
	
	Special attention is also devoted to noise-protected qubit designs, e.g., the $0-\pi$ qubit, or \emph{quiet} qubit \cite{Brooks13,Gyenis21}, or solutions naturally compatible with the QEC overhead requirements. The rich energy-level structure of superconducting circuits, in fact, enables computational encodings that go beyond the two-level qubit approximation.
	This includes \emph{qudit-based} approaches and \emph{bosonic encodings}, where quantum information is stored in multi-level or oscillator-like degrees of freedom \cite{wang2020qudits}. Notable examples are the \emph{cat-state qubits} \cite{MIRRAHIMI2016} which exploit protected superpositions of coherent states to achieve intrinsic error bias and enhanced resilience against certain classes of errors.  Alternatively, the \textit{dual-rail qubit} \cite{Teoh2023} is a hybrid platform where quantum information is encoded within the bosonic single-photon manifold of two high-Q microwave cavities or two weakly coupled transmons. 
	As a result, superconducting platforms offer a diverse design space spanning conventional qubits, multi-level qudits, and hardware-efficient error-corrected logical encodings.
	
	\subsection{Materials engineering for novel hybrid superconducting qubits}
	\label{circuits_materials}
	
	Material design research for qubit architecture is also fundamental to increasing coherence times \cite{Siddiqi2021,MURRAY2021, Bal2024, Brosco2024}. A remarkable milestone that confirms the technological advancement of transmon design is the achievement of millisecond coherence times by materials engineering, without any modifications to the qubit architecture \cite{Bland2025}. In parallel, alternative less mature proposals consider to integrate unconventional materials in superconducting quantum circuits of different type: ferromagnetic $\pi$ - JJs in quiet-qubits to relax on certain geometrical and circuital constraints \cite{Kim2024}; hybrid semiconducting Josephson-based systems \cite{Larsen2015} or carbon nanotubes \cite{Riechert2025} in transmons (e.g. the \emph{gatemon}) to circumvent the use of local magnetic flux lines for flux-tunable transmons leveraging the superconducting proximity effect to tune electrostatically the carrier density within the semiconductor channel. 
	
	The \emph{ferrotransmon} \cite{Ahmad22} proposes to counteract flux-noise in flux tunable transmon by integrating in a transmon architecture a memory-like magnetic hybrid tunnel SIsFS JJs. Especially relevant is the use of a JJ in a series-regime, i.e., where the quantum coherence is dominated by the tunnel SIs counterpart, fabricated using the very same technology as standard transmon designs \cite{Satariano2025,calloni2026}. The sFS component of the series, instead, allows for introducing a qubit frequency tunability dominated by magnetization mechanisms \cite{Ahmad2025_transmon}, allowing for exploiting the residual magnetization of the ferromagnetic barrier to select between different qubit frequency working points without any external magnetic field applied. This reduces connections and electronics overhead and protects the qubit from flux noise arising from static flux signals used for parking. 
	
	\subsection{Engineering qubit couplings}
	\label{circuits_couplings}
	
	Coupling between qubits is also of fundamental importance. One of the main bottlenecks of superconducting hardware is the electrical circuit footprint, which renders connecting multiple qubits together a challenge \cite{Kim16, Wallraff2004}. The coupling architectures are very different, depending on the qubit designs, and a balance between gate fidelity, on-chip footprint, and system scalability is necessary \cite{forati2025}. The most straightforward approach is \emph{direct capacitive coupling} \cite{Krantz19}, widely used in transmons and cat qubits due to its constructive simplicity and the absence of additional active components; the permanently active nature of this interaction generates strong residual ZZ crosstalk and parasitic interactions that limit operation fidelity. To mitigate these undesired, effects \emph{dedicated couplers} are employed: \emph{coplanar bus resonators} exploit virtual photon exchange within cavities to mediate long-range interactions, reducing direct crosstalk at the expense of a significantly larger geometric footprint \cite{DiCarlo09}, whereas \emph{tunable couplers}  \cite{Moskalenko2022}, i.e. SQUID-based, allow for the dynamic modulation of the interaction down to zero ($J \approx 0$), effectively eliminating ZZ crosstalk at the cost of increased circuit complexity and additional flux control lines \cite{yan2018, sete2021, Sanclemente26}. 
	
	The physical nature of the coupling directly impacts the native two-qubit logic gates: while fixed coupling induces dispersive interactions that favor all-microwave driven gates, such as Cross-Resonance (CR), or Controlled-Phase (CPhase or CZ) \cite{Jiang2025}, tunable couplers enable ultra-fast energy-exchange gates, like iSWAP, bSWAP, or fSim \cite{Roth2017}, by bringing the qubits into geometric resonance. Other coupling strategies include, for example, fluxonium qubits paired with a capacitively coupled transmon qubit \cite{Ding23}, or the integration of unconventional hybrid qubits as couplers (e.g. ferromagnetic couplers) \cite{patent_ferro}.
	
	As the qubit count increases, however, the footprint of two-dimensional circuits saturates the available space, exacerbating destructive microwave and ZZ crosstalk. To overcome this geometric bottleneck, 3D modular interconnection approaches have been recently proposed, with a partition of the processor into multi-chip modules integrated via through-silicon vias (TSVs) and vertical microwave links, offering a viable pathway to spatially isolate crosstalk and scale quantum hardware without compromising coherence times \cite{Muthusubramanian_2024, patent_rigetti,patent_quantware}.
	QEC itself imposes a massive overhead in the number of physical qubits required to define a single logical qubit in surface code implementations, leading to recent proposals toward \emph{quantum low-density parity check} (qLDPC) codes: by leveraging long-range couplers, these codes can pack significantly more logical qubits into a given number of physical qubits \cite{Bravyi22, Ruiz2025}. 
	
	A comparative analysis of the different hardware platforms is therefore essential to assess their strengths and limitations, particularly given the current lack of standardized approaches for mitigating errors and decoherence. As discussed in section \ref{gates}, each qubit technology requires dedicated calibration procedures for single- and two-qubit gates, preventing a unified and automated framework.
	
	\section{Computational utility of experimental quantum gates}
	\label{ComputationalUtility}
	
	Within the broad realm of hardware platforms available, a similarly impactful role is played by the design and implementation of algorithms in the NISQ era. Developing quantum algorithms in NISQ devices requires a computational approach fundamentally different from the idealized circuit models commonly considered in theoretical quantum computing. In the NISQ regime, algorithmic performance is strongly constrained by limited coherence times, gate imperfections, restricted qubit connectivity, calibration drifts, and measurement errors \cite{Fu25, Almudevar20, Ahmad26}. As a consequence, algorithm development cannot be decoupled from hardware characteristics, and practical quantum workflows increasingly rely on hardware-aware design principles. Variational methods \cite{Cerezo2021} and hybrid quantum-classical schemes have emerged as prominent strategies for adapting computational tasks to the capabilities and limitations of present day quantum processors \cite{endo2021}.
	
	A realistic assessment of NISQ algorithms additionally requires accurate noise modeling and hardware characterization \cite{Bharti2022}. Since present-day quantum devices operate in intrinsically noisy environments, predictive algorithmic analysis must incorporate experimentally informed descriptions of decoherence processes, gate-dependent errors, readout imperfections, crosstalk effects, and time-dependent calibration fluctuations. \emph{Noise models} constitute a critical interface between hardware physics and algorithmic validation, enabling the study of error propagation, performance degradation, and mitigation strategies under realistic operational conditions \cite{Dahlhauser2021,Loglisci2026}.
	
	Within this context, quantum compilation and \emph{transpilation} play a central role in bridging the gap between abstract algorithmic descriptions and executable hardware instructions \cite{Maronese2022,Rattacaso2025}. Quantum algorithms are typically formulated in terms of logical or platform-independent gate sets, whereas physical quantum processors support only restricted native operations determined by the underlying hardware architecture. The transpilation process then involves gate decomposition, qubit mapping and connectivity adaptation, scheduling and low-level optimization in order to transform an ideal circuit into a form compatible with a target quantum device. Importantly, this transformation is not merely a technical translation step, but can substantially affect computational performance by modifying circuit depth, two-qubit gate count, idle times, and overall noise accumulation as partly discussed in section \ref{Softwarenisq}.
	
	In this review, the experimental implementation and characterization of elementary quantum gates in superconducting architectures are discussed from the perspective of their effective computational utility. After reviewing the physical realization of single- and two-qubit gates in circuit-QED platforms, including microwave-driven operations and flux-activated interactions, we analyze the limitations of widely adopted benchmarking methodologies. Special attention is devoted to pulse engineering, calibration workflows, noise mitigation strategies, gate characterization techniques, including randomized benchmarking and quantum process tomography, and the interplay between device physics, control electronics, and software orchestration.

	\subsection{Universal Quantum computing}
	\label{UniversalComputing}
	
	In quantum computing, the state of the system is represented by a state vector, $\Psi$, a complex linear superposition of all possible binary configurations of its $m$ bits ($x \in \{0, 1\}^m$). Mathematically, this is expressed as \cite{Barenco95}:
	\begin{equation}
		\Psi(t) = \sum_{x \in \{0,1\}^m} \alpha_x |x_1, \dots, x_m\rangle, \qquad \sum_x |\alpha_x|^2 = 1.
	\end{equation}
	The evolution of a quantum system over time $t$ is governed by a unitary operator $U$, i.e.  a linear, bijective mapping that preserves the norm of the state vector. This unitary framework provides the precise physical description of how an isolated system changes according to the laws of quantum mechanics, allowing for logically and thermodynamically reversible computation \cite{Barenco95}.
	In the framework of ``gate-based quantum computing", these unitary operators play the role of quantum logic gates\footnote{{\bfseries Clarification on notation}: Throughout this review, we use the following gate notations interchangeably without loss of generality, employing both radians and degrees, e.g. $X_{\frac{\pi}{2}}$ or $X_{90^{\circ}}$ or $RX(90^{\circ})$. Radians are used for theoretical formulas and circuits depictions, while degrees reflect the actual hardware implementation of the angles.} acting on the qubit system state, leveraging superposition and entanglement features \cite{Pereira23}.
	
	Quantum operations can be classified based on their effect on the qubit's state and whether they are conditioned on a control qubit \cite{albayaty2024}:
	\begin{itemize}
		\item Single-qubit operations: these include gates that allows to exploit the superposition, e.g the Hadamard ($H$) gate, fixed rotation gates such as the $Pauli-X$ (NOT), $ Pauli-Y$, and $Pauli-$Z gates, and customized rotation gates like $RX(\theta)$, $RY(\theta)$, and $RZ(\theta)$, which rotate the qubit by a specific angle $\theta$ in the Bloch sphere framework.
		\item Controlled (multi-qubit) operations: these gates apply a transformation to a target qubit only when the control qubit is in the $|1\rangle$ state. For two qubits, they are categorized into controlled superposition (e.g., $CH$ gate), fixed controlled rotations (e.g., $CNOT$ and $CY$), and customized controlled rotations (e.g., $CRX(\theta)$, $CRY(\theta)$, and $CRZ(\theta)$ gates). The extension to a three-qubit system is given by the Deutsch-Toffoli gate ($CCRX(\theta)$).
		\item Special multi-qubit gates: distinct from standard controlled gates, the $SWAP$ gate involves two target qubits, effectively exchanging their quantum states, while the $CPhase$, or $CZ$, acts on the global phase of the two-qubit system.
	\end{itemize}
	
	Di Vincenzo's criteria 4. states that the hardware must be able to perform a complete set of logic operations, called ``universal gate set". 
	Specifically, a set of quantum gates $G$ is said to be universal for quantum computation if any unitary operator can be approximated with arbitrary precision by a circuit involving only those gates. The key quantum gates required are:
	\begin{itemize}
		\item a base-changing gate (e.g. $H$);
		\item an entangling gate (e.g. $CNOT$ or Toffoli gate) \cite{Shi02}.
	\end{itemize}
	However, the Gottesman-Knill theorem \cite{Gottesman98} implies that any circuit involving CNOT and Hadamard only can be simulated efficiently by a classical circuit. In order to take a real advantage of quantum computation, gates that allow the exploration of non-stabilizer states are necessary, like the $T$ gate (i.e., $Z_{\frac{\pi}{8}}\equiv RZ(45^{\circ})$) and the Deutsch-Toffoli gate (i.e. $CCRX(\theta)$). In particular, the $T$ gate is strictly linked with the non-stabilizerness, also colloquially known as ``magic", claimed to be necessary for quantum utility along with superposition and entanglement \cite{Alioscia2020,Oliviero_Leone_Hamma_Lloyd_2022,Alioscia2025,ahmad2025_magic}.
	
	Therefore, a wide plethora of universal gate sets $G$ have been proposed in time, and are here summarized:
	\begin{itemize}
		\item $G_{0}=\{Toffoli, CNOT, T \}$ shown in \cite{Kitaev97};
		\item $G_{1}=\{CNOT, H, S \}$ shown in \cite{BOYKIN00};
		\item $G_{2}=\{CNOT, g \}$, with $g$ any single-qubit gate, shown in \cite{Barenco95};
		\item $G_{3}=\{X_{\theta}, Y_{\theta}, Z_{\theta}, Ph_{\theta}, CNOT \}$ and $G_{4}=\{H, S, T, CNOT \}$ (also referred to as ``Clifford $+$ T" gates), both shown in \cite{Krantz19}.
	\end{itemize}
	By exploiting one of the universal gate set it is possible to implement any logic operation necessary to perform an algorithm on a quantum computer. However, each quantum computing architecture will have certain gates that are simpler to implement than others at the hardware level, referred to as \emph{native gates} of the architecture.
	
\begin{figure}[h!]
		\centering
		\includegraphics{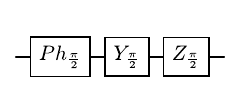} 
		\caption{Gate H decomposition in the $G_{4}$ universal gate base set.}
		\label{Hadamard}
	\end{figure}

	Given for example the gate set $G_{4}$, the $H$ gate can be decomposed as (figure \ref{Hadamard}):
	\begin{equation}
		H=Ph_{\frac{\pi}{2}}Y_{\frac{\pi}{2}}Z_{\pi} = \frac{1}{\sqrt{2}} \left[
		\begin{array}{cc}
			1 & 1  \\
			1 & -1  \\
		\end{array} \right].
		\label{HadamardEq}
	\end{equation}
	
	The $S$ and $T$ gates are, instead, implemented with $RZ(\theta)$ rotations with $\theta=90^{\circ}$ and $45^{\circ}$, respectively \cite{Krantz19}. 
	
\begin{figure}[h!]
		\centering
		\includegraphics[width=\linewidth]{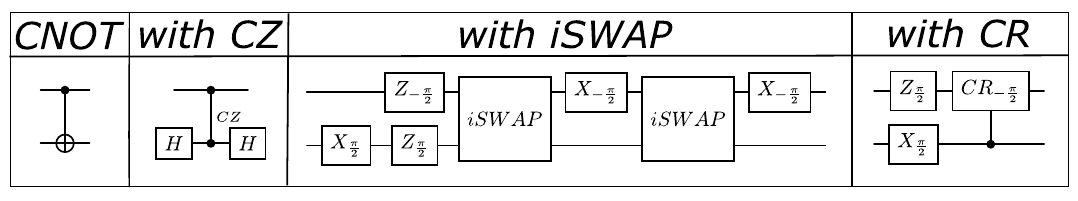}
		\caption{CNOT decomposition with CZ, iSWAP and CR native gates.}
		\label{CNOTdecomposition}
	\end{figure}

	The $CNOT$ can be decomposed as:
	\begin{equation}
		CNOT=\{ I\otimes H\}CZ\{ I\otimes H\}=\left[
		\begin{array}{cccc}
			1 & 0 & 0 & 0 \\
			0 & 1 & 0 & 0 \\
			0 & 0 & 0 & 1 \\
			0 & 0 & 1 & 0
		\end{array} \right]  ,
	\end{equation}
	or as a combination of $X_{\frac{\pi}{2}}$, $Z_{\frac{\pi}{2}}$ and two-qubit gate $iSWAP$ \cite{Krantz19} (figure  \ref{CNOTdecomposition}), depending on the highest-fidelity two-qubit gate one can naturally implement in a specific architecture. 
	
	Especially relevant is the SWAP gate, which is useful for all those cases in which the information stored in a qubit must be moved on another qubit. For example, in processors in which there is not an all-to-all connectivity, a re-routing process can be implemented using a concatenation of multiple SWAP gates \cite{albayaty2024}. If not available as a native gate, the SWAP can be decomposed in terms of atomic gates \cite{Krantz19}, or through a decomposition of three CNOT gates \cite{albayaty2024}, as shown in figure \ref{swap}:
	\begin{figure}[h!]
		\centering
		\includegraphics{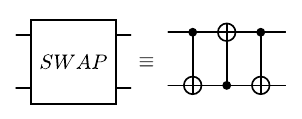}
		\caption{SWAP decomposition in terms of CNOT gates.}
		\label{swap}
	\end{figure}
	
\begin{figure}[h!]
		\centering
		\includegraphics[width=0.5\linewidth]{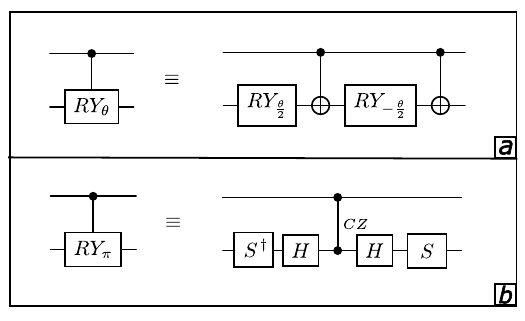}
		\caption{CY gate decompositions: in a $CRY_{\theta}$ decomposed in $G_{4}$ gate base set, in b $CRY_{\pi}$ decomposed in  $G_{1}$ gate base set.}
		\label{RYdecomposition}
	\end{figure}

	Other relevant two-qubit gates to mention are those commonly used in quantum algorithms and applications. Let us consider a practical example: the quantum amplitude estimation (QAE) algorithm \cite{Brassard_2002}, which is used in a variety of applications. Among them, it is worth mentioning the field of quantum finance (e.g., for option pricing \cite{Stamatopoulos2020}, credit risk evaluation \cite{Woerner2019}), and physical systems study (e.g., for hamiltonian ground state estimation). For QAE, a key role is played by the $CRY(\theta)$ gate, which can be decomposed in terms of CNOT and $RY(\theta)$ gates \cite{Barenco95}, e.g., the gate-set $G_4$, as reported in figure  \ref{RYdecomposition}a. 
	
	Alternatively, in case of $\theta=180^{\circ}\equiv\pi$, one may use gate-set $G_1$, thus using $H$ and $SH$ gates as shown in figure  \ref{RYdecomposition}b.
	
	Finally, scaling to n-qubit operations is also essential. Multi-controlled gates, such as Toffoli and Fredkin gates, play a central role in the implementation of reversible classical logic within quantum circuits \cite{nielsen2010quantum}. They are 3-qubit universal gates proposed in arithmetic subroutines (e.g., adders and modular multipliers), which constitute the computational core of algorithms such as Shor’s factoring algorithm \cite{Shor94}. They also enable efficient construction of boolean oracles for Grover-type search \cite{Grover96} and amplitude amplification procedures \cite{Brassard_2002}, and are employed in quantum error-correction \cite{Cory98}, quantum simulation \cite{Georgescu14}, and reversible circuit synthesis \cite{shende08}. 
	
	Among the Fredkin-Toffoli gates, the Toffoli (CCX or CCNOT) and Fredkin (CSWAP) gates \cite{Toffoli80} are canonical examples of universal reversible logic gates. Remarkably, the Toffoli gate, together with the Hadamard gate, forms another universal quantum gate set that adds to the ones previously mentioned. This provides, in principle, the practical advantage of reducing the overhead of gate synthesis: for instance, both the overall gate count and the circuit depth can be reduced by a factor of $\mathcal{O}( \log{n} )$ in Shor’s factoring algorithm. QEC protocols with reduced resource cost can also be constructed using such a gate set, as well as in the oracle functions of Grover’s search algorithm \cite{Kim22}. Even more powerful is the Deutsch gate, often represented as a doubly controlled rotation $CCRX(\theta)$, which stands as a quantum extension of reversible multi-controlled logic. While the Toffoli gate acts as a reversible classical logic operation on computational basis states, the Deutsch gate enables arbitrary controlled quantum rotations and can therefore generate superposition and interference effects. Hence, for suitable values of $\theta$, it is universal for quantum computation.  
	
	In most quantum computing platforms, multi-qubit logical operations such as the Toffoli gate are not implemented as native hardware primitives but are instead decomposed into sequences of one- and two-qubit gates supported by the underlying architecture. This is the case for the majority of superconducting \cite{Fedorov2012}, trapped-ion \cite{Monz2009}, and photonic platforms \cite{Li2022}, where the native gate set typically consists of single-qubit rotations and an entangling two-qubit operation (e.g., CZ, XX, or iSWAP). Notable exceptions have been demonstrated in specific experimental settings. For example, trapped-ion systems have realized efficient Toffoli implementations exploiting collective motional modes \cite{Monz2009}, while neutral-atom platforms based on Rydberg interactions have demonstrated multi-qubit gates that leverage intrinsically collective dynamics and can approach genuinely native implementations of controlled multi-qubit operations \cite{Beterov2018}. These examples, however, remain special purpose realizations rather than the standard paradigm adopted across most quantum computing hardware. Therefore, in most standard situations, implementing a high-fidelity Toffoli gate is experimentally challenging: the operation is usually realized by decomposing it into single- and two-qubit gates. 
	
\begin{figure}[h!]
		\centering
		\includegraphics[width=0.6\linewidth]{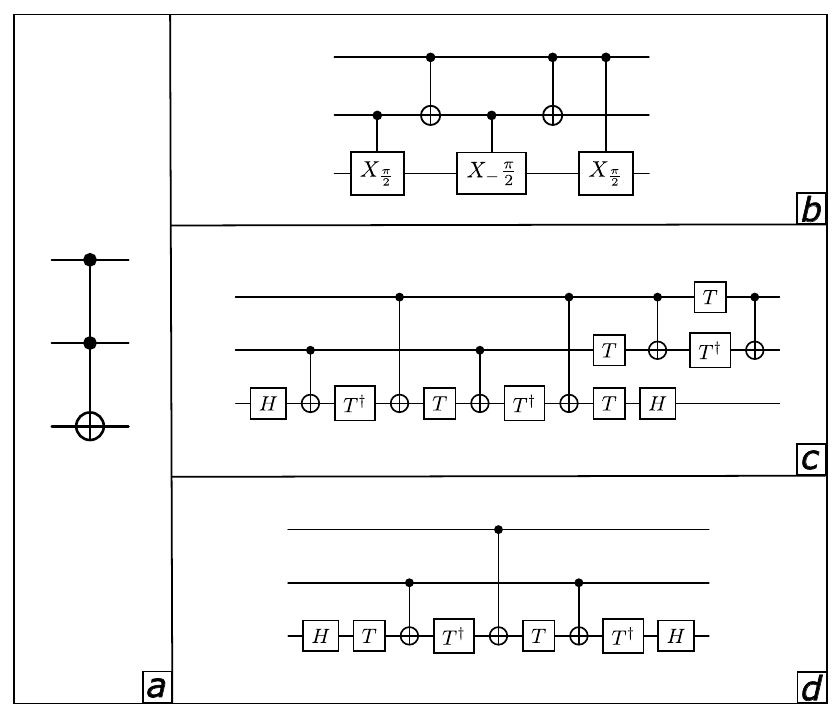}
		\caption{Toffoli gate decomposition {\bfseries a}: in {\bfseries b} Sleator, Weinfurter \cite{Sleator95} and Barenco \cite{Barenco95} decomposition, in {\bfseries c} Shende and Markov  \cite{shende08} decomposition, in {\bfseries d} mirrored structure decomposition \cite{albayaty2024}.}
		\label{Toffoli decomposition}
	\end{figure}

	Known decompositions require at least five two-qubit gates, which negatively impact the fidelity \cite{Yu13,Sleator95,Barenco95,shende08} (figure  \ref{Toffoli decomposition}). Nevertheless, these decompositions require an all-to-all connectivity, which is not always available on the hardware. In reference \cite{albayaty2024}, a layout-aware n-bit Toffoli gate is designed as a quantum circuit of straight mirrored hierarchical structure, then constructed symmetrically: when any of the controls are in the state of $|0\rangle$, i.e., there is a false value of the solution, then all quantum circuit collapses to generate a state of $|0\rangle$ for the target qubit (figure  \ref{Toffoli decomposition}d). This structure is more hardware-friendly by eliminating the crossing connections among the qubits, avoiding circuit overhead due to the SWAP gates. 
	
	Finally, the core of the layout-aware 3-bit Toffoli gate is the main quantum circuit that is utilized to iteratively build the layout-aware n-bit Toffoli gate. For instance, the core of the layout-aware 3-bit Toffoli gate is mirrored and repeated to construct the core of the layout-aware 4-bit Toffoli gate. Subsequently, the core of the layout-aware 4-bit Toffoli gate is mirrored and repeated to construct the core of the layout-aware 5-bit Toffoli gate, and so on \cite{albayaty2024}.
	
	\subsection{Software in NISQ era}
	\label{Softwarenisq}
	When synthesizing an algorithm, one wants to keep the overall number of time steps in which gates are applied (i.e., the circuit depth) as low as possible, and one wants to use as many of the native gates as possible. Quantum compilation represents a critical bridge between hardware-agnostic quantum algorithms and physical hardware. Given the severe constraints of NISQ, compilation cannot merely rely on standard abstraction layers. Instead, modern workflows leverage \textit{transpilation}, an optimization process that breaks down abstractions by rewriting a given input circuit to match the specific topology and native basis gate set of the target hardware. 
	Designed for modularity and extensibility, modern transpilation encompasses several interdependent layers, including processor mapping for layout selection, routing, gate synthesis, and pulse-level scheduling \cite{Shi20}:
	\begin{itemize}
		\item {\bfseries Processor mapping (layout and routing)}: this stage governs the allocation of virtual qubits to physical registers and resolves hardware connectivity constraints via topological routing. By evaluating real-time device calibration data, it identifies optimal physical configurations to minimize operational error rates \cite{Siyuan2020}.
		\item { \bfseries Gate synthesis and footprint optimization}: This stage decomposes abstract quantum operations into the hardware's native basis gate set. It reduces overall gate counts and compresses the circuit's structural footprint \cite{Davis2020}.
		\item { \bfseries Gate scheduling and timing}: This final stage maps the optimized circuit onto a precise timeline based on physical gate durations and hardware clock cycles. By utilizing advanced cross-layer techniques, such as SABRE \cite{Li2019}, it is possible to reduce idle times, swapping operations and squeezing the total circuit execution well within the hardware's coherence limits to mitigate phase decoherence.
	\end{itemize}
	Recognizing the vital importance of these compilation pipelines, numerous industrial and academic organizations have developed specialized software environments \cite{IBMQuantumTranspile,GoogleCirq,RigettiPyQuilDocs,QuantinuumTKET,JavadiAbhari2015}.
	\ignore{Qiskit by IBM \cite{IBMQuantumTranspile}, Cirq by Google \cite{GoogleCirq}, PyQuil by Rigetti \cite{RigettiPyQuilDocs}, TKET by Quantinuum \cite{QuantinuumTKET}, ScaffCC \cite{JavadiAbhari2015} and so on.}
	However, these compilation techniques sometimes may not be sufficient, as they do not fully account for all hardware noise sources. Because certain noise channels are not completely modeled within the backend, they can still degrade circuit performance. Consequently, alternative strategies, such as developing quantum circuits through variational experiments tailored specifically to the hardware, offer a complementary approach to enhancing circuit performance \cite{Ahmad26}.
	
	\subsection{Physical Implementation}
	\label{gates}
	
	A quantum gate can be decomposed in different ways while always giving the same output, i.e., the overall process implemented on the system is the same.  
	Also in the framework of sQPUs, decompositions can change in various ways depending on the hardware design. Following previous discussions, we here summarize the most relevant constraints: 
	\begin{itemize}
		\item the circuit design: flux-tunable or fixed transmons, fluxonium, and other circuit design type allows for different control, readout and coupling mechanisms, as described in section \ref{circuits_engineering};
		\item the hardware topology and connectivity, which defines how many qubits are coupled together within a constrained geometry. It is highly relevant especially in the framework of QEC codes, which all rely on different qubit connectivities. However, it is also important for applications in the NISQ regime, since most quantum algorithm development may require specific qubit connectivity.
		\item implementable native gates and their calibration capability: if it is conceptually possible to realize different types of quantum gates within the same processor design, but the reliability of calibration protocols and performances of these gates are reduced, it is in general impractical to acknowledge their use.
	\end{itemize}
	
	As follows, we start by detailing on the physical mechanisms behind single- and two-qubit gates implementation in transmon devices, while also giving comparisons with other circuital design whenever required.

	\paragraph{How to physically implement single-qubit gates.}
	\label{PhysicalImplemetation}
	
	Single-qubit $X$ and $Y$ gates, namely $RX(\theta)$ (or $X_{\theta}$) and $RY(\theta)$ (or $Y_{\theta}$) rotations, are reproduced via microwave pulses. In weakly anharmonic superconducting circuits, in fact, the time-dependent control (or drive) hamiltonian in the rotating-wave approximation reads as \cite{Krantz19}:
	\begin{equation}
		H_{\textit{drive}}(t) = \frac{\hbar}{2} \Omega(t) [(\cos(\phi)\cos(\delta\omega t) -\sin(\phi)\sin(\delta\omega t))\hat{\sigma}_x -(\cos(\phi)\sin(\delta\omega t) - \sin(\phi)\cos(\delta\omega t))\hat{\sigma}_y ], 
	\end{equation}
	where the drive strenght $\Omega(t)$ depends on the capacitive coupling between the qubit and the control line $C_d$, the total capacitance to the ground of the transmon island $C_{\Sigma}$, the circuit impedance $Z$ and the pulse parameters (e.g., the amplitude $V_0$ and the shape $s(t)$):
	\begin{equation}
		\Omega(t)=\sqrt(\hbar/2Z)C_d/C_{\Sigma}V_0s(t).
	\end{equation}
	Here, $\delta\omega$ is the difference between the drive signal and the qubit frequency. The control pulse is often implemented via heterodyne up-conversion of complex baseband envelopes, synthesized by arbitrary waveform generators (AWGs), and modulated onto a microwave carrier through balanced IQ mixers to drive the time-dependent hamiltonian. 
	
	In order to efficiently trigger the desired operation on the qubit for an algorithm, or to prepare the initial state to eventually evolve (preparation), or to readout the output of a quantum circuit, the physical parameters of these signals must be accurately calibrated, i.e.:
	\begin{itemize}
		\item the pulse shape;
		\item the power (e.g., the amplitude and duration of the pulses);
		\item the frequency of the control signals;
		\item the phase of the microwave signal.
	\end{itemize}
	For example, when the control signal is perfectly on resonance with the qubit frequency, $\delta\omega=0$, the drive hamiltonian reduces to:
	\begin{equation}
		H_{\textit{drive}}(t) = \frac{\hbar}{2} \Omega(t) [\cos(\phi)\hat{\sigma}_x + \sin(\phi)\hat{\sigma}_y ], 
	\end{equation}
	which means that one may select to perform an $X_{\theta}$ by maximizing the $\cos(\phi)$ term (or in-phase component, $I(t)$), or to perform a $Y_{\theta}$ by maximizing the $\sin(\phi)$ term (or quadrature-phase component, $Q(t)$). The rotation angle, instead, is regulated by the power of the pulse. A poor calibration of these physical parameters can introduce detrimental effects: the shape or the power of the pulses may introduce undesired over- or undershoot in the rotation angle with respect to the Bloch sphere, leakage and dephasing; dephasing can be also induced by the detuning of control signals from the qubit frequency \cite{Reed2013}. 
	
	The presence of spurious microwave tones induced by standard analogue upconversion techniques, especially relevant if resonating with higher-order energy transitions, or with the energy transitions of other qubits in the processor, must also be addressed. Currently, room-temperature control electronics can employ digital signal synthesis (DSS) techniques to improve microwave signal integrity, thereby avoiding the imperfections typically associated with analog IQ-mixer-based upconversion \cite{Kalfus20}. However, these approaches remain constrained by the bandwidth and Nyquist-zone limitations of the underlying digital-to-analog conversion hardware. The integration of tunable couplers and flux-tunable SQUID-based qubits allows processors to dynamically shift operational frequencies, parking qubits at safe, at highly detuned frequencies during idle states and bringing them into resonance only during gate execution, thus also contributing to frequency crowding mitigation \cite{sung2021realization}. Other solutions at the hardware level employ crosstalk-aware software compilers and mixed-integer linear programming algorithms that optimize the spatial allocation of frequencies and schedule gate operations to maximize parallelism while strictly avoiding simultaneous pulses on structurally vulnerable channels \cite{Ding20,Wagner25}.
	
	The final fundamental single-qubit operation is related to the rotation around the $z$ axis, i.e., $RZ(\theta)$. In contrast to resonant $X$ and $Y$ control, which necessitate the physical synthesis of microwave pulse envelopes, $z$ axis rotations can be implemented as \textit{virtual Z} gates by leveraging the phase-degree of freedom in the microwave drive \cite{Krantz19}. This technique involves an instantaneous shift of the local oscillator phase $\phi$ of the subsequent quadrature pulse, effectively rotating the equatorial frame of reference relative to the qubit’s state vector \cite{Krantz19}. This implementation allows for getting an $RZ(\theta)$ gate achieving near-perfect fidelity with zero temporal overhead, thereby avoiding the detrimental effects of the counterparts $RZ(\theta)$ gates implemented via flux-tuning or off-resonant Stark shifts \cite{kjaergaard2020superconducting}.
	
	\paragraph{Single-qubit gate calibration.}
	
	Key pulse parameters for $RX(\theta)$ and $RY(\theta)$ gates in superconducting qubits, such as the the power of the signal (the amplitude at a fixed duration) can be calibrated by experimental protocols, as Rabi oscillations and flipping (or error amplification) protocols (see supplementary of reference \cite{Ahmad26}). Ramsey interferometry is used, instead, for the estimation of the correct frequency of the drive signal, i.e., the qubit transition frequency $f_{0,1}$ \cite{Ahmad2023,Ahmad26}.
	
	In superconducting quantum computing, the choice of microwave pulse envelopes is also a critical factor in optimizing gate fidelity and suppressing leakage into non-computational states, such as the $\vert{}2\rangle$ state in weakly anharmonic transmon qubits \cite{Krantz19}. Gaussian pulses are widely adopted as the default standard due to their unique mathematical property of scaling smoothly in both time and frequency domains, which minimizes off-resonant high-frequency tails that would otherwise excite spectator transitions \cite{Boradjiev2013}. However, because a theoretical Gaussian function extends infinitely and must be artificially truncated by the control hardware, many leading industrial architectures frequently employ cosinusoidal or raised-cosine pulses \cite{Polat2026}. These sinusoidal variants provide a smooth, strictly finite temporal duration that naturally vanishes at the boundaries, eliminating truncation discontinuities while maintaining strong spectral confinement. Another way to intervene on the pulse shape itself to carefully calibrate the in-phase $I$ and out-of-phase $Q$ components of a control signal is the derivative removal by adiabatic gate (DRAG) \cite{Krantz19}. This technique is employed to maximize gate speed while mitigating leakage into the non-computational state (e.g. $|2\rangle$), as a consequence of phase accumulation. It consists in applying an extra signal in the out-of-phase component as $Q(t) = -\beta \dot{I}(t)$ \cite{Motzoi09,Krantz19}. The phase calibration becomes possible thanks to the calibration of the so-called Motzoi parameter $\beta$, which modifies the waveform envelope $s(t)$ according to:
	\begin{equation}
		s(t)\rightarrow s'(t)= 
		\begin{cases}
			s(t) \qquad on \quad I \\
			\beta \frac{\dot{s}(t)}{\alpha} \qquad on \quad Q 
		\end{cases}.
	\end{equation}
	Here $\beta$ is a dimensionless scaling parameter, and $\alpha$ is the anharmonicity. For $\beta=0$, no \textit{DRAG} correction is applied. The theoretical optimal choice for reducing phase error is $\beta = 0.5$, meanwhile, for reducing leakage error, it is $\beta = 1$ \cite{Krantz19}. Therefore, a trade-off value is needed. However, it is possible to reduce both errors simultaneously by introducing a frequency detuning parameter $\delta f$ to the waveform ($\delta f=0$ corresponds to standard qubit frequency):
	\begin{equation}
		s'_{\delta f}(t)=s'(t)e^{i2\pi\delta ft} ,
	\end{equation}
	and choosing $\beta$ to minimize leakage errors \cite{Krantz19}. An accurate estimation of the $\beta$ Motzoi parameter can be obtained through the Motzoi protocol \cite{Reed2013}, in which two pulse sequences exhibiting the opposite error sign syndrome (for instance, $Y_{\pi}X_{\frac{\pi}{2}}$ and $X_{\pi}Y_{\frac{\pi}{2}}$) are performed: from the measured population for each sequence as a function of the $\beta$ derivative correction, the optimal values are obtained at the intersection of the two fitted lines (supplementary materials of \cite{Ahmad26}).
	
\begin{figure}[h!]
		\centering
		\includegraphics[width=0.8\linewidth]{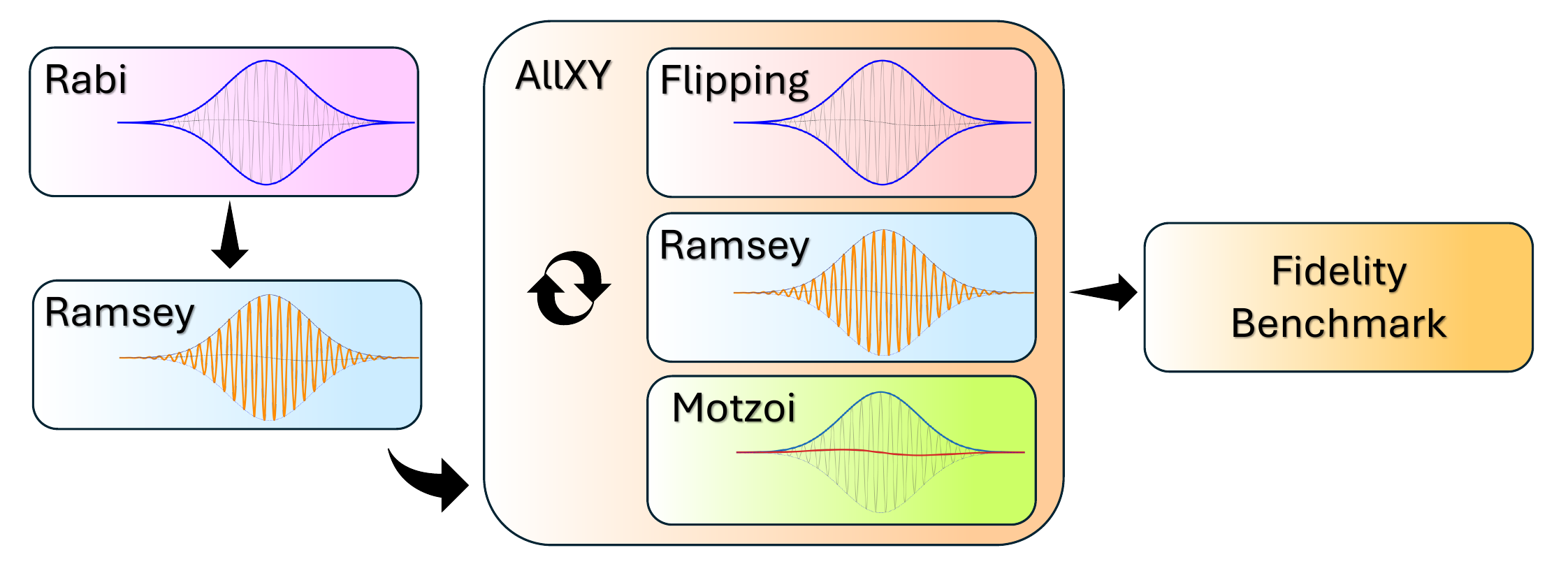}
		\caption{Calibration workflow for DRAG pulse: Rabi and flipping protocols act on the pulse amplitude, Ramsey protocol acts on the pulse frequency, Motzoi protocol acts on the IQ components. The AllXY protocol can be exploited to detect the presence of a syndrome and then to choose the best correction protocol \cite{Reed2013}. Finally a benchmarking method, as randomized benchmarking or gate set tomography, is performed to get a fidelity value.}
		\label{fig:DragCal}
	\end{figure}

	A summary of a common calibration workflow for single-qubit gates is shown in figure \ref{fig:DragCal} for convenience. Please note that other circuit designs can bypass some of the calibration nodes. For example, the fluxonium does not need DRAG pulse-shaping techniques due to the intrinsic robustness of the separation of the energy levels \cite{Rower2024}. 
	
	\paragraph{Hardware-dependent two-qubit gates implementation.}
	\label{two-qubit}
	
	The two-qubit gates available in the transmon architectures can be divided into two categories: 
	\begin{enumerate}
		\item those that require local magnetic fields to tune the transition frequency of qubits, e.g., in case flux-tunable transmons or tunable couplers are available;
		\item those that rely on all-microwave control, e.g., in case of fixed-frequency transmons.
	\end{enumerate}
	In both cases, one can always define in a pair the \emph{control} and \emph{target} qubits.
	In the case of fixed-frequency transmons, the common choice is to use the cross resonance (CR) gate as an entangling gate. Among the industrial realities that exploit such gates, we mention IBM \cite{Wei2024}. The CR gate is implementable only using microwave control and exploits the static coupling between qubit pairs due to the link of quantum buses \cite{Rigetti10}. Experimentally, it is obtained by sending a microwave pulse to the control qubit with a frequency of the target qubit and with a duration and amplitude sufficient to implement an effective $Z \otimes X$ gate. Since the simple CR gate relies on qubit-qubit interaction for a long time ($>300-400 \; ns$), it is prone to unwanted crosstalk. For this reason, the implementation of the echo-CR gate is a better option \cite{ward2026}. The echo-CR gate is obtained by adding an active cancellation $\pi$-pulse on the control qubit and by splitting the $CR(\theta)$ drive in two equal contributes with opposite phase. In reference \cite{Sheldon16}, a tune-up procedure for faster and high fidelity echo-CR gate to implement an effective $Z \otimes X_{\frac{\pi}{2}}$ is shown. The authors generated a $CNOT$ gate using single-qubit Clifford gates and by suppressing unwanted drive-hamiltonian terms, achieving a fidelity $>99\, \%$, and show the protocol required to achieve this result. Finally, by choosing the right phase and amplitude for the echo-CR gate, not only one can implement $CNOT$ gates, but also $CY$ gates by producing the right components of the drive hamiltonian \cite{Krantz19}.
	
	In case of flux-tunable transmons, entangling CNOT gates are operated by moving qubits on resonance to exchange excitations through baseband flux pulses \cite{DiCarlo09}. Among flux-based two-qubit gates thanks to which one can decompose a CNOT gate, we recall the $iSWAP$ and the CPhase gates. The CPhase gate can also be specialized to the $CZ$ gate whenever the conditional phase of the gate is $\pi$ ($180^{\circ}$) \cite{Krantz19}. CPhase, CZ and iSWAP are typically identified as the native gates for these architectures.
	
	The iSWAP gate is strictly related to the electrodynamic properties of the circuit, i.e., by hamiltonian engineering, and it is obtained by tuning the qubits into resonance for a time $\tau=\pi/2J$, with $J$ the the qubit-qubit coupling strength. Direct capacitive coupling or capacitive interaction mediated through resonator bus \cite{Majer07}, leads to a swapping of excitations between the two qubits,  also referred as ‘XY’ interaction \cite{Schuch03} and adding a phase of $i = e^{i\frac{\pi}{2}}$.
	
	From an operative point of view, the \textit{$i$SWAP} is performed at the nearby region called avoided-level crossing (ALC) and it consists in fixing the lower frequency qubit in the pair in its sweetspot and prepared in its ground state $|0\rangle$ (for this reason, the lower frequency qubit is also called \emph{spectator}), and then tuning the higher frequency qubit (also called as \emph{tuned qubit}), initially prepared in its excited state $|1\rangle$, through a flux pulse. An example of ALC spectroscopy for qubits coupled through high-frequency resonators is also reported in the supplementary material of reference \cite{Ahmad2024_mit}. 
	
	The two qubits are kept on resonance for a time $\tau$ long enough for the excitation to be swapped, and the corresponding transformation is: 
	
	\begin{equation}
		U\left(\frac{\pi}{2J}\right)=\left[
		\begin{array}{cccc}
			1 & 0 & 0 & 0 \\
			0 & 0 & -i & 0 \\
			0 & -i & 0 & 0 \\
			0 & 0 & 0 & 1
		\end{array} \right] \equiv iSWAP.
		\label{swapmatrix}
	\end{equation}
	
	This transition must be adiabatic with respect to the time-scale set by the transverse exchange interaction between the qubits of strength $J=\frac{g_{1}g_{2}}{2}(\frac{1}{\Delta_{1}}+\frac{1}{\Delta_{2}})$, with $\Delta_{1,2}=\omega_{1,2}-\omega_{r}$ the detuning of qubit $i^{th}$ to the resonator, and $g_{1,2}$ resonator couplings to the qubit $i^{th}$ \cite{Krantz19}.
	Although the computation itself is carried out taking into account only two energy levels, i.e., the ground and the first excited states, the implementation of some gates may also exploit higher energy levels outside the computational basis \cite{Krantz19}, e.g., the CPhase stands out. Indeed, the same approach allows to operate entangling gates by modulating the flux bias of a flux-tunable transmon in such a way to obtain a resonance with higher energy levels of the other neighbors qubits, e.g., the state $|11\rangle$ and the $|02\rangle$ states for $CZ$ implementation, which transformation is \cite{Krantz19}: 
	\begin{equation}
		U\left(\frac{\pi}{2J'}\right)=\left[
		\begin{array}{cccc}
			1 & 0 & 0 & 0 \\
			0 & 1 & 0 & 0 \\
			0 & 0 & 1 & 0 \\
			0 & 0 & 0 & -1
		\end{array} \right]  \equiv CZ .
		\label{CZmatrix}
	\end{equation}
	
	When two qubits are set on resonance through flux pulses starting from the $|11\rangle$ state, in terms of occurrence, the first encounter is the ALC between $|11\rangle-|02\rangle$ at lower flux pulse amplitude, while at higher flux pulse amplitude the crossing between $|01\rangle-|11\rangle$ occurs, allowing for SWAP-like gates. Moreover, the coupling strength $J'$ between  $|11\rangle-|02\rangle$ states is larger than $J$ between $|01\rangle-|10\rangle$ states by a factor $\sqrt{2}$, resulting in a faster CZ gate compared to the SWAP one \cite{Negirneac21}.
	However, given the intrinsic nature of the CZ, which exploits non-computational levels, makes the gate very vulnerable to leakage of population (information) from the computational state.
	A more general definition of the unitary operator that implements the CPhase can be found in \cite{Negirneac21}: 
	\begin{equation}
		U_{CPhase}=\left[
		\begin{array}{ccccc}
			1 & 0 & 0 & 0 & 0  \\
			0 & e^{i\phi_{01}} & 0 & 0 & 0  \\
			0 & 0 & e^{i\phi_{10}} & 0 & 0 \\
			0 & 0 & 0 & \sqrt{1-4L}e^{i\phi_{11}} & \sqrt{4L}e^{i\phi_{02,11}} \\
			0 & 0 & 0  & \sqrt{4L}e^{i\phi_{11,02}} & \sqrt{1-4L}e^{i\phi_{02}} \\
		\end{array} \right] ,
		\label{UCPHASE}
	\end{equation}
	in the $\{ |00\rangle , |01\rangle, |10\rangle, |11\rangle, |02\rangle \}$ subspace. Here $\phi_{01}$, $\phi_{10}$ single qubit phases and the leakage $L$ is introduced, key parameters for the two qubit gate calibration we will define later in paragraph \ref{PulseShapeTwo}.
	Even in the case of flux-tunable transmons, the way to implement both CZ and SWAP gates may change. Besides the application of baseband flux-pulses \cite{Negirneac21}, another mechanism involves \emph{parametric RF flux pulsing} \cite{Caldwell18}. In the parametric RF flux pulsing, the stimulated Rabi oscillations can drive population between states $|11\rangle-|02\rangle$, and the phase accumulates as the population undergoes a cycle, obtaining a $CZ$ gate, i.e., a CPhase with a fixed phase of $\pi$. The frequency of the first qubit $\omega_{1}$ is set at its DC flux bias sweetspot $\omega_{1,ss}$, and driven away from it by an AC flux modulation: $\omega_{1}(t)=\omega_{1,ss}+A\cos{(\omega_{m}t+\theta_{m})}$, with $A$ and $\omega_{m}$ the amplitude and frequency modulation \cite{didier2018analytical}. However, the modulation frequency and phase can be tuned to bring into resonance whatever desired pair of states, depending on the effective coupling between energy levels, giving rise to stimulated Rabi oscillations that can drive population between energy levels, and phase accumulation, as the population undergoes a cycle, also obtaining $iSWAP$. In the case of parametric pulsing, the desired entangling phase $\phi$ of a CPhase can be calibrated by measuring the interaction as a function of the frequency and the duration of the flux pulse modulation near the resonance to estimate the appropriate gate parameters. Finally, Ramsey protocol allows for checking for $CZ$ gates implemented via parametric pulsing, featuring fidelities $>99 \, \%$ with a duration $\sim200 \, ns$ \cite{sung2021realization}.
	
	The critical drawback for the iSWAP and CPhase gates in flux-based two-qubit gates compared to all-microwave two-qubit gates is the introduction of flux noise channels, which causes a detrimental effect on the phase of the system.  However, one point of strength of parametrically driven flux-based gates is that under modulation around the DC sweetspot, the $1/f$ noise is dynamically decoupled. Also, for specific modulation amplitudes, where the average frequency is flat, the qubit becomes first-order insensitive to $1/f$ noise, allowing to define also AC sweetspots \cite{Hong20}. Moreover, even if baseband flux pulsing produces the fastest two-qubit gates ($\sim 30 -45 \; ns$), in most cases it is not the best candidate from an operative point of view, mainly for two reasons: the big displacement from the sweetspot caused by the pulse, which causes the qubit to be sensitive to flux noise with detrimental effects in the fidelity, and second the non-atomicity of the pulse which can cause the calibration to be hard.

	\paragraph{The role of pulse shape engineering in two-qubit gates.}
	\label{PulseShapeTwo}
	
	As for single-qubit gates, by engeneering the pulse shape allows for mitigating the leakage of quantum information and the impact of flux noise, obtaining better performances in terms of fidelity. In reference \cite{Martinis14}, the unipolar baseband pulse has been replaced by a fast-adiabatic pulse. In reference \cite{Rol19}, a zero-average bipolar flux-pulsing method, known as \emph{net zero} (NZ) pulse, has been introduced. The NZ pulse consists of two back-to-back unipolar pulses of half the duration ($t_{pulse}/2$) and opposite amplitude $\pm A$ and a rising and falling sharpness $\lambda$. The first strong flux pulse on the higher frequency qubits moves $|11\rangle$ into the ALC with $|02\rangle$, and than the second opposite pulse moves it back in place, allowing the system to acquire conditional phase $\phi_{2Q}$, defined as $\phi_{2Q}=\phi_{11}-\phi_{01}-\phi_{10}$. Given this shape, the NZ pulse presents a built-in echo effect which protects against low-frequency flux noise \cite{Rol19}. 
	
	In fact, the flux symmetry of the transmon hamiltonian makes the phases acquired by the pulsed qubit first-order insensitive to low-frequency flux noise, increasing fidelity compared to a unipolar pulse, while the zero average characteristic makes NZ insensitive to long-time scale distortions that originate in the flux-control line from limited waveform-generator bandwidth, bias tees, filters, impedance mismatches, couplings, and so on. Real-time precompensation due to the elimination of the DC component of the pulse \cite{johnson2011} has been the key to achieving high fidelities, indicating that the correct implementation of flux pulses requires prior knowledge of the latency and distortions of the flux control originating. We refer to reference \cite{Rol2020} for further details on specific distortion correction mechanisms.
	
\begin{figure}[h!]
		\centering
		\begin{subfigure}{0.52\textwidth}
			\centering
			\includegraphics[width=\linewidth]{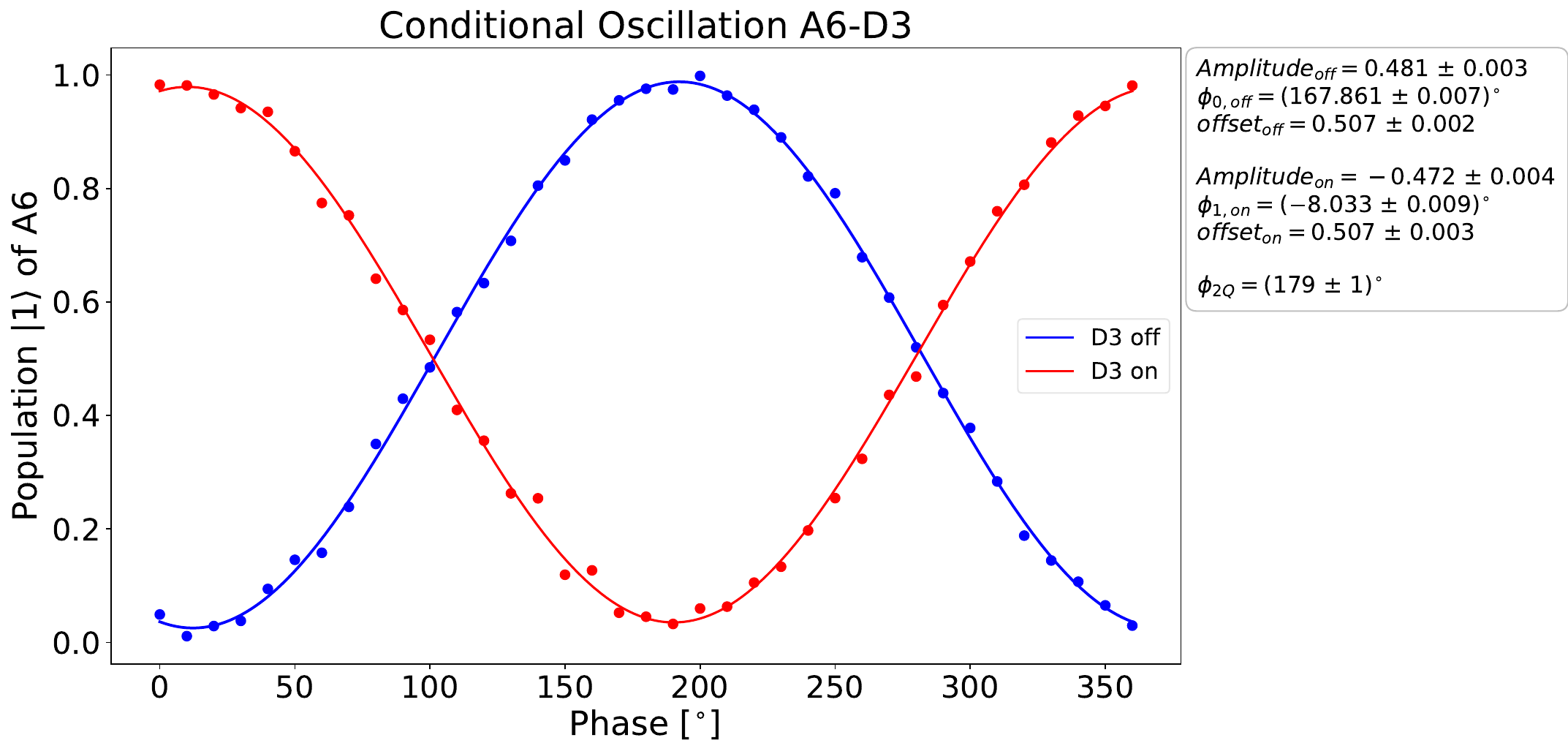}
			\caption{}
			\label{}
		\end{subfigure}
		\hfill
		\begin{subfigure}{0.4\textwidth}
			\centering
			\includegraphics[width=\linewidth]{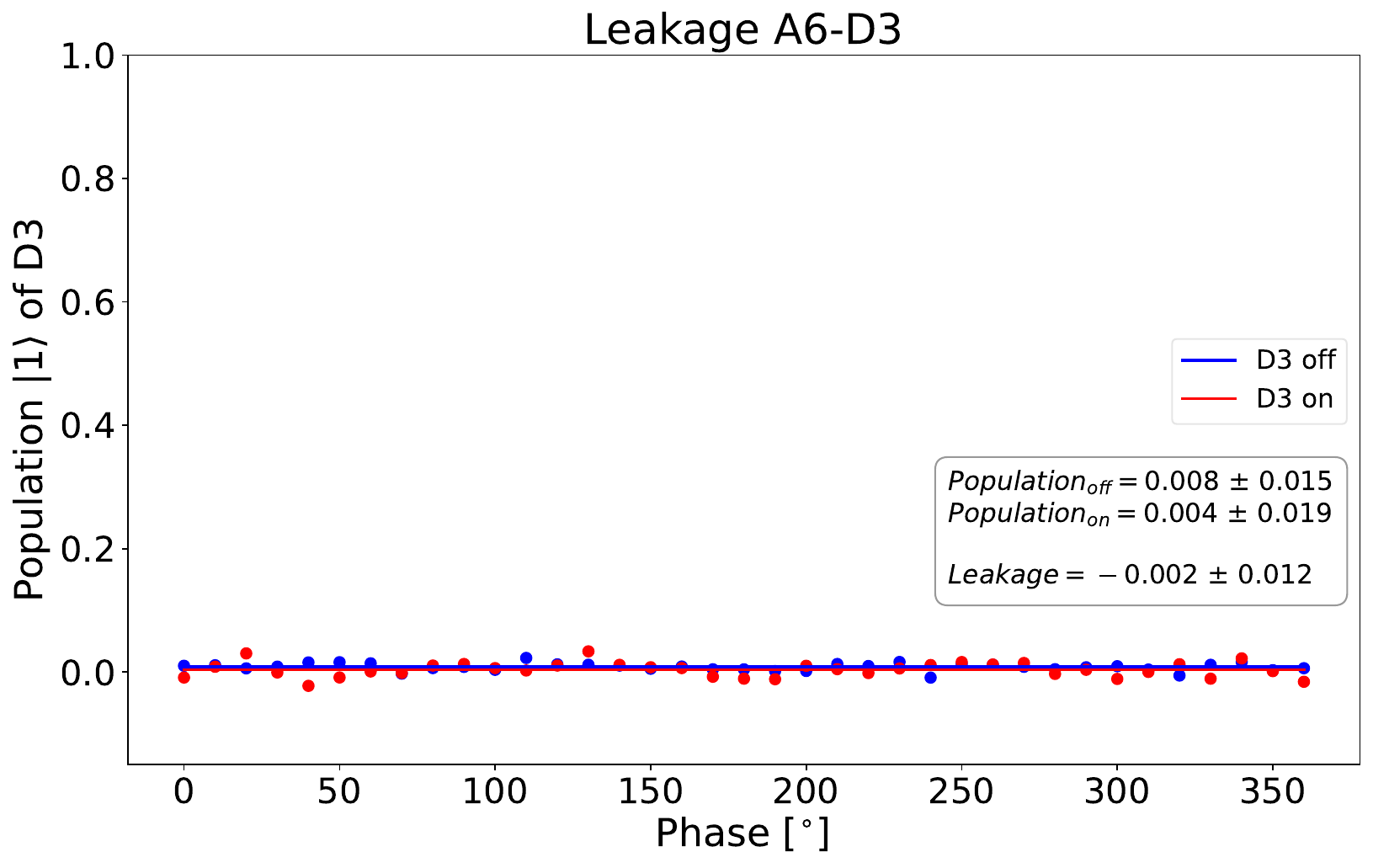}
			\caption{}
			\label{}
		\end{subfigure}
		\caption{Conditional oscillations experiment on A6-D3 register (see the supplementary material for more information on processor connectivity): in {\bfseries a} the control qubit population as function of the applied phase where oscillations shift rigidly of $180^{\circ}$ between the ``on" and ``off" configuration; in {\bfseries b} the target qubit $|1\rangle$ population as function of the applied phase, it appears to be near-zero in both ``on" and ``off" configuration, meaning the leakage $L$ is ideal.}
		\label{CO}
	\end{figure}

	To implement an optimal $CZ$ gate, the pulse shaping is then intended to minimize the leakage $L$, while obtaining a conditional phase $\phi_{2Q}=180^{\circ}$, and single-qubit phases $\phi_{01}=\phi_{10}=0^{\circ}$ \cite{Negirneac21}. The key protocol used for the calibration is the conditional oscillation experiment, which consists of a Ramsey-like experiment in which $\phi_{2Q}$ and $L$ are estimated. We define the gate $R_{XY}(\theta,\varphi)$ as a rotation of $\theta$ degrees around an axis in the equatorial plane of the Bloch sphere, tilted by an angle $\varphi$ with respect to the $x$ axis, i.e., $R_{XY}(\theta,0^{\circ})\equiv RX(\theta)$ and $R_{XY}(\theta,90^{\circ})\equiv RY(\theta)$.  The protocol involves the comparison of two variants: the first variant, denoted as ``off" consists in preparing $|00\rangle$ state, then performing a $R_{XY}(90^{\circ},\varphi)$ gate with $\varphi=0^{\circ} or \, 90^{\circ}$ on the the ``control" qubit, followed by a $NZ$ flux pulse on the pulsed qubit, and then again a $R_{XY}(90^{\circ},\varphi)$ gate for different phases $\varphi\in [0^{\circ},360^{\circ}]$. The outcome is the measurement of $|1\rangle$ state population for both qubits and for different gate phases $\varphi$. The second variant, denoted as ``on", consists of the addition of a $X$ gate on the ``target" qubit, before and after the implementation of the flux pulse, allowing for the $CZ$ to occur. From these experiments, it is possible to estimate $\phi_{2Q}$, $\phi_{01}$, $\phi_{10}$ and $L=(p_{|1\rangle,on}-p_{|1\rangle,off})/2$, with $p_{|1\rangle}$ the target qubit population measured after the two variants experiments. An example is reported in figure \ref{CO}, where the leakage is ideally negligible.
	\begin{figure}[h!]
		\centering
		\includegraphics[width=\linewidth]{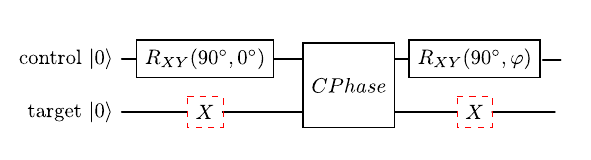}
		\caption{Conditional Oscillation, in red the gates added only in the ``on" version.}
	\end{figure} 
	
\begin{figure}[h!]
		\centering
		\includegraphics[width=0.8\linewidth]{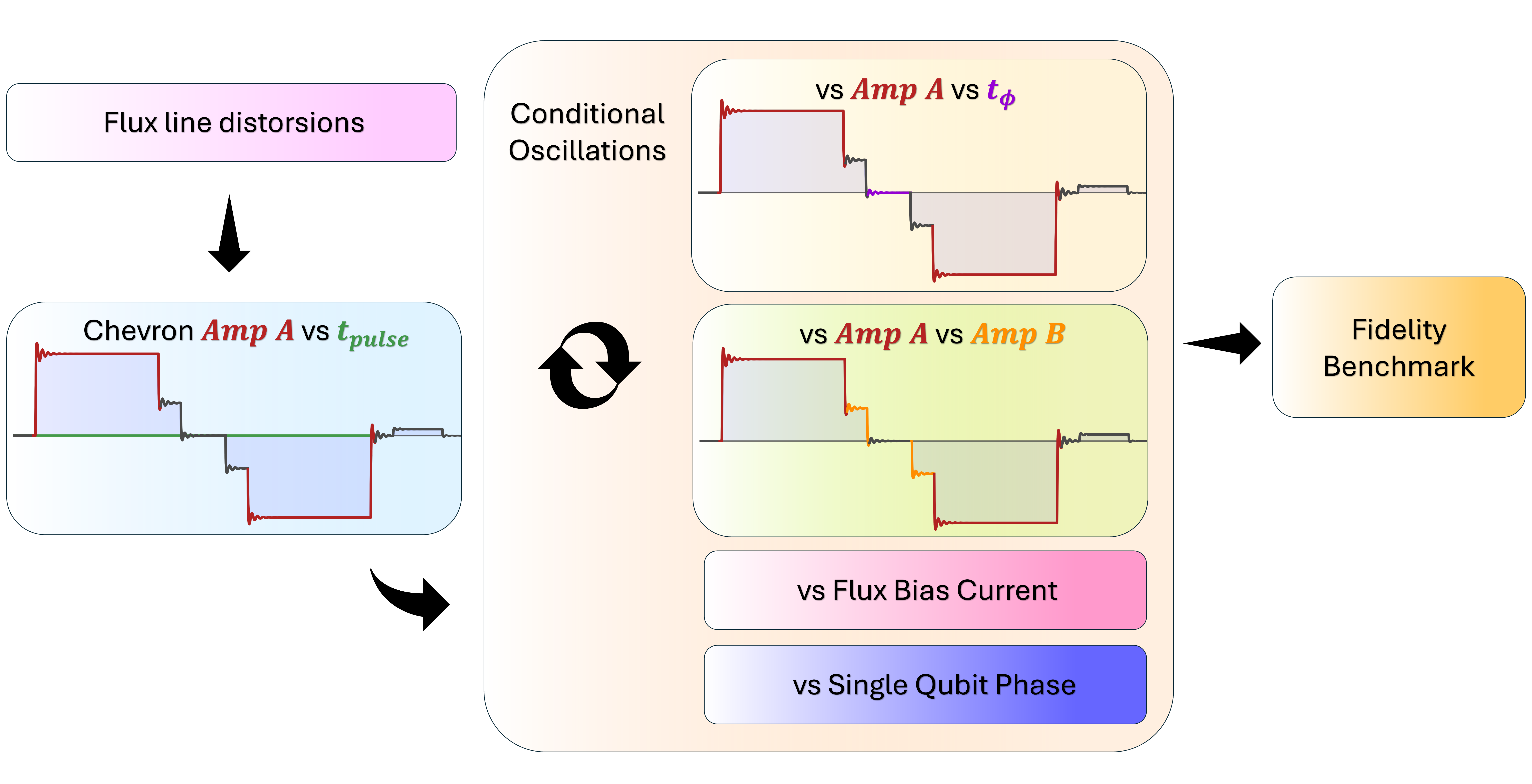}
		\caption{Calibration workflow for SNZ: after correction of flux line distortions via flux latency and cryoscope protocols, a first chevron experiment is performed to identify a raw value for the amplitude $AmpA$ and the duration $t_{pulse}$. Therefore, Conditional Oscillation as a function of $AmpA$ and $t_{\phi}$ and as a function of $AmpA$ and $AmpB$ allow to retrieve information through landscapes of conditional phase and leakage.
		}
		\label{two-qubit_cal}
	\end{figure}

\begin{figure}[h!]
		\centering
		\includegraphics[width=0.4\linewidth]{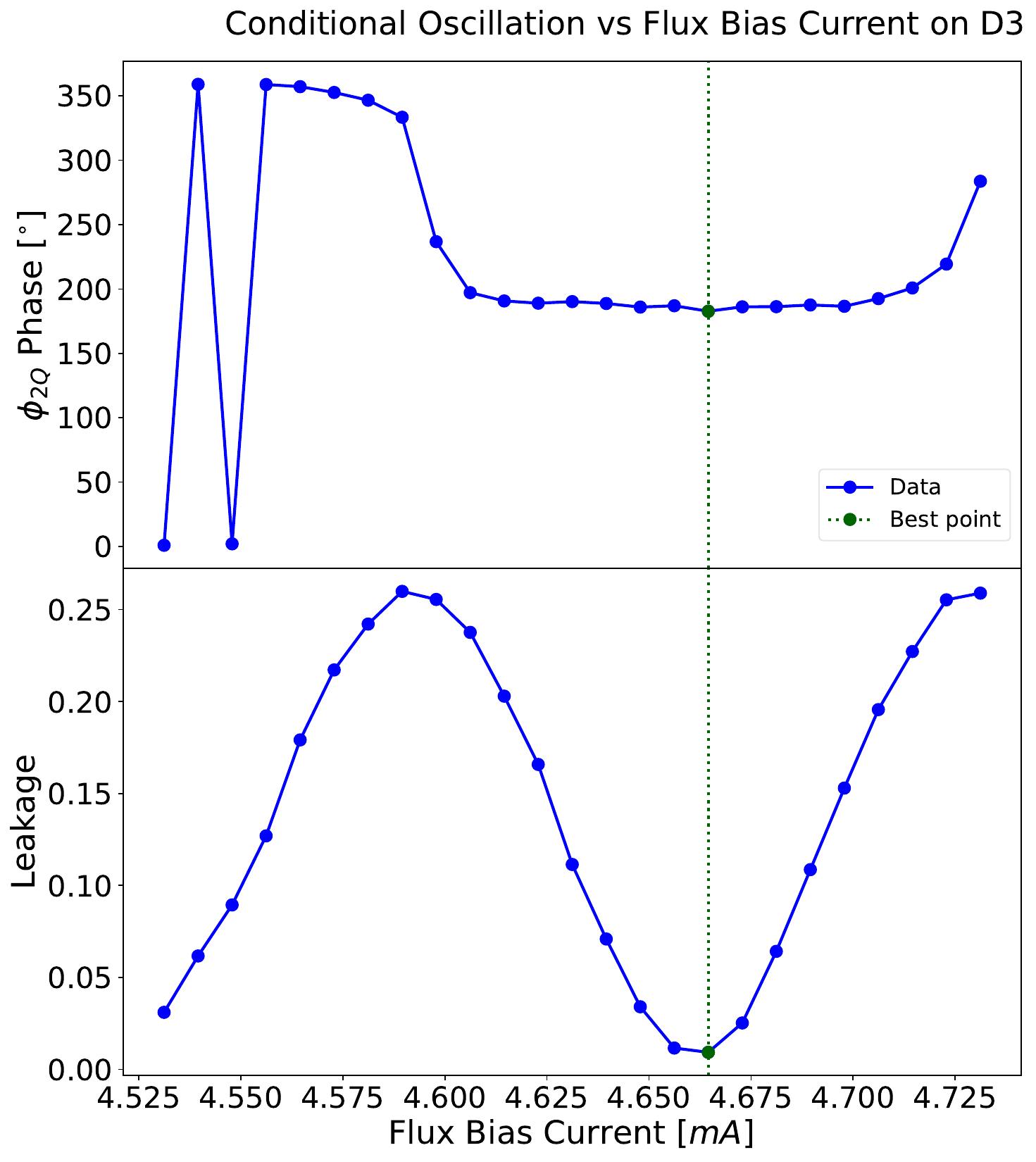}
		\caption{Conditional oscillation on D3-C4 register vs flux bias current on pulsed qubit D3.}
		\label{flux-co}
	\end{figure}

	In reality, obtaining a CZ gate with both $180^{\circ}$ phase and minimal leakage comes after careful calibration of the flux-pulse parameters. In the calibration protocol proposed in \cite{Rol19}, the conditional oscillation are performed as function of pulse rising and falling sharpness $\lambda$, and the flux-dependent angle $\theta=\arctan \left( \frac{2J'}{\epsilon}\right)$, where $\epsilon=\omega_{|02\rangle}-\omega_{|11\rangle}$ with $\theta=\frac{\pi}{2}$ at $\epsilon=0$, resulting in landscapes in which the best configuration can be obtained. However, due to the complex dependency of these parameters on the conditional phase and leakage, the calibration process becomes particularly challenging. For this reason, in reference \cite{Negirneac21}, the \emph{sudden net zero} (SNZ) pulse variant is introduced. As for the NZ pulse, the SNZ CZ gate is realized with two square half pulses with equal and opposite amplitude $\pm A$ and duration $t_{pulse}/2$ each: the difference resides in the presence of an additional idling time $t_{\phi}$ between the half pulses which allows for an accrual of the relative phase between $|11\rangle-|02\rangle$ states. Moreover, the amplitude parameter $\pm B$ is introduced to provide fine phase control beyond the discrete timing resolution of the AWG, enabling more precise destructive interference of intermediate leakage and thereby improving the symmetry of the SNZ pulse. This variant preserves the built-in echo mechanism and the destructive leakage interference, while providing the advantages of operating at the transverse-coupling speed limit by intentionally maximizing transient leakage into the non-computational state and offering a simpler tune-up procedure. As a matter of fact, the landscapes of the conditional phase $\phi_{2Q}$ and leakage $L$ as functions of these new pulse parameters have regular structures from which the best combination is easily found, and provides a practical calibration protocol. One possible calibration workflow for CZ gates in flux-tunable fixed-frequency coupler architectures employing SNZ flux pulses is reported in figure \ref{two-qubit_cal}, and follows reference \cite{Negirneac21} in most of its steps. Additionally, conditional oscillations experiments also allow fine-tuning of the flux sweet spot of the control qubit. As reported in figure \ref{flux-co}, the leakage follows an oscillatory behavior with the external flux. A minimum of leakage occurs for a conditional two-qubit phase of nearly $180^{\circ}$. By sweeping the external field close to this minimum (sensitivity of a few microampere on the current through the flux line), one can achieve leakages of the order of $<1\, \%$. 

	\paragraph{Two-qubit gates with tunable couplers.}
	
	Finally, another possible way to implement two-qubit gates in superconducting architectures is to exploit tunable couplers, rather than fixed coupling schemes. Tunable couplers are superconducting circuit elements whose effective coupling strength can be dynamically controlled through an external magnetic flux \cite{yan2018}. Rather than directly coupling neighboring qubits with a fixed interaction, then, the coupler acts as an intermediate nonlinear element that enables the interaction to be activated, suppressed, or even sign-reversed on demand. Tunable coupling architectures also use flux-pulsing to control the effective exchange interaction between two qubits, i.e., generating coherent population transfer between the $|10\rangle$ and $|01\rangle$ states. This allows for implementing SWAP-like operations, including partial exchanges such as the iSWAP and $\sqrt{\mathrm{iSWAP}}$ gates, or to enhance the state-dependent interaction between the computational states and the higher excited levels of the coupled system, producing a conditional phase accumulation that realizes a CZ gate. In this case, the interaction is engineered so that one computational basis state undergoes an additional dynamical phase while population leakage is minimized through careful pulse shaping and adiabatic control. The ability to dynamically tune the interaction strength allows high-fidelity entangling gates while simultaneously suppressing residual couplings during idle periods, thereby mitigating crosstalk and frequency-collision constraints in large-scale sQPU.
	
	Among the big players that use tunable coupling schemes, we mention Google and Rigetti, and several different circuit designs have been proposed in time to further enhance the performance of two-qubit gates \cite{Chow2012,yan2018,Foxen20}. What is more relevant is that also in this case flux-pulsing is essential, and some calibration procedures are shared with fixed-coupling flux pulses calibration previously mentioned. Examples can be seen in \cite{Chen14}, which proposed the \emph{Gmon} planar architecture, composed of two Xmon coupled via a fast tunable inter-qubit coupler, allowing for protection from the frequency crowding arising from fixed coupling. In this architecture, the authors implement an alternative efficient two-qubit gate, the $fSim(\theta,\phi)$ gate, 
	\begin{equation}
		fSim(\theta,\phi)=\left[
		\begin{array}{cccc}
			1 & 0 & 0 & 0 \\
			0 & \cos{\theta} & -i\sin{\theta} & 0 \\
			0 & -i\sin{\theta} & \cos{\theta} & 0 \\
			0 & 0 & 0 & e^{-i\phi}
		\end{array} \right],
		\label{CZmatrix}
	\end{equation}
	which combines CPhase and SWAP operations into one smooth process \cite{Foxen20}, and is useful for fermionic simulations in chemistry problems \cite{Arute2020HartreeFock}. However, the multitude of circuit schemes, entanglement control, and calibration workflow procedures do not allow for a generalized framework, complicating the picture of a hardware-agnostic quantum computing approach. 
	
	\section{Toward gate validation: application-specific gate benchmarking}
	
	Accurate characterization of quantum operations is also essential for assessing processor performance, validating hardware improvements, and estimating the resources required for fault-tolerant quantum computing \cite{Koch2026}.  
	
	A common strategy is to represent a processor's performance benchmarking through coherence and fidelity metrics. Such parameters are evaluated in extreme conditions: qubits in the isolated regime, or few at a time, soon after a calibration run has been performed, or where only single- and two-qubit gate operations are the main target. Evaluating fidelities and coherence in parallel settings over multiple qubits \cite{Gambetta2012}, at run-time during an algorithm \cite{Majumder2020, Proctor2022}, or including in the benchmarking set complicated multi-qubit gates, not only able to exploit standard quantum resources such as superposition and entanglement, but also more complex features like the non-stabilizerness in non-Clifford gates \cite{Oliviero_Leone_Hamma_Lloyd_2022}, is unusual. Most importantly, to preserve data memory and fast determination of such metrics, smart protocols have been proposed in time. 
	
	\emph{Randomized Benchmarking} (RB) protocols, for example, have been developed to efficiently estimate gate fidelities, or the average error rate of quantum gates, while remaining largely insensitive to state-preparation and measurement (SPAM) errors and with a reduced overhead cost of resources and experimental measurements \cite{Hashim2025}. RB techniques form a hierarchy of increasingly sophisticated characterization techniques, progressing from average gate-set performance to individual gate fidelities, and finally to the benchmarking of complete computational cycles in multi-qubit processors. The first one developed historically is the \emph{Standard RB}, which consists of applying random sequences of Clifford gates of increasing length, followed by an inversion gate that ideally returns the system to its initial state. The survival probability exhibits an exponential decay as a function of sequence length, from which the average gate fidelity can be extracted. However, standard RB provides only an average error metric over the entire gate set and does not directly identify the fidelity of individual gates \cite{Hashim2025}. \emph{Interleaved RB} extends the RB protocol to characterize a specific target gate, which is inserted between successive random Clifford operations. Many other declinations of RB techniques have been developed in time, e.g., \emph{cycle} or \emph{mirror RB}, and a comprehensive review can be found in reference \cite{Hashim2025}. 
	
	However, standard RB and IRB gate quality metrics do not necessarily predict algorithmic performance in realistic quantum computations, and the interpretation of RB results requires caution.  Despite its robustness and scalability, RB can mask several important error mechanisms that are relevant for practical quantum computations. RB provides an estimate of the average gate fidelity by analyzing the exponential decay of survival probabilities over random gate sequences \cite{Hashim2025}. The protocol relies on the assumption that the randomization effectively twirls the underlying noise channel, transforming it into an equivalent depolarizing process characterized by a single decay parameter. Under this approximation, the measured decay can be directly related to the average error rate of the implemented gates. Since RB reduces the effect of noise to an average decay parameter, coherent unitary errors such as systematic over- or under-rotations are often randomized into an apparently stochastic contribution, potentially underestimating their algorithmic impact. Similarly, leakage outside the computational subspace may not be accurately captured by standard RB models and can lead to misleading fidelity estimates if not explicitly monitored.

	A natural solution is to fully characterize a universal gate set by \emph{standard quantum process tomography} (QPT)\cite{Nielsen2021}. QPT consists of a state tomography over all bases of a quantum process through the application of an overcomplete set of analysis pulses \cite{Chow2012}.

	Considering $\rho$ as the density matrix, for a pure state $\rho = |\psi\rangle \langle\psi|$ and for a mixed state $\rho = \sum_{i} p_{i}|\psi_{i}\rangle \langle\psi_{i}|$, where $p_{i}$ represent the classical probabilities of an ensemble $\{\vert{}\psi_{i}\rangle\}$. Any density matrix of an n-qubit system can be expanded in the basis of Pauli operators
	$P_{k}$. For a single qubit system, this basis is made of the four Pauli matrices $\{I,X,Y,Z\}$; for a two-qubit system, the basis is constructed from the $16$ tensor products of these matrices (e.g., I $\otimes$ I, I $\otimes$ X, X $\otimes$ I, etc.), and this principle extends to n-qubit systems, resulting in a basis of $4^{n}$ operators.
	Since the density matrix can be written as:
	\begin{equation}
		\rho =\frac{1}{d} \sum_{k=0}^{d^{2}-1} p_{k}P_{k},
	\end{equation}
	where the coefficients $p_{k}=Tr(P_{k}\rho)$ are the real-valued expectation values of the Pauli operators, the vector $\vec{p}$ with components $p_{k}$ is the Pauli vector representation of the state $\rho$. 
	
	A general quantum process $\mathcal{E}$ on a finite d-dimensional Hilbert space $H_{d}$ ($d = 2n$ for qubit systems) is a completely positive trace-preserving (CPTP) map \cite{Merkel13} that transforms an input density matrix $\rho_{in}$ into an output matrix $\rho_{out}= \mathcal{E}(\rho_{in})$.

	In the Pauli basis, this action corresponds to a linear transformation on the Pauli vectors \cite{Chow2012}:
	\begin{equation}
		\vec{p}_{out}=M\vec{p}_{in},
		\label{pout}
	\end{equation}
	where $M$ is the \emph{Pauli transfer matrix} (PTM), a $4^{n}\times4^{n}$ real matrix that completely characterizes the quantum process.
	Knowing the action of the map, the elements of $M$ are:
	\begin{equation}
		M_{i,j}=\frac{1}{d}Tr[P_{i}\mathcal{E}(P_{j})].
	\end{equation}

	Two-qubit QPT is, instead, accomplished through the compilation of an overcomplete set of $36 \times 36$ measurements. Therefore the state tomography as described above is performed for the 36 different input states generated by $\{I, X_{\pi}, X_{\pm\frac{\pi}{2}}, Y_{\pi}, Y_{\pm\frac{\pi}{2}}, Z_{\pi}, Z_{\pm\frac{\pi}{2}}\}^{\otimes2}$. Here, $M$ maps an input Pauli state vector $\vec{p_{in}}$ to an output Pauli state vector $\vec{p_{out}}$, $\overrightarrow{p}_{out}= M\cdot\overrightarrow{p}_{in}$.

	The structure of the PTM provides a direct method for calculating the gate fidelity, $F_{g}$, a crucial metric that quantifies how close an experimental process $\mathcal{E}$ is to the ideal unitary operation $U_{ideal}$, averaged over all possible input pure states.  
	The fidelity can be calculated directly from the experimental and ideal Pauli transfer matrices, $M_{exp}$ and $M_{ideal}$, using the formula: 
	\begin{equation}
		F_{g}=\frac{Tr[M_{ideal}^{T}M_{exp}]+d}{d^{2}+d},
		\label{FidelityTomo}
	\end{equation}
	where $d=2^{n}$ is the Hilbert space dimension for $n$ qubits.
	
	A major bottleneck in validating multi-qubit processors through process tomography, although elegant and complete, is the unscalable overhead of self-consistent characterization techniques. While one- and two-qubit characterizations are routinely performed to calibrate gate sets, extending these frameworks to larger subsets of qubits rapidly becomes computationally and experimentally intractable, as can be seen for the extensive three-qubit iToffoli gate PTM in the supplementary material of \cite{Kim22}. Indeed, while the dimensionality of the density matrix ($\rho$) for an $n$-qubit system scales as $4^n$, characterizing a quantum operation via QPT requires determining an associated PTM or superoperator: because this superoperator maps a $4^n$-dimensional operator space onto itself, the resulting process matrix is of size $4^n \times 4^n$. Consequently, this squaring of the operator space means that the number of independent, a priori unknown parameters for a single global gate or cycle scales as $\mathcal{O}(16^n)$. This extreme exponential growth restricts standard, full-vehicle QPT to at most three qubits, driving the recent paradigm shift toward alternative scalable tomographic frameworks designed to handle multi-qubit architectures. Among the others, linearized, localized, or shadow tomography \cite{Gu21, Chen21, Vinas2025}, where hybrid computational approaches can also come in hand. For example, machine learning methods have been proposed to improve qubit characterization, device calibration, quantum control, and noise-aware compilation in superconducting architectures \cite{Huang2021}.  
	
	In summary, average gate fidelities provide a convenient estimate of the overall performance of a quantum operation, but they offer limited insight into the specific error mechanisms affecting the implemented gate. QPT provides a complete characterization of the gate, since it reconstructs the full quantum channel and enables a detailed analysis of coherent and incoherent error sources, but the experimental overhead associated with QPT grows rapidly with system size, making it impractical as a routine benchmarking tool for large-scale sQPU.
	
	However, sometimes one may not want to reconstruct the underlying quantum channel but rather to assess the impact of different gate decompositions on the execution of a target quantum circuit. An alternative method is to adopt a truth-table-based analysis. This method consists of applying the sequence of quantum gates to a quantum register and performing single-shot readout measurements in the computational basis states set.
	
	The readout of superconducting transmon qubits is commonly implemented within the circuit quantum electrodynamics (cQED) framework through dispersive measurement techniques \cite{Krantz19}. In the dispersive regime, where the qubit-resonator detuning significantly exceeds their coupling strength, the interaction induces a qubit-state-dependent shift in the resonance frequency of the readout resonator. Consequently, the quantum state of the transmon can be inferred by probing the amplitude and phase response of a microwave signal transmitted through or reflected from the resonator. This approach enables high-fidelity, quantum non-demolition (QND) single-shot measurements of the computational states. For one qubit, the computational basis set is $\{|0\rangle$, $|1\rangle\}$. For multi-qubit registers, particularly when several qubits are coupled to readout resonators connected to a common feedline, simultaneous qubit measurement is typically achieved through frequency-division multiplexed readout \cite{Stasino2025}. In this approach, a multi-tone microwave signal, with each tone resonant with a specific readout resonator, is injected into the shared feedline, enabling the concurrent readout of multiple qubits. Such simultaneous measurement capability is essential for accurately discriminating the quantum state of entangled multi-qubit systems and for characterizing correlated qubit dynamics, such as those involved in the calibration of two-qubit gates including the $CZ$ gate. 
	
	The readout measurement procedure consists of repeatedly performing single-shot measurements to calibrate the state-dependent response distributions associated with the computational states $|0\rangle$ and $|1\rangle$ of each qubit. Thresholds are then defined independently for each readout channel to discriminate the corresponding qubit states. Finally, the conditional probabilities associated with the quantum circuit measured outcomes are evaluated to reconstruct and identify the complete multi-qubit state of the quantum register, thereby enabling high-fidelity characterization also of entangled states while preserving the scalability advantages of multiplexed architectures \cite{Stasino2025}.
	
	The truth-table approach sacrifices the completeness of tomographic methods but offers a practical application-level benchmark, capturing the cumulative effect of calibration imperfections, coherent control errors, and decomposition choices on the observable computational result. In a sense, tomographic-based benchmarking answers to the question: \emph{how well can we implement a quantum process?} The truth-table answers to a different question: \emph{how well can we decompose a quantum circuit to solve the problem for which it has been designed?} These are two different goals that earn equal importance, though one must be aware of the underlying differences.
	
	Towards practical examples, we here aim to provide an informed perspective and technical details on how to construct the most important single- and two-qubit gates for quantum computing, outlining, whenever relevant, the strategies pursued to implement them efficiently. We compare the QPT and the truth-table methods on two- and three-qubit registers of a Contralto processor from Quantware at the UNINA Superconducting Quantum Computing Center (see supplementary material for more details), giving as metric the gate fidelity $F$ in equation  \ref{FidelityTomo} and the Hellinger fidelity $F_{H}$\cite{Aktar22}, respectively:
	\begin{align}
		F_H(p, q) &= \left( 1 - d_H^2(p, q) \right)^2 = \left( \sum_{i} \sqrt{p_i q_i} \right)^2 \\[10pt]
		d_H(p, q) &= \sqrt{1 - \sum_{i} \sqrt{p_i q_i}},
	\end{align}
	where $p_{i}$ represents the probability of the $i$-th outcome of the ideal distribution, $q_{i}$ the experimental one, and $d_H$ the Hellinger distance. The latter is bounded between $0$ and $1$, and quantifies the discrepancy between two discrete probability distributions. \newline

	\subsection{Single-qubit gate set}
	
	At the UNINA Superconducting Quantum Computing Center, the control electronics is a Qblox cluster, interfaced with Python packages to build quantum circuits and schedules at the pulse-level. Single-qubit gate calibration follows the workflow in figure \ref{fig:DragCal}: Ramsey interferometry for frequency calibration, Flipping experiment for amplitude calibration and Motzoi protocol for DRAG shape calibration. We extensively use the AllXY protocol \cite{Reed2013} to detect specific error syndromes to identify which calibration node should be performed in an informed way. Details on the software packages used to develop the calibration schedules can be found in supplementary material.
	
	\paragraph{The Hadamard gate.}
	
\begin{figure}[h!]
		\begin{subfigure}{0.33\textwidth}
			\centering        
			\includegraphics[width=\linewidth]{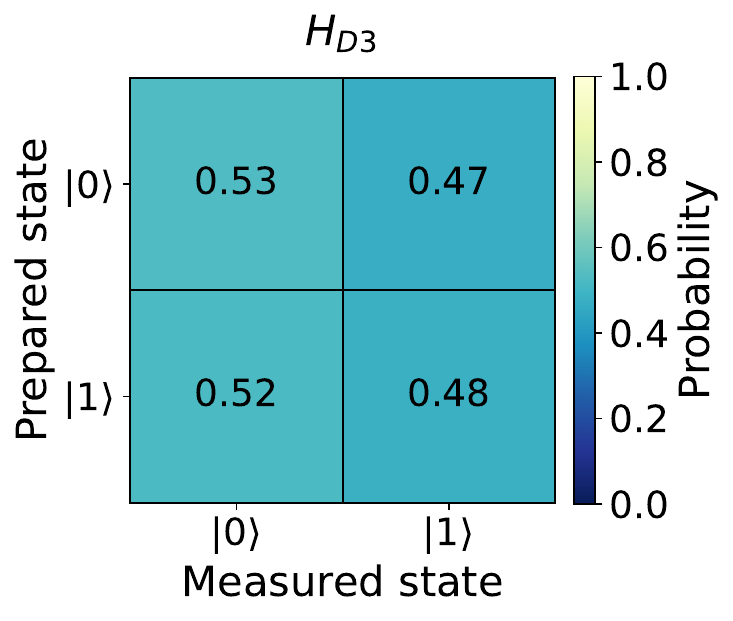}
			\caption{}
			\label{}
		\end{subfigure}
		\begin{subfigure}{0.33\textwidth}
			\centering
			\includegraphics[width=0.9\textwidth]{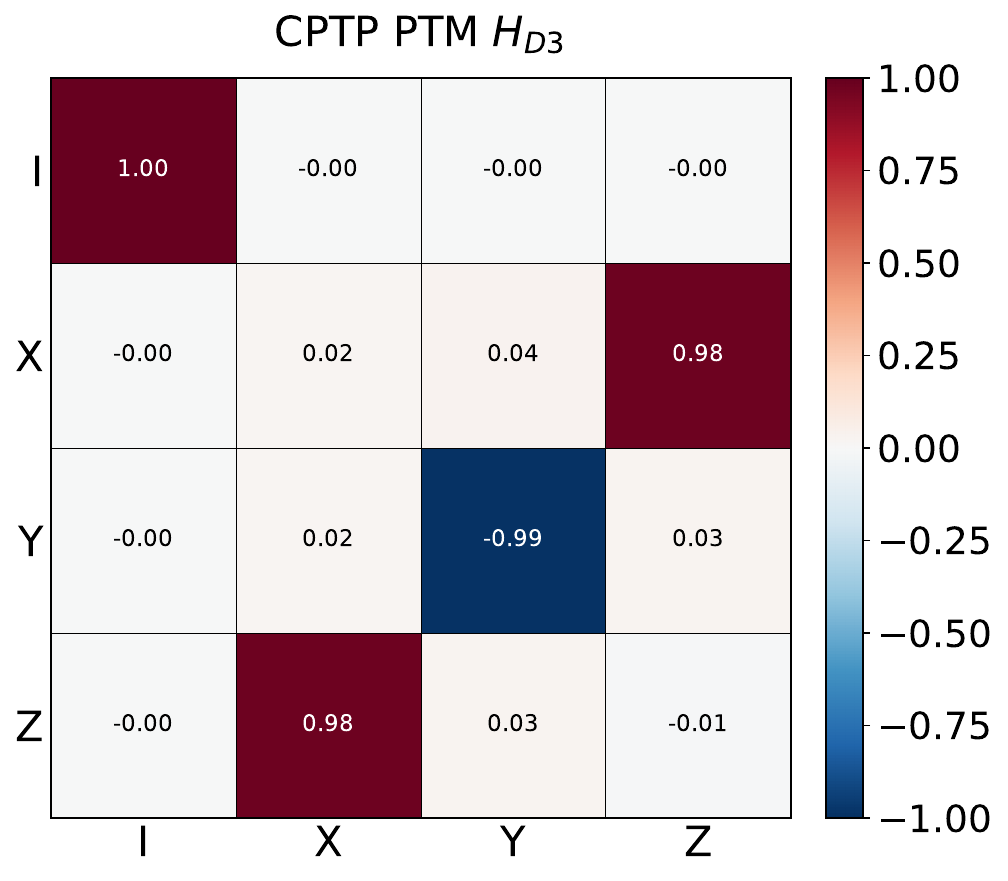}
			\caption{}
			\label{}
		\end{subfigure}
		\begin{subfigure}{0.33\textwidth}
			\centering
			\includegraphics[width=0.9\textwidth]{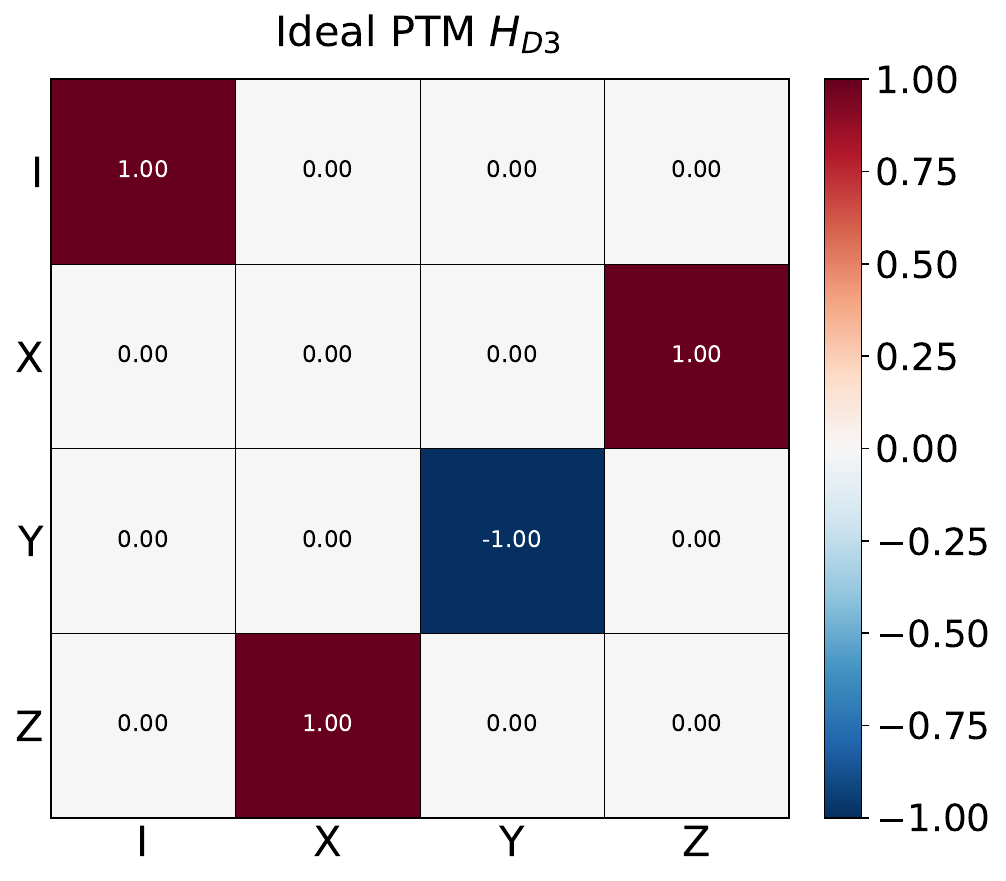}
			\caption{}
			\label{}
		\end{subfigure}
		\caption{In (a), Hadamard gate truth table with $F_{H}=(99.94\pm0.04)\, \%$; in (b), gate tomography with $F=(99.2\pm 0.2)\, \%$ in comparison with the ideal PTM matrix in (c). In (b) and (c), on x axis the input Pauli operator, on the y-axis the output Pauli operators.}
		\label{HadamardExp}
	\end{figure}

	The Hadamard gate is one of the most fundamental single-qubit gates, able to implement single-qubit state superposition. In equation \ref{HadamardEq} and figure \ref{Hadamard}, we show one possible decomposition for this gate built on top of native gates in flux-tunable transmon processors, which use $Y_{\frac{\pi}{2}}$ and $Z_{\frac{\pi}{2}}$ rotations. 
	As shown in figure  \ref{HadamardExp}a, the truth table shows the H gate inducing an equally distributed probability in $|0\rangle$ and $|1\rangle$ in either preparations with a $F_{H}=(99.94\pm0.04)\, \%$, which would be in principle the same result one can obtain through a $RX(90^{\circ})$ or $RY(90^{\circ})$ gate. The effective induced process can be captured by the gate tomography: figure  \ref{HadamardExp}b in the experimental PTM shows a good agreement with the theoretical one with a fidelity $F=(99.2\pm 0.2)\, \%$. \newline
	
\begin{figure}[h!]
		\centering
		\begin{subfigure}{0.4\textwidth}
			\centering
			\includegraphics[width=\linewidth]{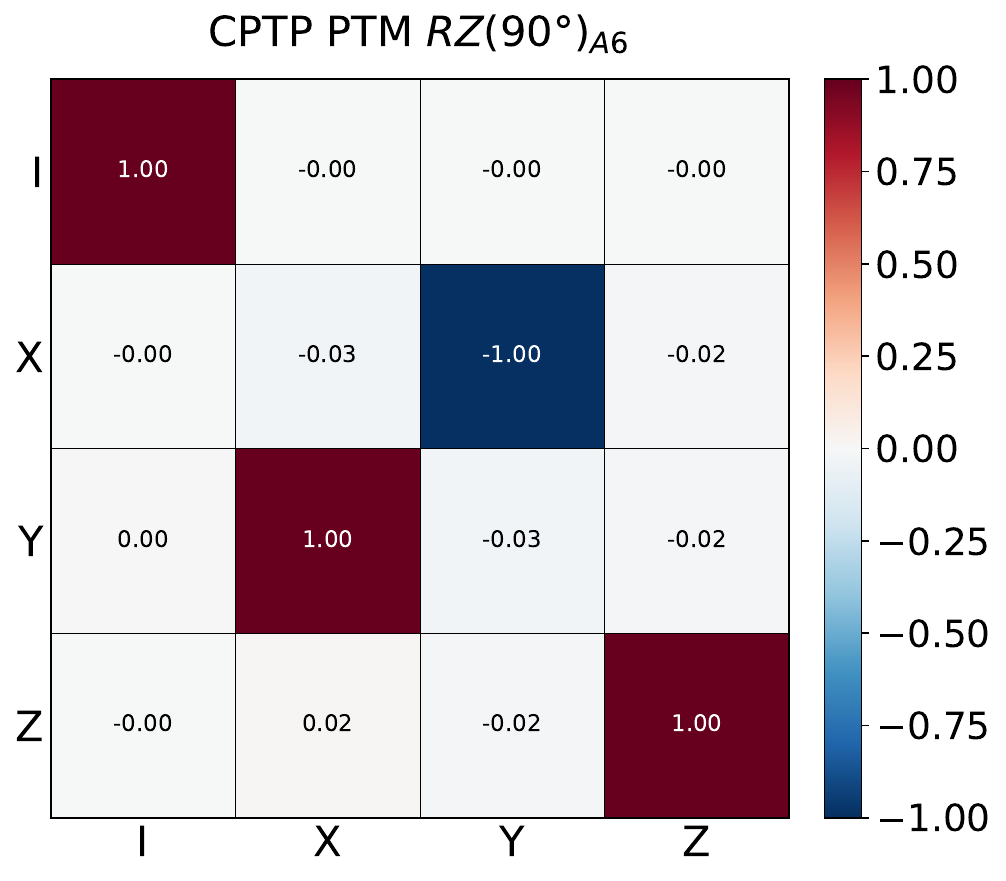}
			\caption{}
			\label{}
		\end{subfigure}
		\hfill
		\begin{subfigure}{0.4\textwidth}
			\centering
			\includegraphics[width=\linewidth]{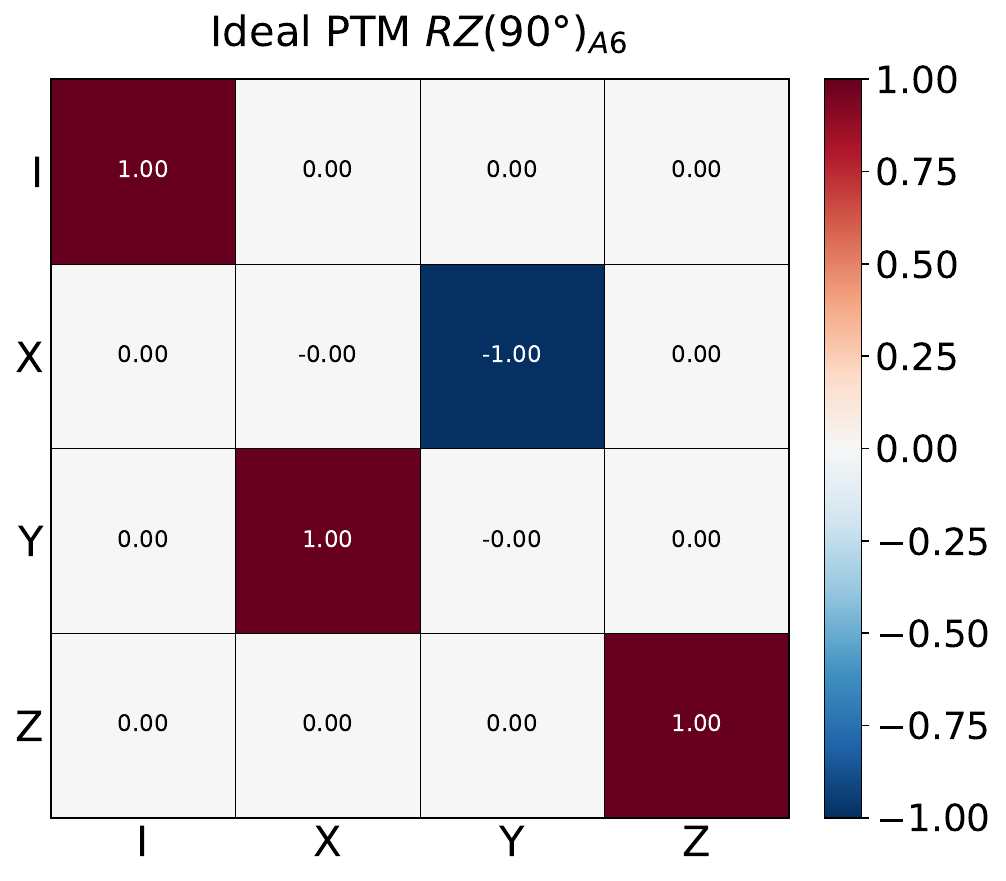}
			\caption{}
			\label{}
		\end{subfigure}
		\caption{Gate tomography of virtual-$RZ(90^{\circ})$ on $A6$: in {\bfseries a} experimental PTM with $F=(99.9\pm 0.2)\, \%$; in {\bfseries b} ideal PTM as comparison.}
		\label{Z90gate}
	\end{figure}

\begin{figure}[h!]
		\centering
		\begin{subfigure}{0.4\textwidth}
			\centering
			\includegraphics[width=\linewidth]{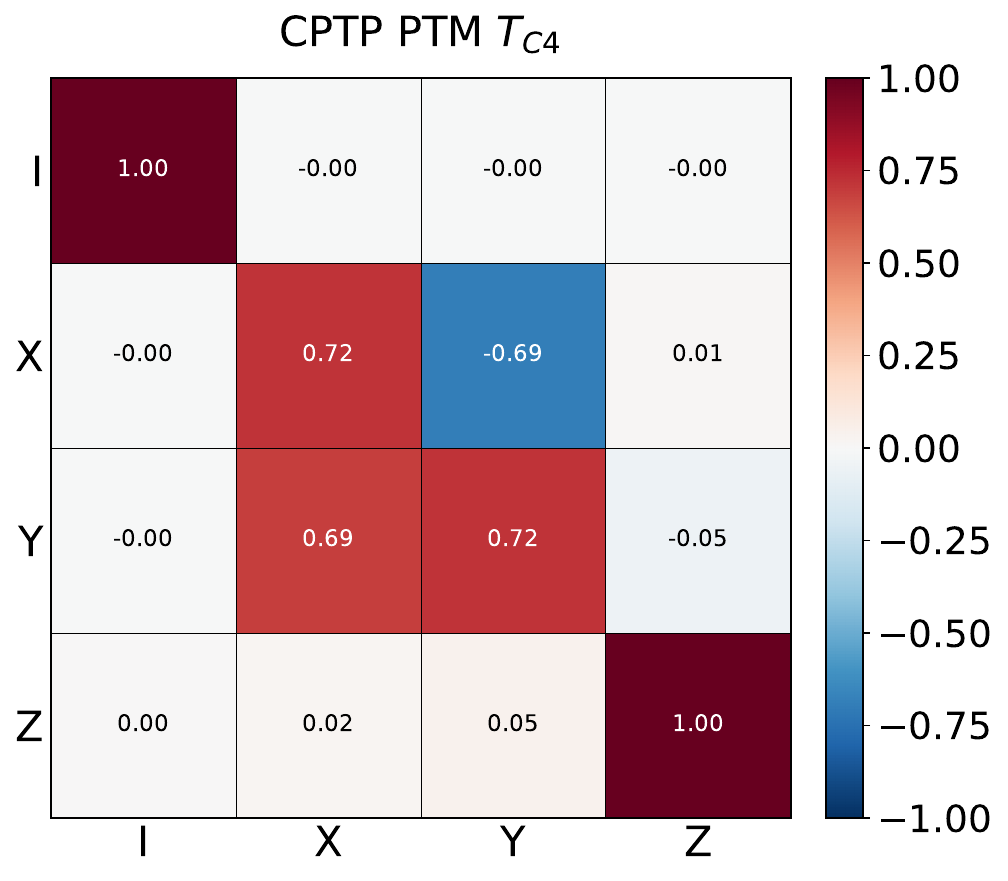}
			\caption{}
			\label{}
		\end{subfigure}
		\hfill
		\begin{subfigure}{0.4\textwidth}
			\centering
			\includegraphics[width=\linewidth]{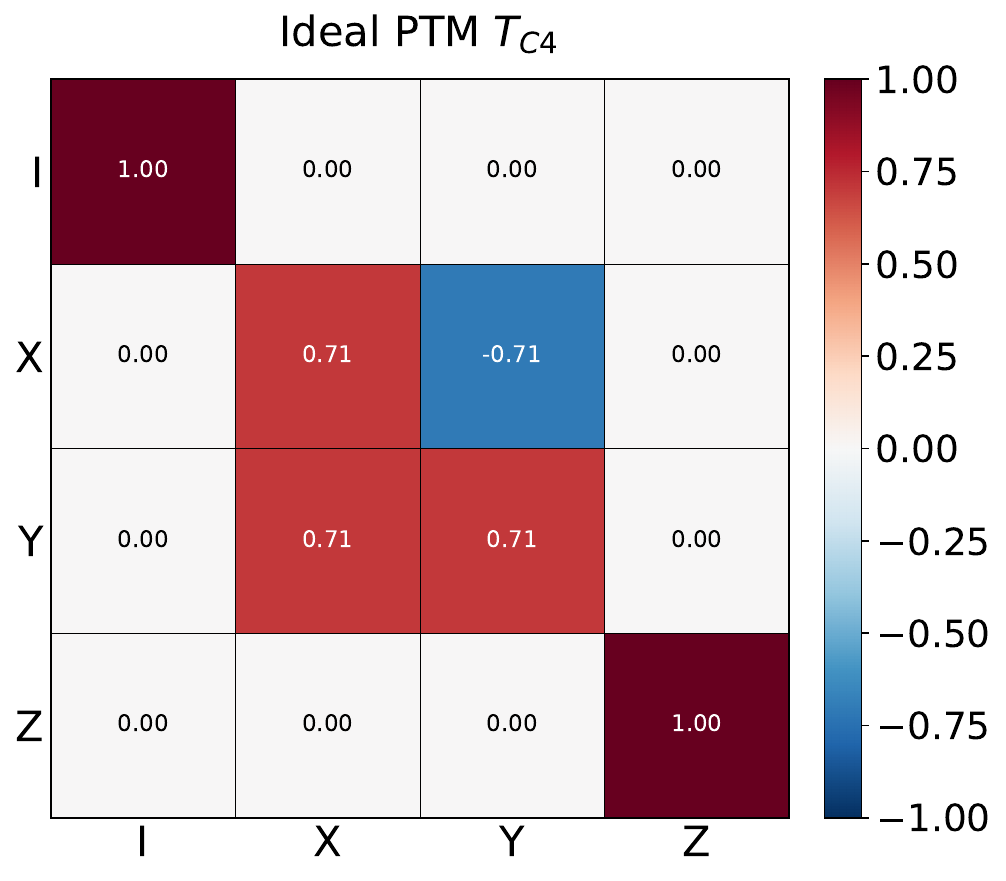}
			\caption{}
			\label{}
		\end{subfigure}
		\caption{Gate tomography of virtual-$RZ(45^{\circ})$, i.e., $T$ gate, on $C4$: in {\bfseries a} the experimental PTM with $F=(99.9\pm 0.7)\, \%$; in {\bfseries b} the ideal PTM as comparison.}
		\label{Z45gate}
	\end{figure}

	\paragraph{The virtual-Z gate.}
	The Qblox electronics in use at the UNINA Superconducting Quantum Computing Center implement virtual $Z$-gates via internal clock phase shifts, which influence subsequent operations and are not generally calibrated within the single-qubit gate workflow in figure \ref{fig:DragCal}. Because measurement occurs along the $z$ axis, this phase effect is not captured by a standard truth table. Therefore, quantum process tomography provides a more appropriate benchmark, demonstrating an effective implementation with near-perfect fidelity: as shown in figure \ref{Z90gate} the $RZ(90^{\circ})$, i.e. $S$ gate, exhibits $F=(99.9\pm 0.2)\, \%$, while in figure \ref{Z45gate} the $RZ(45^{\circ})$, i.e. $T$ gate, exhibits $F=(99.9\pm 0.7)\, \%$. Any remaining errors are primarily introduced by SPAM errors, specifically during the projection and de-projection operations within the tomography protocol itself. These gates, belonging to native gate-set of our hardware, are the building blocks for the implementation of other composite gates, like $H$, $CNOT$,  $CY$ and $Toffoli$, meaning that their flawless implementation decreases the number of noise sources to take care. As an example, $\theta=90^{\circ}$ is used in the Hadamard decomposition, and $\theta=45^{\circ}$, which is crucial to explore the non-stabilizerness in the magic distillation process \cite{ahmad2025_magic}, and is also used for the $Toffoli$ decomposition. \newline
	
	\paragraph{Two-qubit gates: from the CZ to the CNOT}
	
	After the accurate calibration in figure \ref{two-qubit_cal}, the CZ gate can be implemented and tested.
	As discussed in section \ref{PhysicalImplemetation}, the $CZ$ is implemented by pulsing the higher resonance frequency qubit on the pair to the lower one through a SNZ flux pulse inducing a symmetric global phase exchange of $180^{\circ}$.
	As for the $Z$ gate, the CZ benchmark can be done via gate tomography (shown in figure \ref{CZtomography}): the gate implemented on register $A6-D3$ of the processor (see the schematics in supplementary material) shows a fidelity $F=(94.1 \pm 0.5) \, \% $, compatible with the state-of-art values \cite{Yamamoto10}.
	This gate represents our native entangling gate, from which the other two- and three- qubit gates can be build: a faulty $CZ$ gate inevitably degrades the performance of all downstream composite gates. Therefore, achieving high $CZ$ fidelity is crucial, as it sets an upper limit on the fidelity of these derivative gates. Most of the errors arising from its physical implementation are due to flux-noise-induced dephasing, since the pulsed qubit accumulates an unintended phase when shifted away from the sweet-spot, as well as readout errors, which can be in future mitigated by implementing cryogenic TWPA amplifiers \cite{Esposito2021}. \newline
	
	\begin{figure}[h!]
		\begin{subfigure}{0.4\textwidth}
			\centering
			\includegraphics[width=\linewidth]{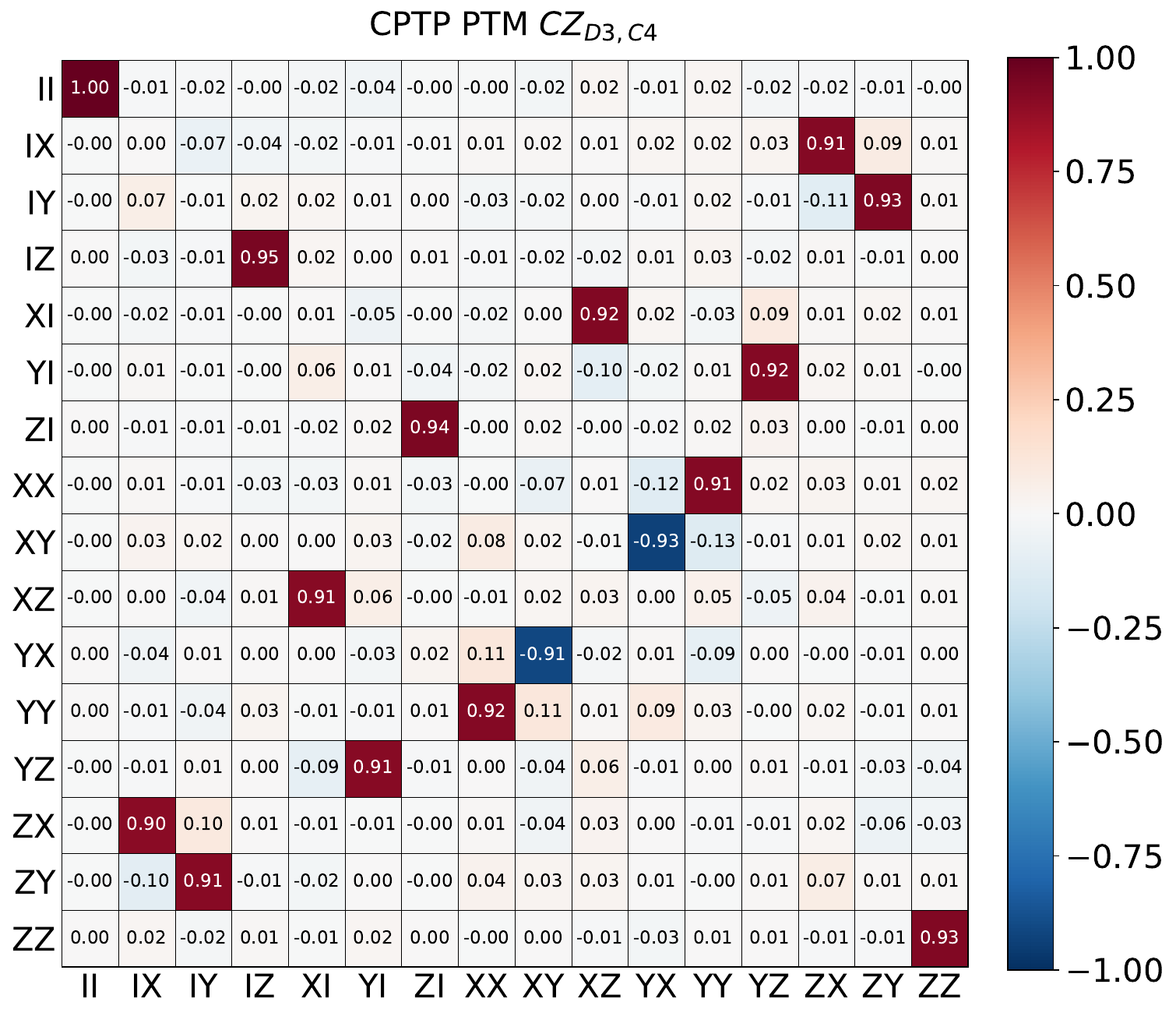}
			\caption{}
			\label{}
		\end{subfigure}
		\hfill
		\begin{subfigure}{0.4\textwidth}
			\centering
			\includegraphics[width=\linewidth]{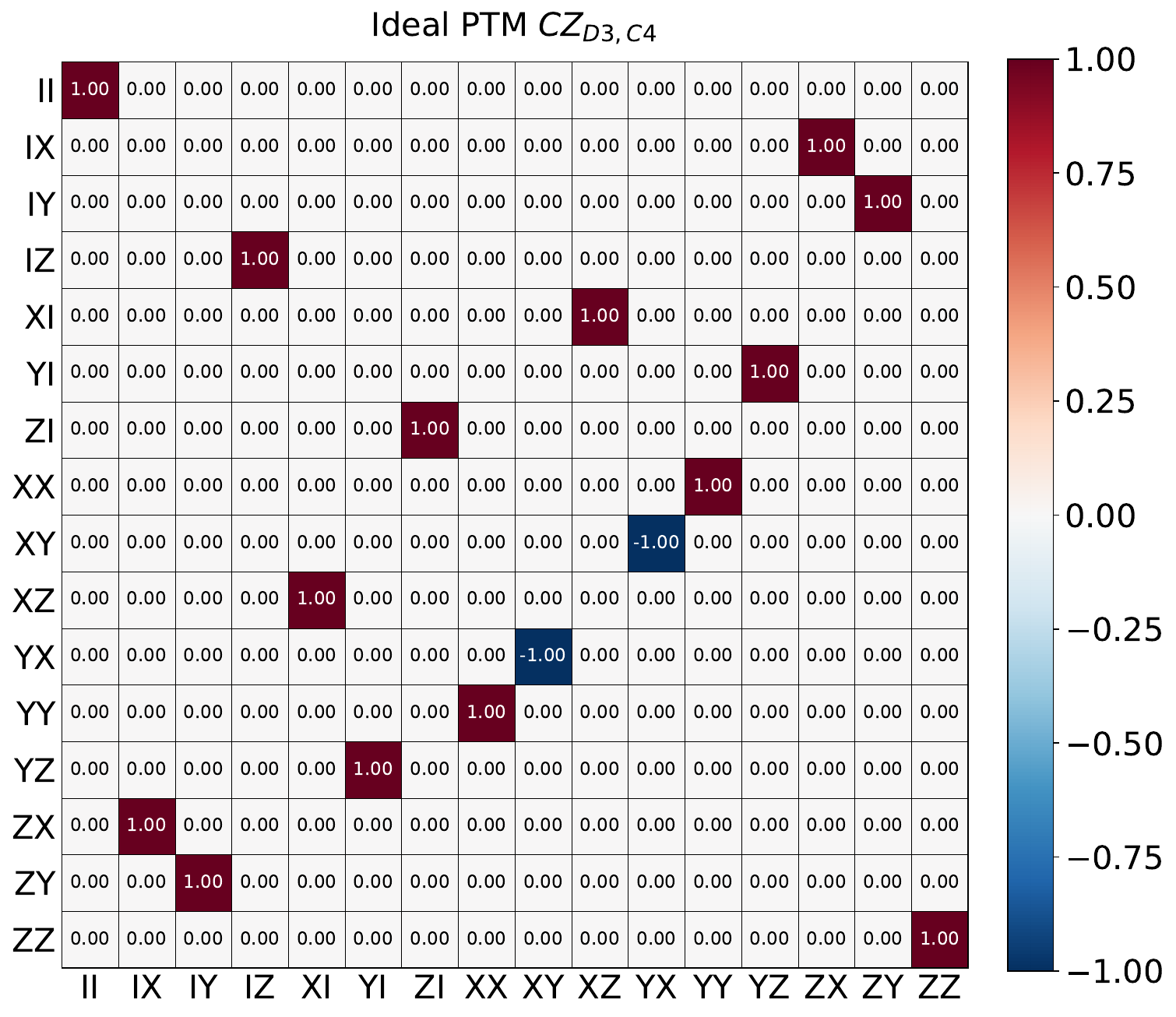}
			\caption{}
			\label{}
		\end{subfigure}
		\caption{$CZ$ gate tomography implemented on $D3-C4$ register: in {\bfseries a} experimental PTM with $F=(94.1 \pm 0.5) \, \% $, in comparison with the ideal PTM in {\bfseries b}.}
		\label{CZtomography}
	\end{figure}
	
\begin{figure}[h!]
		\begin{subfigure}{0.32\textwidth}
			\centering
			\includegraphics[width=\linewidth]{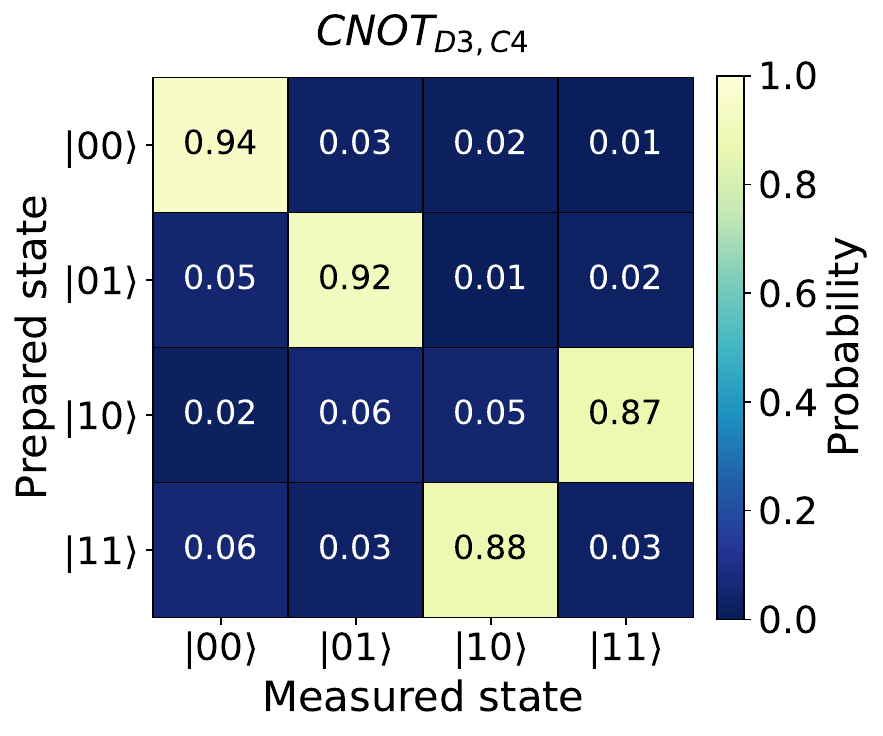}
			\caption{}
			\label{}
		\end{subfigure}
		\hfill
		\begin{subfigure}{0.32\textwidth}
			\centering
			\includegraphics[width=\linewidth]{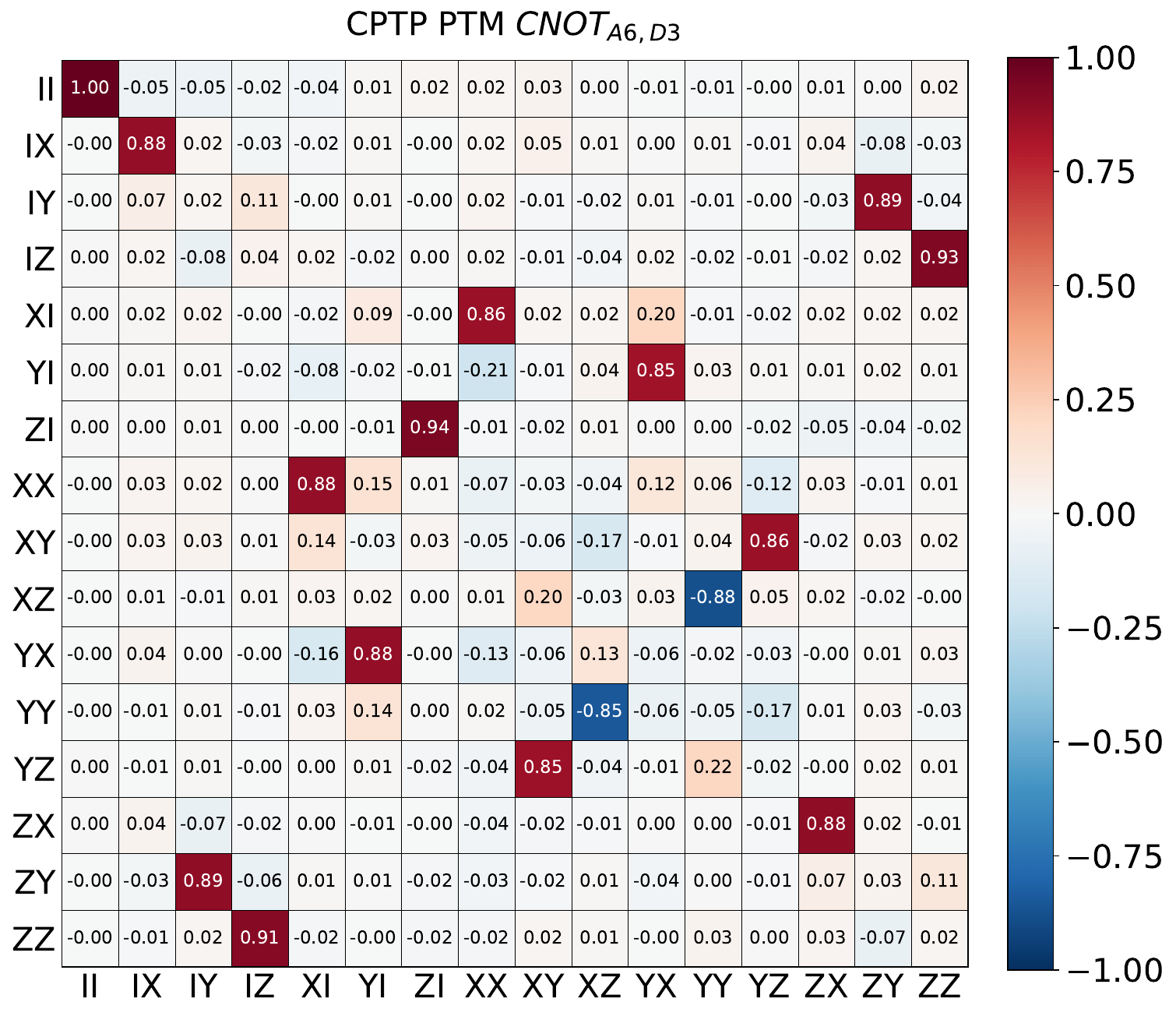}
			\caption{}
			\label{}
		\end{subfigure}
		\hfill
		\begin{subfigure}{0.32\textwidth}
			\centering
			\includegraphics[width=\linewidth]{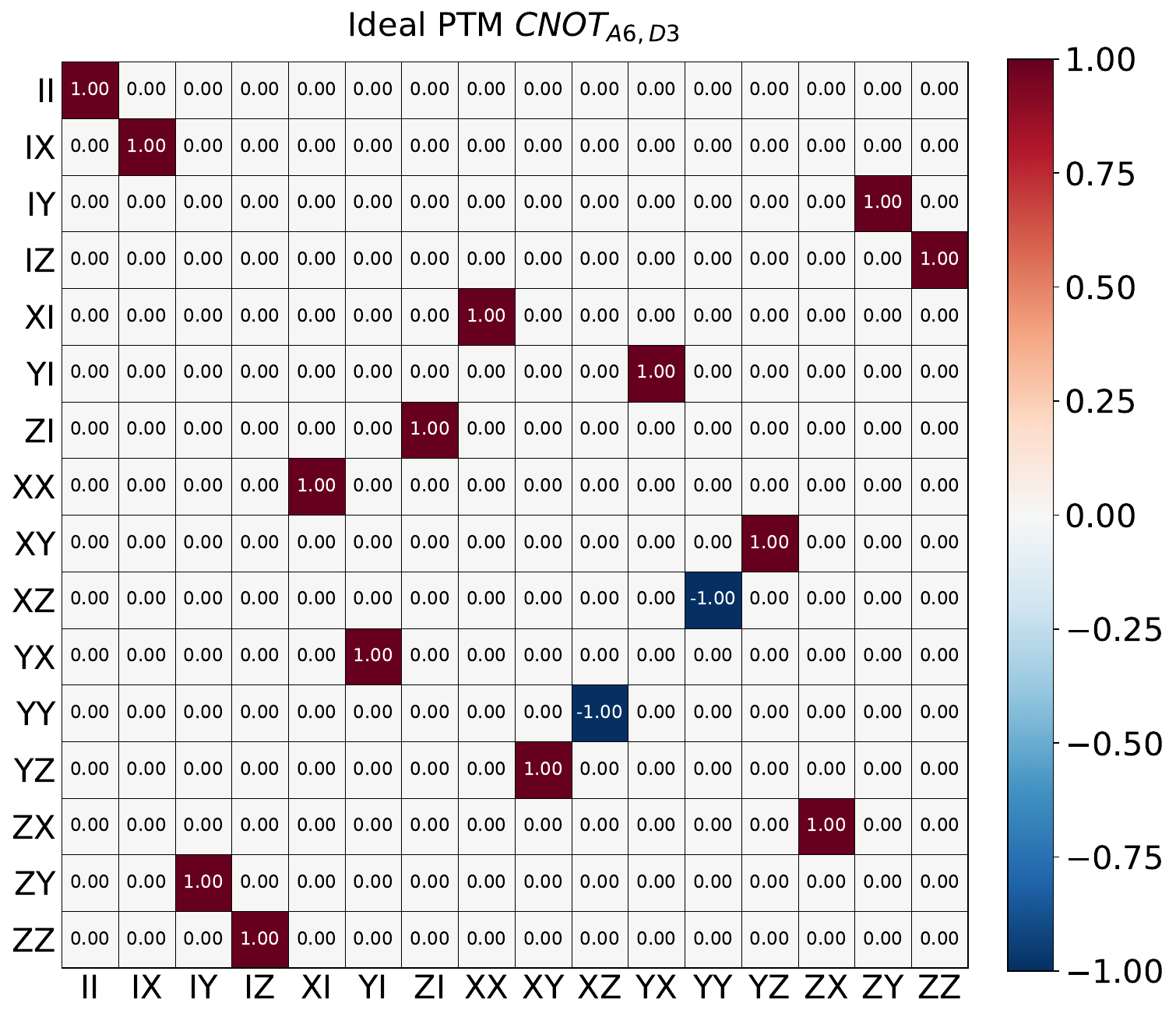}
			\caption{}
			\label{}
		\end{subfigure}
		\caption{$CNOT$ implemented as $(I\otimes H)CZ(I\otimes H)$: in {\bfseries a} truth table on $D3-C4$ register with $F_{H}=(90.2\pm0.7)\, \%$, in {\bfseries b} gate tomography $A6-D3$ register showing  $F=(91.1\pm0.05)\, \%$; in {\bfseries c} the ideal PTM for comparison.}
		\label{CNOTFig}
	\end{figure}

	In our hardware the $CNOT$ can be implemented via $(I\otimes H)CZ(I\otimes H)$ decomposition: as shown in figure \ref{CNOTFig}a, the truth table captures the effect of $CNOT$ the target qubit $D3$ when the control qubit in in $|1\rangle$, with a fidelity $F_{H}=(90.2\pm0.7)\, \%$. Whereas, in figure \ref{CNOTFig}b is shown the experimental PTM matrix exibiting $F=(91.1\pm0.05)\, \%$.  \newline
	
\begin{figure}[h!]
		\centering
		\begin{subfigure}{0.4\textwidth}
			\centering
			\includegraphics[width=\linewidth]{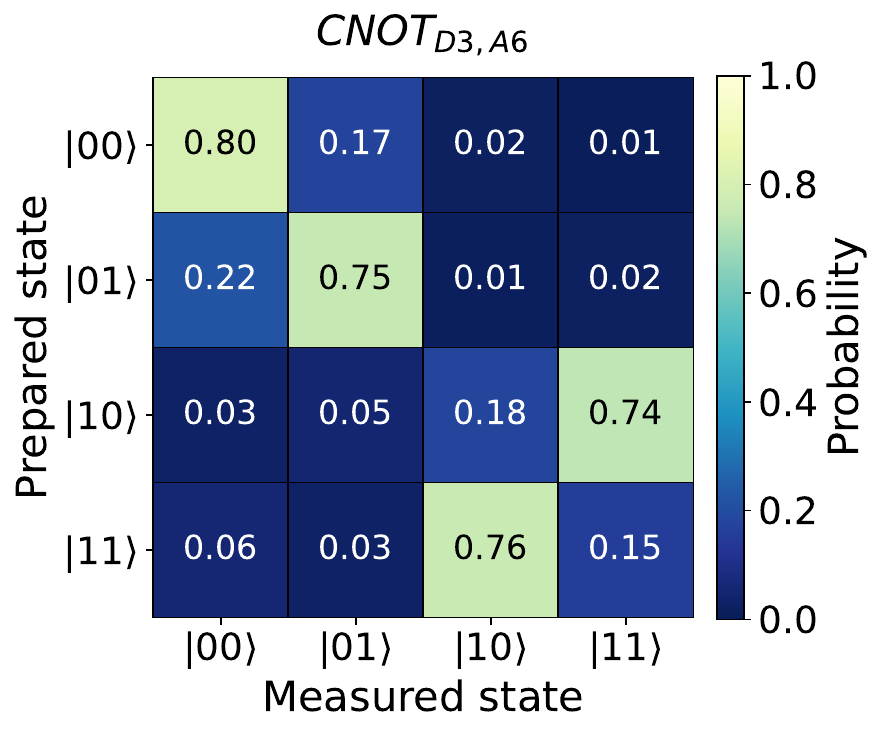}
			\caption{}
			\label{}
		\end{subfigure}
		\hfill
		\begin{subfigure}{0.4\textwidth}
			\centering
			\includegraphics[width=\linewidth]{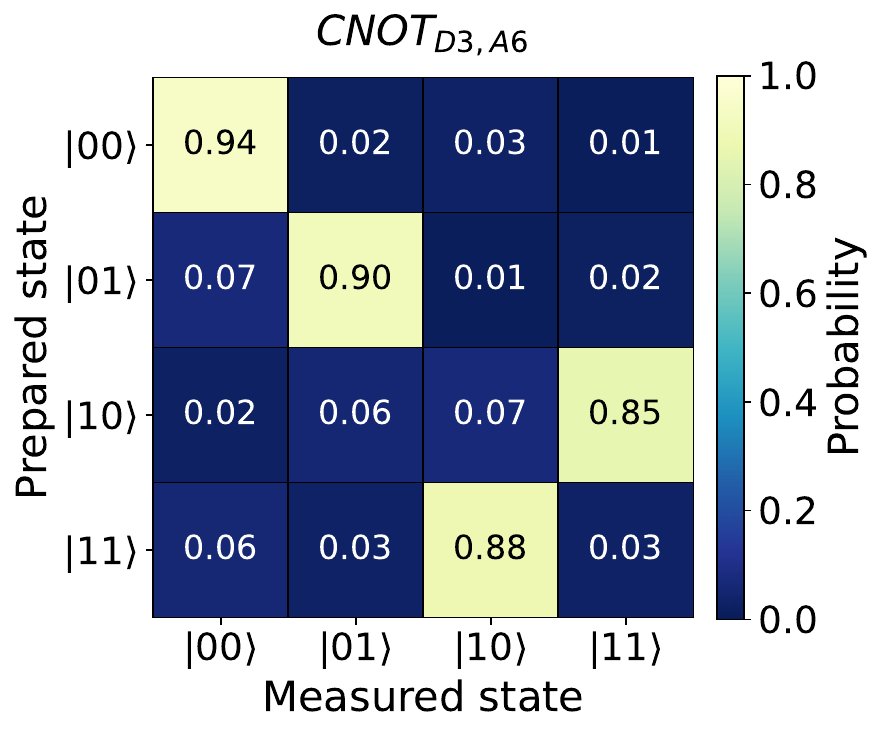}
			\caption{}
			\label{}
		\end{subfigure}
		\caption{Truth tables of $CNOT$ gate on $A6-D3$ register with pulsed qubit $A6$ as target: in {\bfseries a} truth table of bare $CNOT$ with $F_{H}=(76.1\pm 0.7)\, \%$; in {\bfseries b} truth table of $CNOT$ with a counter phase  $RZ(-135^{\circ})$ gate correction on $A6$ increasing the fidelity to $F_{H}=(89.4\pm 0.7)\, \%$.}
		\label{CNOT_counter}
	\end{figure}

\begin{figure}[h!]
		\centering
		\includegraphics[width=\linewidth]{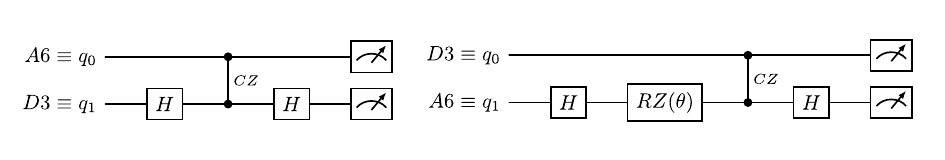}
		\caption{CNOT circuits on A6-D3 register: in {\bfseries (a)} the implementation of $CNOT_{A6,D3}$, in {\bfseries (b)} the implementation of $CNOT_{D3,A6}$ with counter phase $RZ(\theta)$ gate, i.e. when the tuned qubit is the targeted one. }
		\label{Counter_circuit}
	\end{figure}

	As shown in the CZ calibration workflow in figure \ref{two-qubit_cal}, the protocol requires also to correct single-qubit phases. This is practically achieved measuring conditional oscillations first in the standard configuration, and then inverting the control and the target qubit. By fitting the phase-offsets, one can compensate the single-qubit phases to mantain the two-qubit conditional phase equal to $180^{\circ}$. If single-qubit phases are not corrected, the implementation of $CNOT$ gates can fail, especially when the target qubit is also the pulsed one, accumulating dephasing errors during the process. One way to counteract for it, as shown in figure \ref{Counter_circuit}, is to add a counter-phase $RZ$ gate to correct the output. In figure  \ref{CNOT_counter}a is shown the truth table of a $CNOT$ gate implemented on $A6-D3$ register with $A6$ as a target qubit, resulting in a faulty output with $F_{H}=(76.1\pm 0.7)\, \%$. The output can be restored by adding a counter phase $RZ(-135^\circ)$ on the pulsed qubit (Fig \ref{CNOT_counter}b). Then, in  figure  \ref{CNOT_counter}b the truth table after the $RZ(-135^\circ)$  has been implemented, exhibiting $F_{H}=(89.4\pm 0.7)\, \%$. \newpage
	
\begin{figure}[h!]
		\begin{subfigure}{0.4\textwidth}
			\centering
			\includegraphics[width=\linewidth]{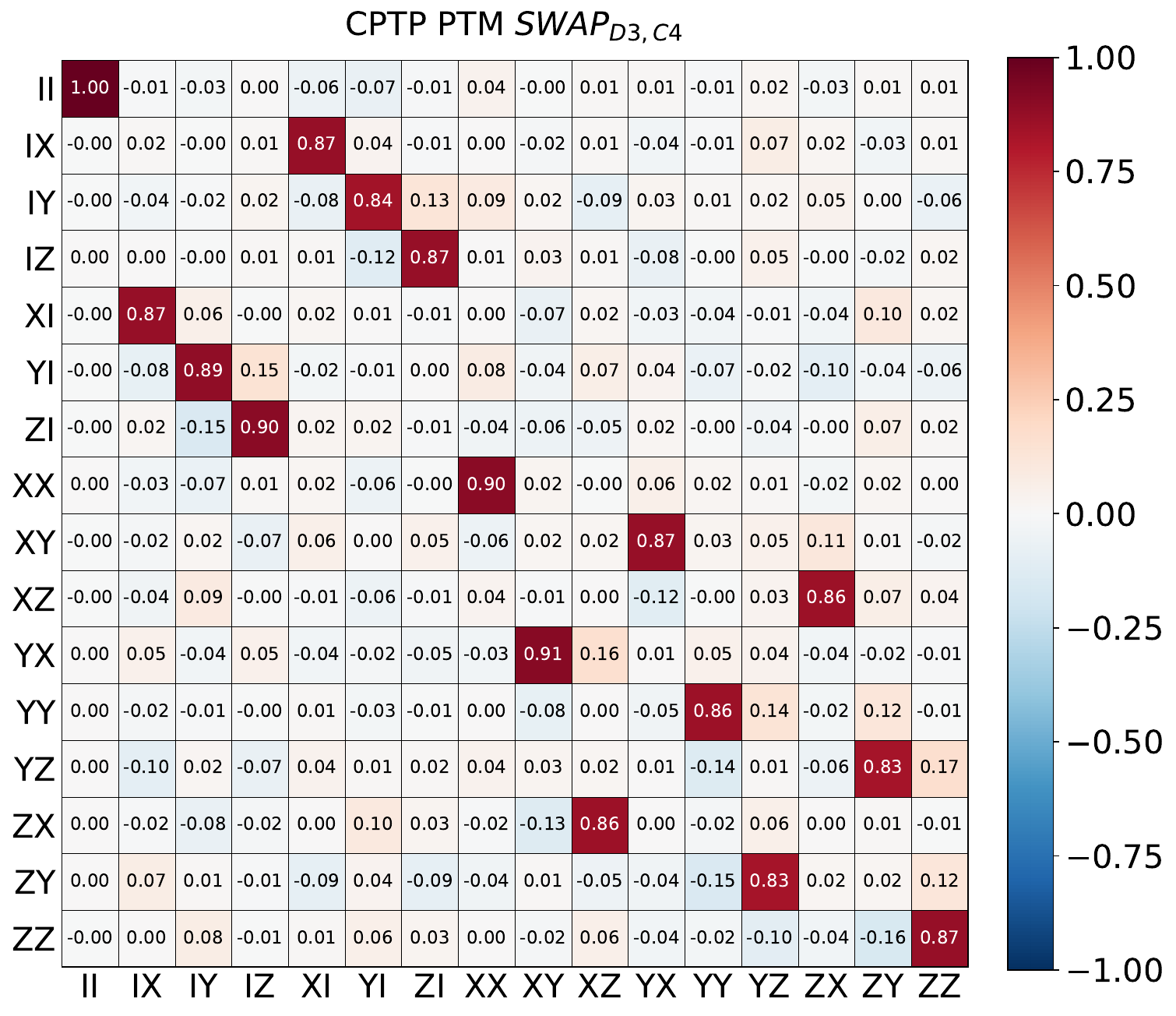}
			\caption{}
			\label{}
		\end{subfigure}
		\hfill
		\begin{subfigure}{0.4\textwidth}
			\centering
			\includegraphics[width=\linewidth]{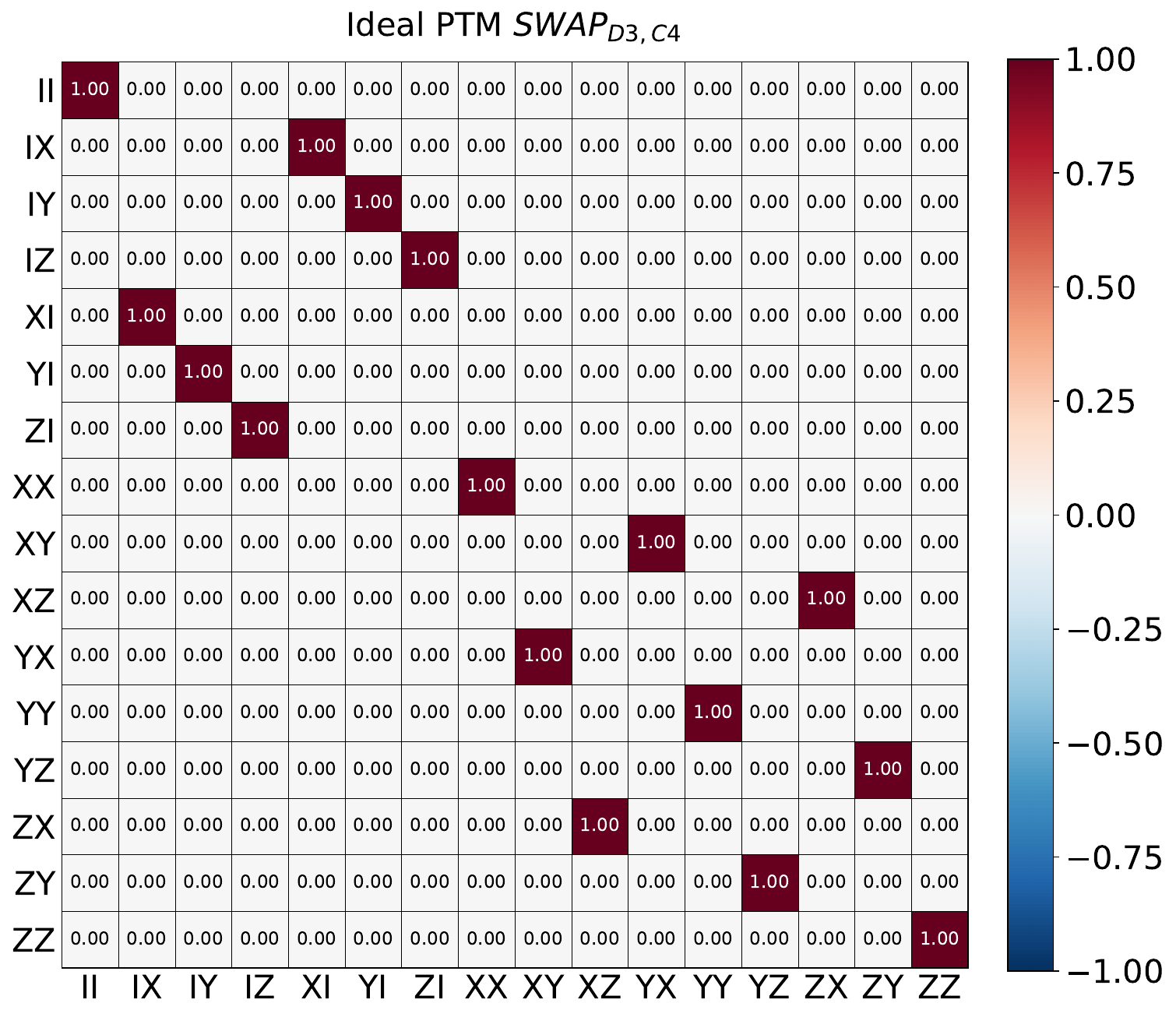}
			\caption{}
			\label{}
		\end{subfigure}
		\caption{Gate tomography of $SWAP$ as three $CNOT$ gates on $D3-C4$ register: in {\bfseries a} experimental PTM with $F=(90.2\pm0.05)\, \%$; in {\bfseries b} ideal PTM of a $SWAP$ process as reference.}
		\label{TomoSWAP}
	\end{figure}

	\paragraph{Non-native gates overview: the SWAP and the CY.} 
	As introduced in section \ref{ComputationalUtility}, the SWAP gate can be natively implemented exploiting the interaction between the state $|01\rangle$ and $|10\rangle$, or by a three-CNOT decomposition (figure \ref{swap}). In figure \ref{TomoSWAP}, we show the experimental PTM matrix obtained by implementing a SWAP gate with the latter decomposition on the register $D3-C4$, reaching a fidelity $F=(90.2\pm0.05)\, \%$.
	This gate turns out to be very useful, allowing to execute circuits re-routing the information to qubits that are not physically connected among the processors.\newline
	
\begin{figure}[h!]
		\begin{subfigure}{0.32\textwidth}
			\centering
			\includegraphics[width=\linewidth]{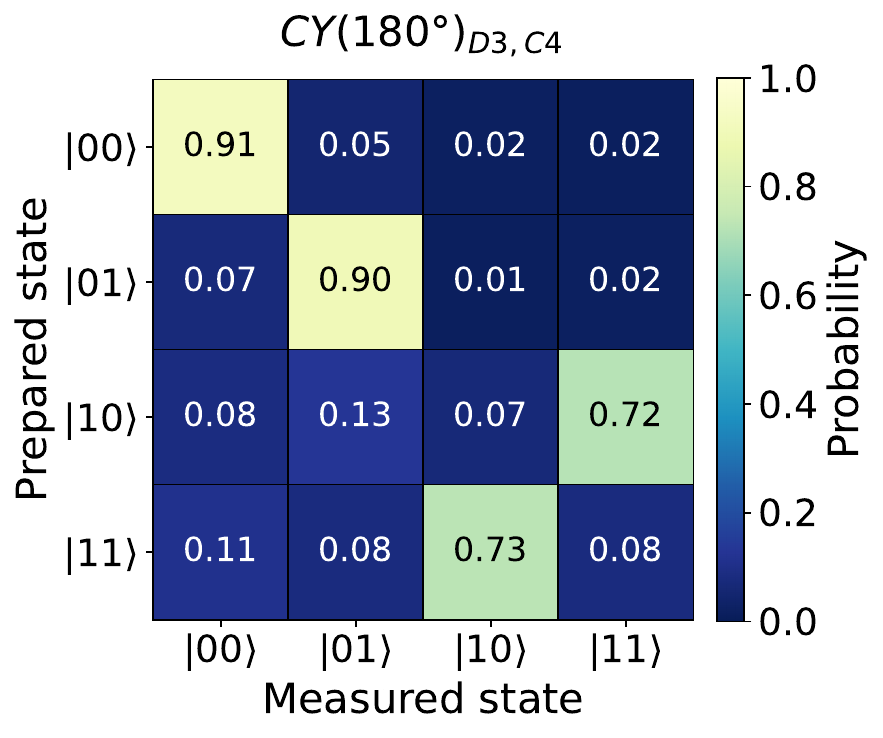}
			\caption{}
			\label{}
		\end{subfigure}
		\hfill
		\begin{subfigure}{0.32\textwidth}
			\centering
			\includegraphics[width=\linewidth]{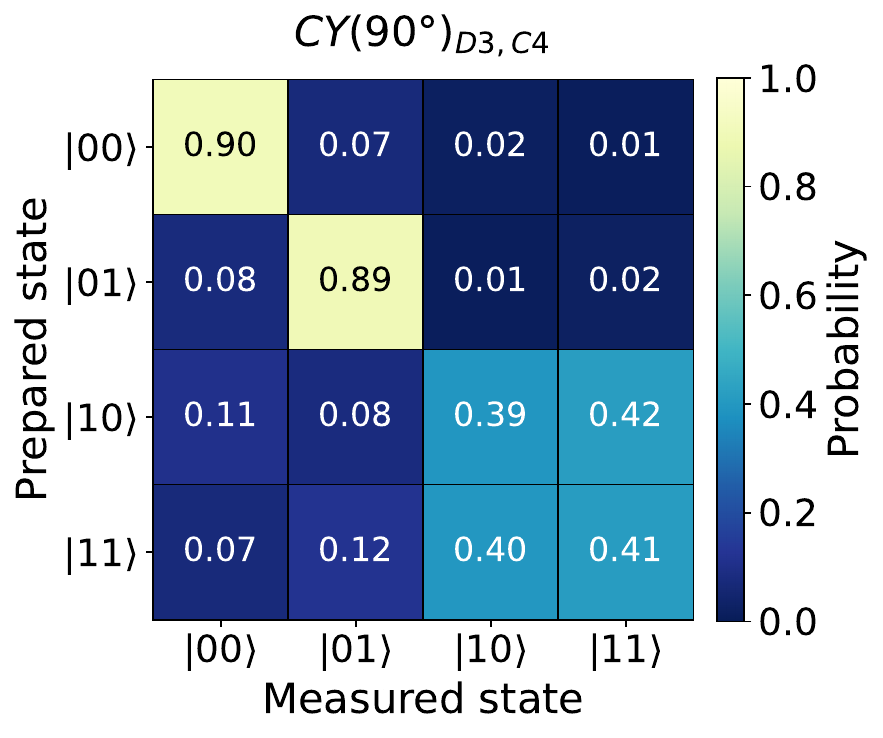}
			\caption{}
			\label{}
		\end{subfigure}
		\hfill
		\begin{subfigure}{0.32\textwidth}
			\centering
			\includegraphics[width=\linewidth]{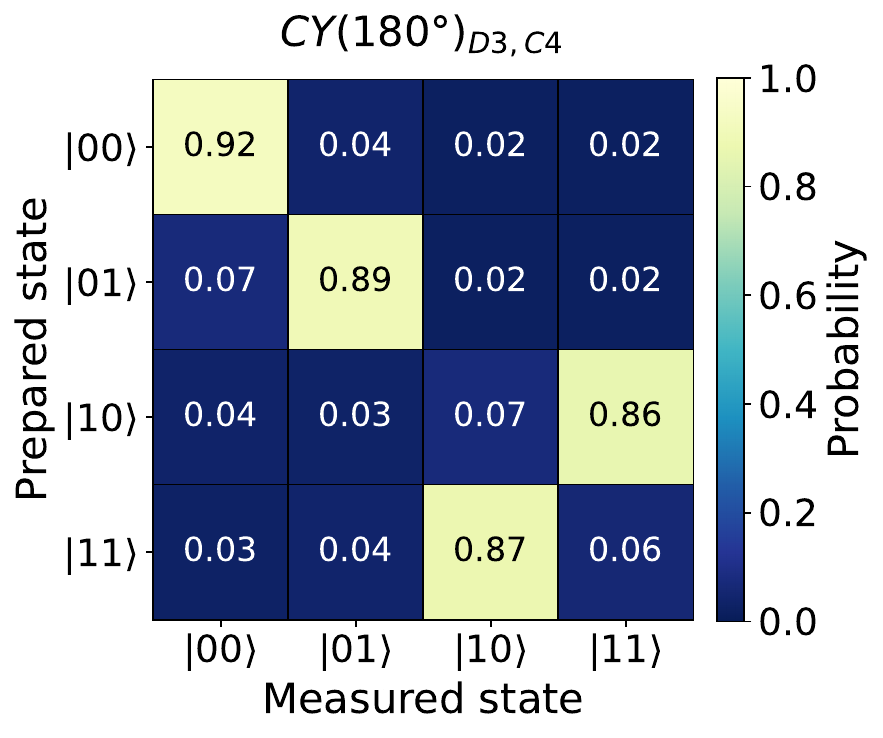}
			\caption{}
			\label{}
		\end{subfigure}
		\caption{Truth tables of $CY$ gate on $D3-C4$ register: in {\bfseries a} and {\bfseries b} truth tables of Tomography of $CY(180°)$ and $CY(90°)$ implemented with Barenco decomposition (figure  \ref{RYdecomposition}a \cite{Barenco95}) with respectively $F_{H}=(81.5\pm 0.7)\, \%$ and $F_{H}=(85.2\pm 0.7)\, \%$; in {\bfseries c} $CY$ as $(I\otimes S)CNOT(I\otimes S^{\dagger})$ (figure  \ref{RYdecomposition}b, with $F_{H}=(88.6\pm0.7)\, \%$}
		\label{TTCY}
	\end{figure}

\begin{figure}[h!]
		\centering
		\begin{subfigure}{0.32\textwidth}
			\centering
			\includegraphics[width=\linewidth]{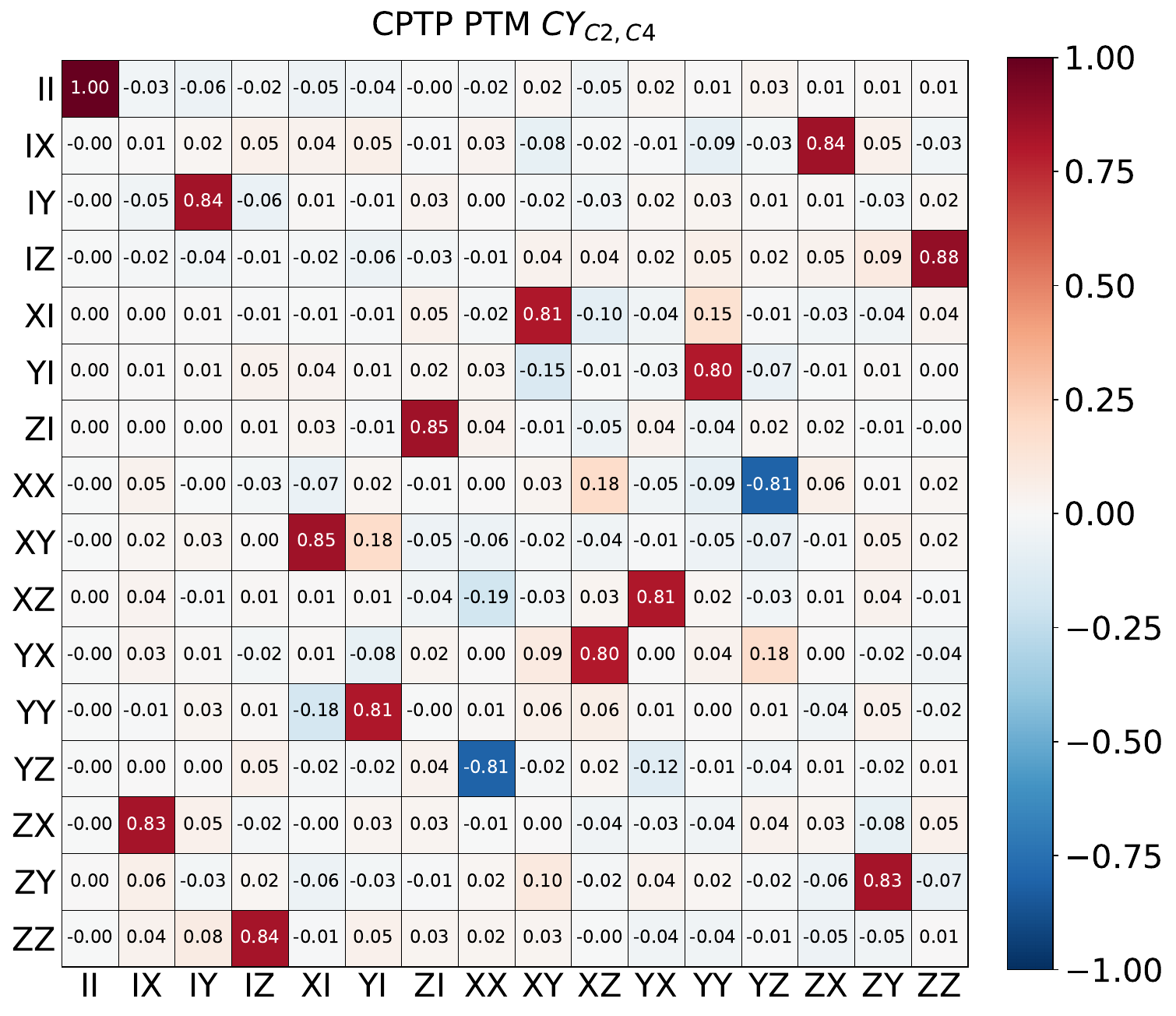}
			\caption{}
			\label{}
		\end{subfigure}
		\hfill
		\begin{subfigure}{0.32\textwidth}
			\centering
			\includegraphics[width=\linewidth]{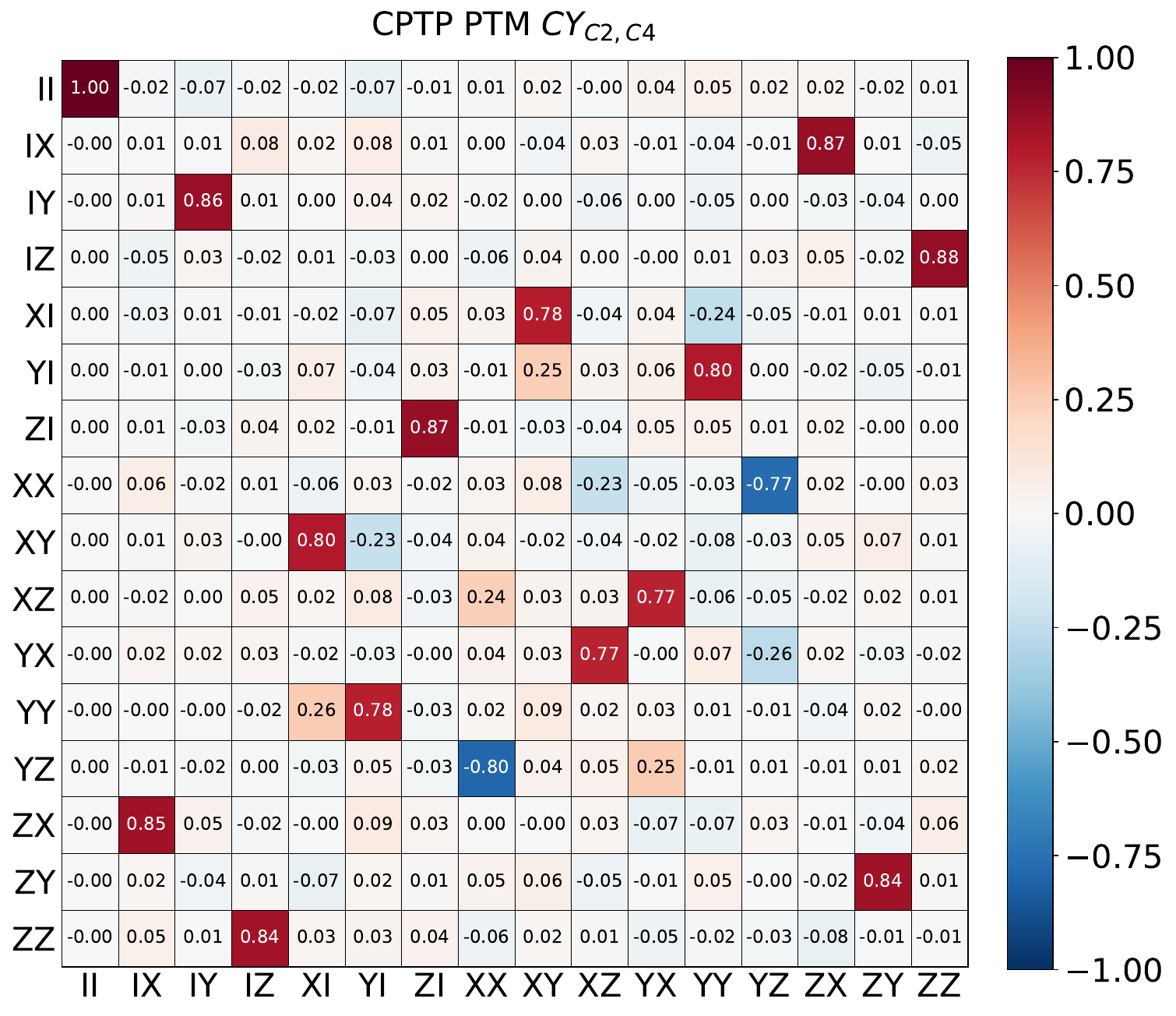}
			\caption{}
			\label{}
		\end{subfigure}
		\hfill
		\begin{subfigure}{0.32\textwidth}
			\centering
			\includegraphics[width=\linewidth]{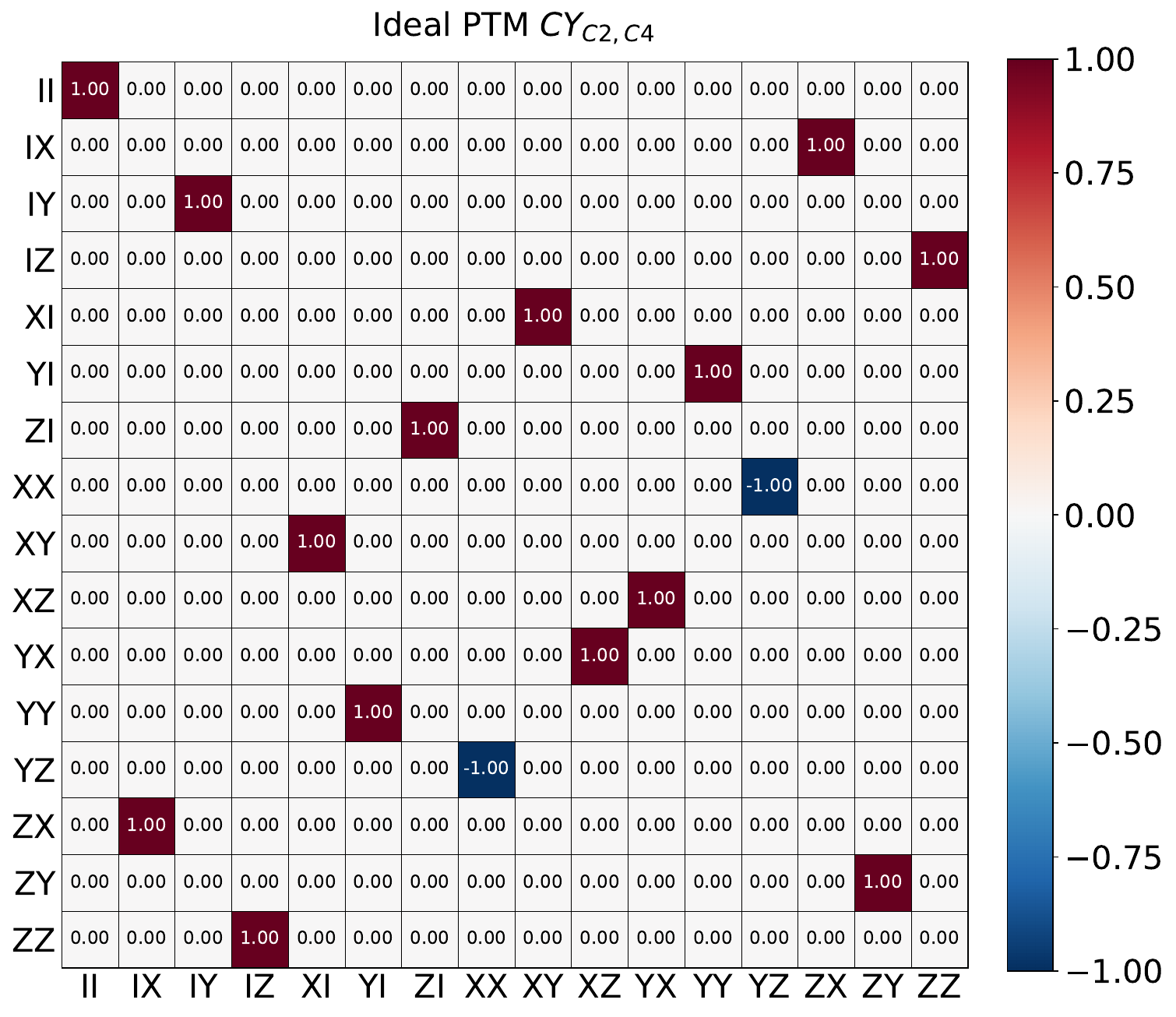}
			\caption{}
			\label{}
		\end{subfigure}
		\caption{Tomography of CY gate on $C2-C4$ register: in {\bfseries a} CY implemented with Barenco decomposition (figure  \ref{RYdecomposition}a \cite{Barenco95}) plus a final counter-phase $Z(45^{\circ})$ gate on the pulsed qubit $C2$ showing $F=(87.0\pm0.7)\, \%$; in {\bfseries b} $CY$ as $(I\otimes S)CNOT(I\otimes S^{\dagger})$ (figure  \ref{RYdecomposition}b, showing $F=(86.4\pm0.7)\, \%$; in {\bfseries c} the ideal PTM matrix of $CY$ gate as comparison.}
		\label{TomoCY}
	\end{figure}

	Finally, the $CY$ gate completes the set of two-qubit controlled gates: in particular, we implemented two different decompositions of the gate, showing they resort to the same process. In figure \ref{TTCY}a and b, we show the truth table of $CY(180^{\circ})$ and $CY(90^{\circ})$, respectively, obtained following Barenco et al. decomposition (figure \ref{RYdecomposition}a \cite{Barenco95}), achieving $F_{H}=(81.5\pm 0.7)\, \%$ and $F_{H}=(85.2\pm 0.7)\, \%$, respectively. In figure \ref{TTCY}c, we implement the $CY(180^{\circ})$ as $(I\otimes S)CNOT(I\otimes S^{\dagger})$ (figure \ref{RYdecomposition}b), achieving $F_{H}=(88.6\pm0.7)\, \%$. The truth tables of $CY(180^{\circ})$ show the same effect on the target qubit when the control qubit is in $|1\rangle$ as for the $CNOT$. For this reason, a complete benchmark requires measuring the difference in the implemented phase through QPT. In figure \ref{TomoCY}a, we show the CY implemented with Barenco decomposition (figure  \ref{RYdecomposition}a \cite{Barenco95}) plus a final counter-phase $Z(45^{\circ})$ gate on the pulsed qubit $C2$ to correct for accumulated dephasing, showing $F=(87.0\pm0.7)\, \%$. In figure \ref{TomoCY}b, the $CY$ is implemented via the second approach (figure  \ref{RYdecomposition}b). The fidelity $F=(86.4\pm0.7)\, \%$ is consistent with Barenco's decomposition, therefore demonstrating that the two decompositions allow for comparable fidelities, leaving the choice in the use-case one wants to implement. The Barenco decomposition allows for $\theta$ angle control, but requires more scheduling time to be implemented (each $RY(\theta)$ gate lasts, in our experiments, $60\;ns$), and can introduce more flux-phase errors due the presence of two $CNOT$s, i.e. two $CZ$ SNZ flux pulses. This is the reason why a counter-phase gate has been required. The $(I\otimes S)CNOT(I\otimes S^{\dagger})$ decomposition is faster, due to the presence of virtual-Z gates, but allows for implementing only a $180^{\circ}$ rotation. These considerations go in the direction of need a pulse level control of the hardware to optimize the desired output from an algorithm, in contrast to agnostic algorithm synthesis.
	
	\paragraph{Building entangled states with CNOT gates and scaling to three qubits: the Toffoli gate and GHZ states}
	
\begin{figure}[h!]
		\centering
		\includegraphics[width=0.5\textwidth]{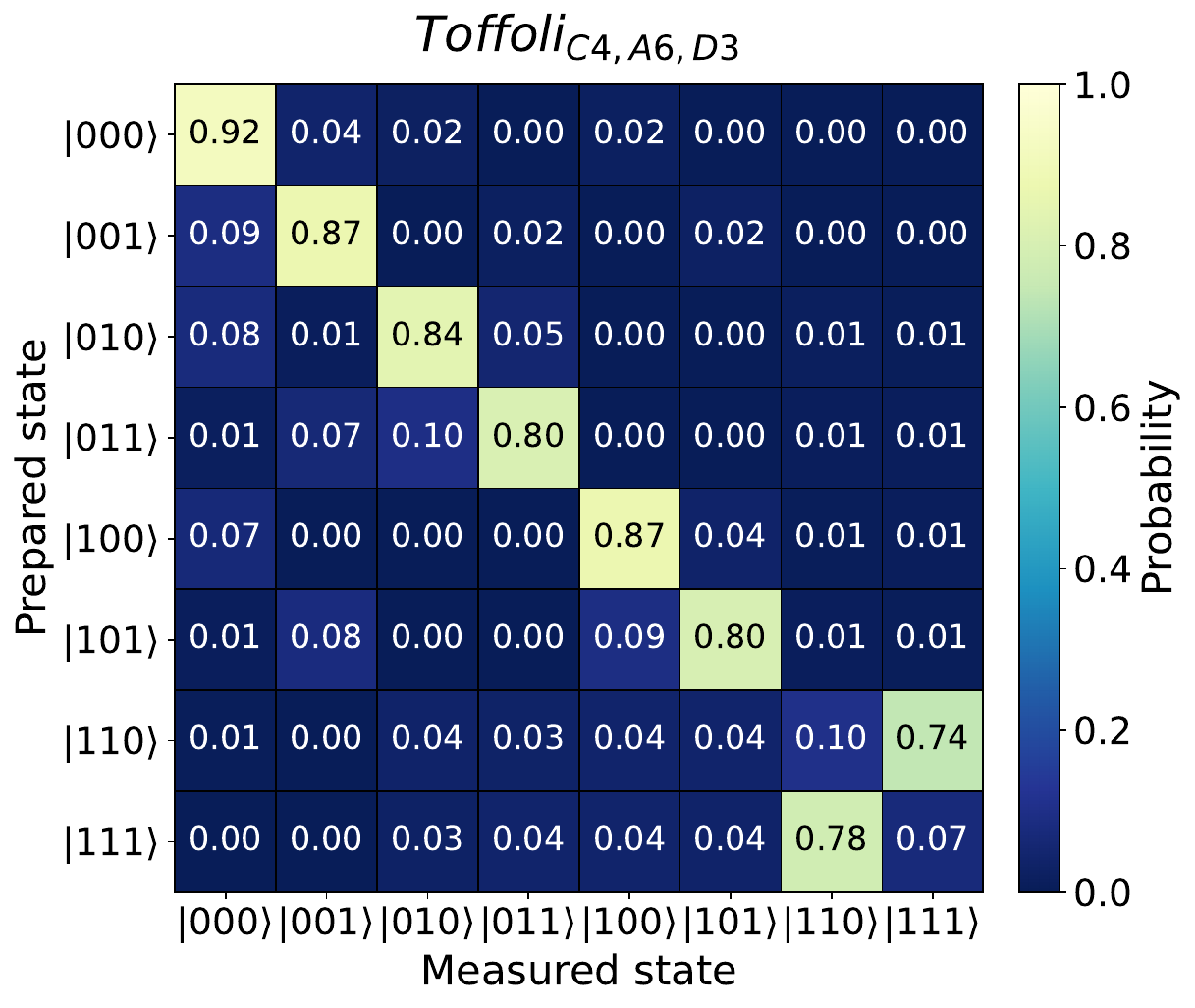}
		\caption{Truth table of Toffoli gate implemented on $C4-D3-A6$ register with $F_{H}=(82.8\pm0.7)\, \%$.}
		\label{TToffoli}
	\end{figure}

	As shown in section \ref{UniversalComputing}, two-qubits gates are the building blocks for three-qubits gates, as the $CCX$ gate, i.e. the $Toffoli$ gate. When scaling to more than two qubits, QPU's topology and connectivity begin to be non negligible, since the implementation of three-qubit gates is strongly dependent on the available physical connections between the pairs on the processor. Given the connectivity of our QPU, the mirrored decomposition $Toffoli$ (figure \ref{Toffoli decomposition}c \cite{albayaty2024}) appears to be the only implementable one. In particular, in figure \ref{TToffoli} we show the truth table for a Toffoli gate implemented in $C4-D3-A6$ register with $F_{H}=(82.8\pm0.7)\, \%$. Also in this case, we required a counter phase $RZ(-90^{\circ})$ for $CNOT$ gates implemented on $C4-D3$, being $D3$ both the pulsed and the target qubit.

\begin{figure}[h!]
		\centering
		\includegraphics[width=\linewidth]{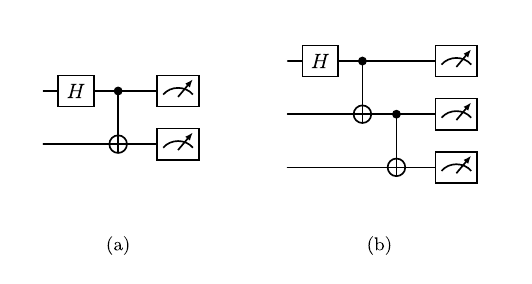}
		\caption{Bell State circuit in {\bfseries a} and GHZ state circuit in {\bfseries b}.}
		\label{BellGHZCircuit}
	\end{figure}

\begin{figure}[h!]
		\centering
		\begin{subfigure}{0.4\textwidth}
			\centering
			\includegraphics[width=\linewidth]{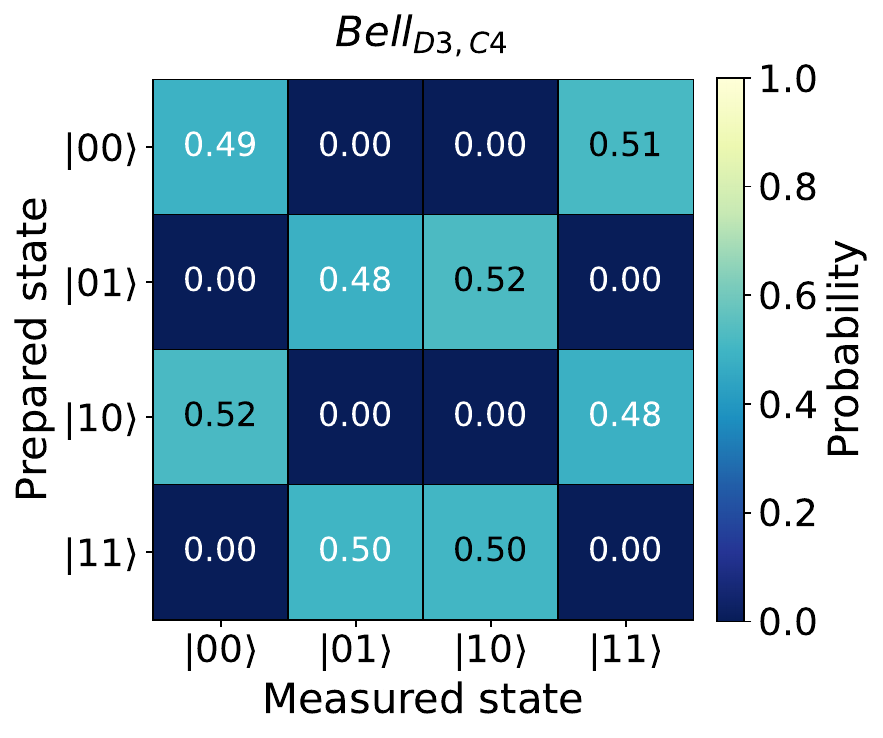}
			\caption{}
			\label{}
		\end{subfigure}
		\hfill
		\begin{subfigure}{0.4\textwidth}
			\centering
			\includegraphics[width=\linewidth]{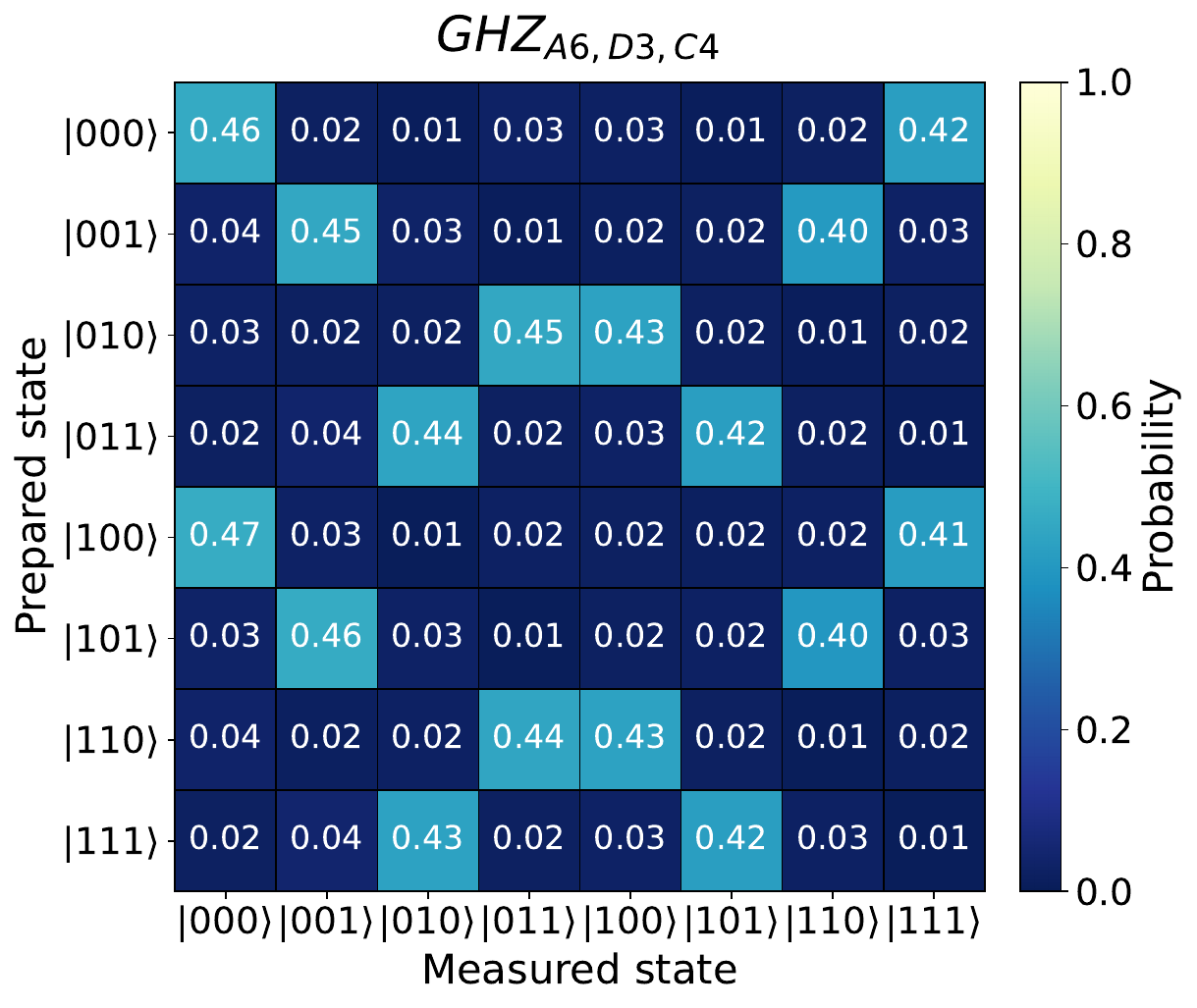}
			\caption{}
			\label{}
		\end{subfigure}
		\caption{Truth table of entangling states: in {\bfseries a} Bell states with $F_{H}=(99.8\pm0.7)\, \%$, in {\bfseries b} GHZ states with $F_{H}=(86.8\pm0.7)\, \%$.}
		\label{TTBellGHZ}
	\end{figure}

	Truth-tables finally comes in hand also to retrieve information on the ability to build entangled states. The $CNOT$ can be used to construct two-qubit entangling states, e.g. the Bell states (figure \ref{BellGHZCircuit}a), and three-qubit entangling states, e.g. $GHZ$ state (figure \ref{BellGHZCircuit}b) \cite{Neeley10}. The ability to leverage entanglement is mandatory for all-around computation. In figure \ref{TTBellGHZ}a, we show the truth table for Bell States with $F_{H}=(99.8\pm0.7)\, \%$ and in figure \ref{TTBellGHZ}b for three qubits GHZ states with $F_{H}=(86.8\pm0.7)\, \%$. 
	 \newline
	
	\newpage
	
	\section{Conclusions}
	
	In this review, we have discussed some aspects between high-level algorithmic synthesis in quantum computing and the underlying quantum physical hardware. As quantum computing transitions from the NISQ era toward practical utility, the traditional decoupling of software abstraction from device physics has emerged as a significant bottleneck. At this stage, hardware-software co-design is not merely advantageous, but mandatory. 
	
	Our study underscores that achieving fault-tolerant quantum computation requires a deeply holistic approach. A critical prerequisite for building a scalable fault-tolerant quantum computer is not only the capability of implementing high-fidelity quantum gates, but also retrieving possible errors as extensively as possible. Currently, global efforts across academia and industry are fragmented across multiple fronts: material science optimizations to mitigate decoherence, architectural innovations to improve connectivity, sophisticated software stacks for automated calibration, and the theoretical synthesis of promising algorithms. Because these domains are deeply interconnected, addressing the scalability challenge requires a unified effort across the entire stack. Navigating the unique trade-offs of distinct qubit designs and the operational divergence between different architectures demands integrated frameworks where physical hardware constraints directly guide the choice of pulse engineering and gate decomposition strategies. In this review, we discuss on the systematization of pulse engineering and calibration workflows, covering both single-qubit microwave pulses and two-qubit flux pulses. By demonstrating how abstract quantum gates are translated into compositions of precisely shaped physical waveforms, we have illustrated a viable path to maximize gate fidelities while accounting for the unique operational footprints of different qubit geometries.
	
	Another key message of this review is that high gate fidelity alone does not constitute a sufficient indicator of computational usefulness. 
	Several error mechanisms can remain largely hidden in average fidelity metrics while becoming significant during the execution of deep quantum circuits. A primary example is coherent error accumulation, where small systematic control imperfections, such as over- or under-rotations, can constructively build up over many gate applications, leading to circuit-level errors substantially larger than those suggested by individual gate fidelities.  
	In superconducting architectures, leakage outside the computational subspace is particularly relevant due to the weak anharmonicity of transmon devices, allowing population to accumulate in higher excited states that are not accounted for in standard qubit error models. Addressing the minimization of leakage errors through two-qubit entangling gates calibration, for example, is fundamental. Moreover, non-markovian effects and temporal drift can cause the noise properties observed during calibration to differ significantly from those encountered during actual computation, thereby reducing the predictive power of characterization experiments. Finally, context-dependent errors arise when the performance of a gate depends on surrounding operations, simultaneous gate executions, or the specific state of neighboring qubits, violating the assumption that gates can be characterized independently. Collectively, these phenomena highlight the gap between isolated gate metrics and true algorithmic performance, motivating the development of characterization and benchmarking techniques capable of capturing realistic error behavior in large-scale quantum processors.
	
	We conclude that a realistic assessment of quantum hardware must account for how gates are physically implemented, synthesized into logical operations, compiled within architectural constraints, and embedded into executable algorithms, and provided examples of single-, two- and three-qubit gates and states implementation,  benchmarked through both QPT and truth-tables experiments. We believe that providing a broader perspective on the topic, and in the form of a guide on how to practically build and validate fundamental multi-qubit gates and states, improvements at the device level are more likely to translate into practical computational advantage in the near term.
	
	Furthermore, as the number of qubits scales, traditional benchmarking methodologies face severe scalability limits. Our discussion on characterization techniques, supported by experimental truth tables and quantum process tomography, emphasizes that benchmarking cannot be one-size-fits-all. It is critical to select tailored, resource-efficient benchmarking metrics that align precisely with the specific evaluation aspects of interest in a given context.
	
	\ack{Halima Giovanna Ahmad, Francesco Tafuri, and Davide Massarotti thank the SUPERQUMAP project (COST Action CA21144), Hany Ali and Alessandro Bruno from Quantware, and Johannes Hermann and Christian Junger from YQuantum for their insights on the two-qubit characterization workflow, and Medhi El Bakraoui from G2Q for the discussions on transpilation and gates decomposition.}
	
	\funding{The work was supported in part by the National Recovery and Resilience Plan (PNRR) Ministero dell’Università e della Ricerca under Grant CN00000013-ICSC and Grant PE0000023-NQSTI, in part by the Superconducting European Quantum Pilot Line HORIZON-JU-Chips-FPA-QAC-1 SUPREME, in part by the Project PRIN 2022-Advanced Control and Readout of scalable Superconducting NISQ Architectures (SuperNISQ) under Grant CUP E53D23001910006, and by the project “Soluzioni Hardware e Software per l’Ingegnerizzazione di una piattaforma Quantistica di Computazione – SHINE QC” (CUP B69J25000260005).}

	\data{Data are available upon reasonable request.}

	\suppdata{Detailed informations on experimental setup (cryogenics, electronic, sQPUs) at UNINA Quantum Computing Center are provided in the supporting information.}

	\section{Supporting data for: Quantum Gate Engineering in Superconducting NISQ Platforms: Experimental Methods and Emerging Directions}
	
	\subsection{UNINA Quantum Computing Center}
	
	\subsubsection{Cryogenics and electronics: Ligea and Partenope}
	
	\begin{figure}[h!]
		\centering
		\includegraphics[width=0.8\linewidth]{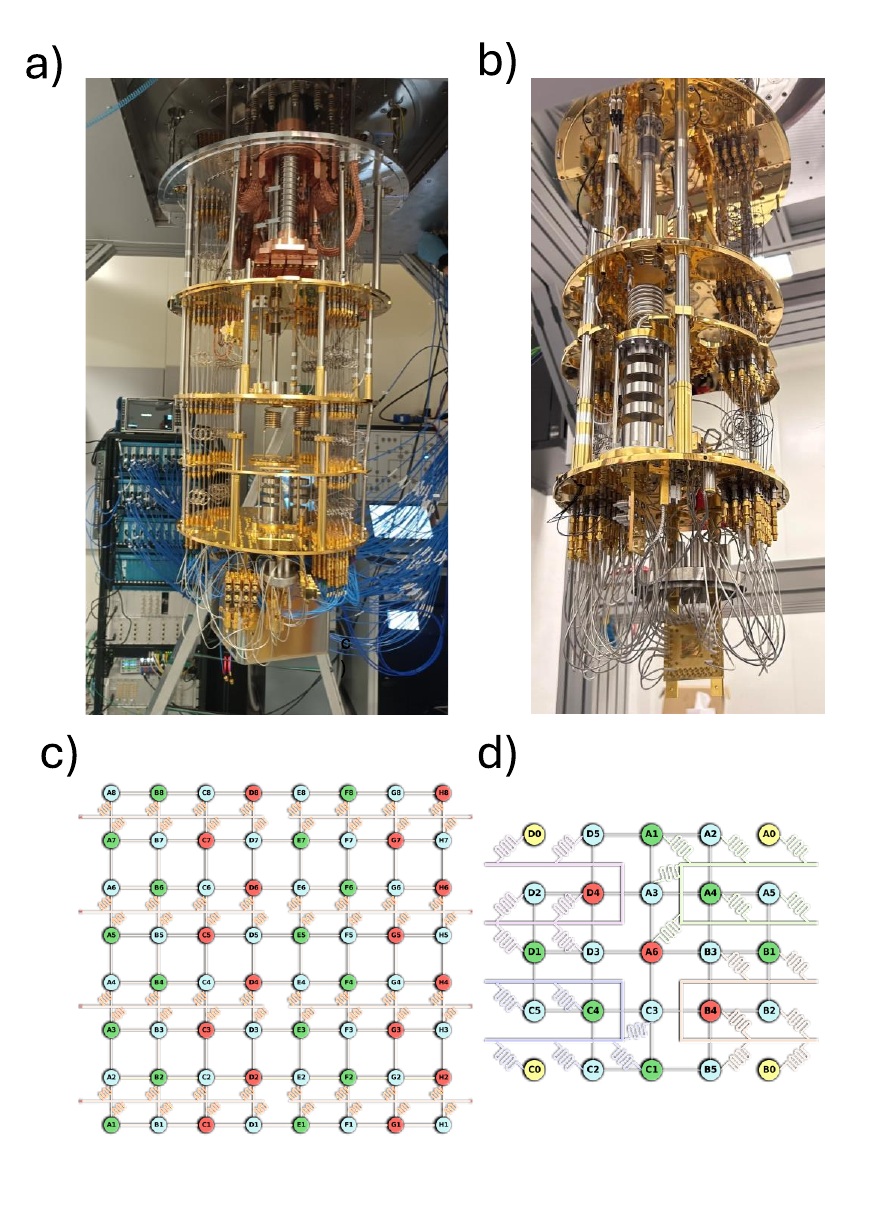}
		\caption{Inner view of the UNINA Superconducting Quantum Computing Center nodes. Partenope (panel (a)) hosts a $64$-qubit processor (Tenor from Quantware), in (c), while Ligea (panel (b)) hosts a Quantware Contralto-D sQPU (panel (d)) of $21$-qubits. Both the chip use symmetric, flux-tunable transmon qubits arranged in a rectangular 2D matrix and interconnected via fixed high-frequency bus resonators. The qubits are structurally distributed across three distinct frequency bandwidths (red: high-frequency; blue: medium-frequency; green: low-frequency; yellow: isolated fixed-frequency qubits).}
		\label{partenope}
	\end{figure}
	
	The UNINA superconducting quantum center now counts two superconducting quantum computers, namely \emph{Partenope} and \emph{Ligea} (figure \ref{partenope}). Both systems have been engineered and assembled with the perspective of having modular and upgradable systems able to keep up with the rapid advancement of superconducting quantum technologies and processors. 
	\begin{figure}[h!]
		\centering
		\includegraphics[width=\linewidth]{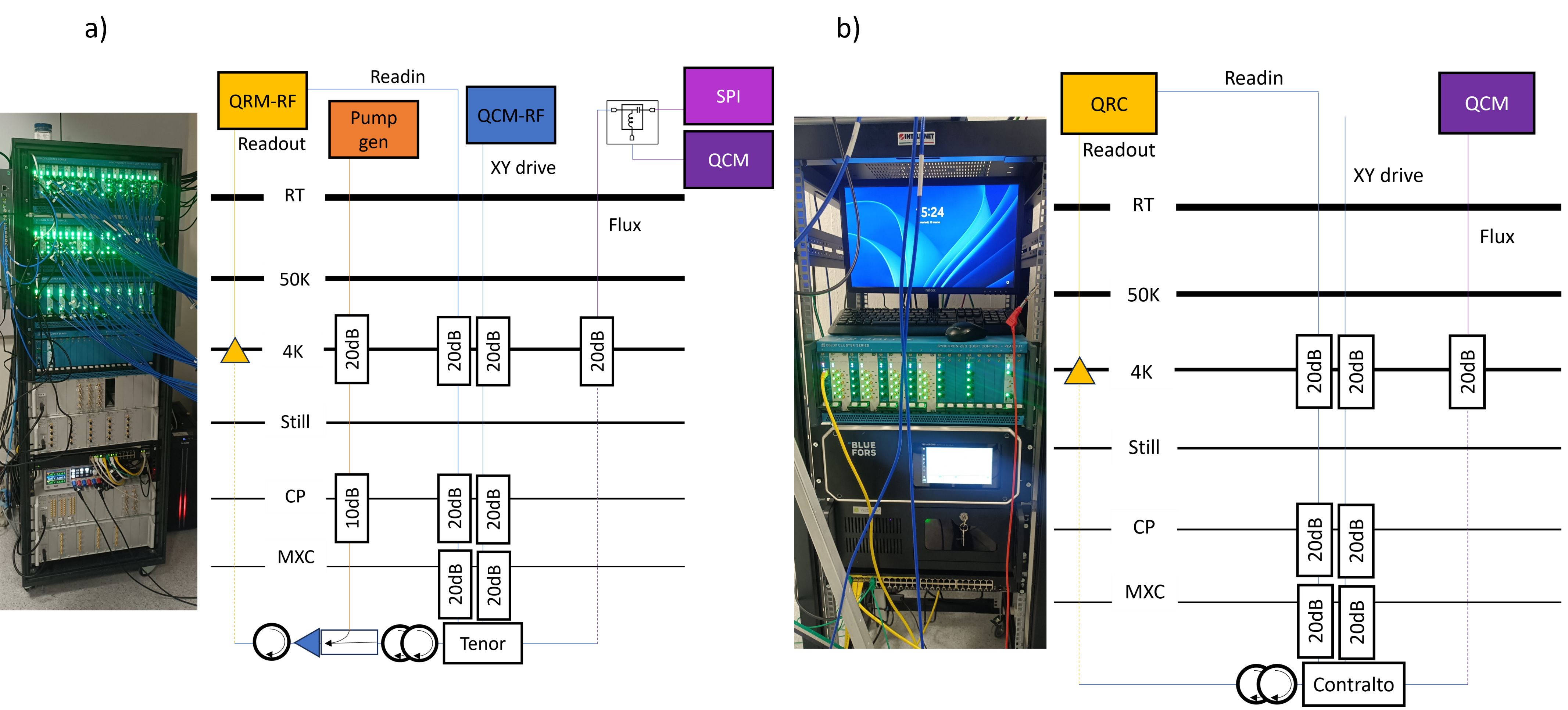}
		\caption{Room-temperature electronics for Partenope (a) and Ligea (b).}
		\label{electronics}
	\end{figure}
	Partenope uses a Bluefors XLD1000SL dilution refrigerator with a cooling power of $1000 \, \mu W$ and $5$ available sideloaders for cryogenic lines (panel (a) of figure \ref{partenope}). Currently, each of them hosts a set of $32$ microwave coaxial cables for drive, flux-tuning, readout, and travelling-wave parametric amplifiers (TWPAs) pumping, for a total of $64$ attenuated drive lines, $64$ superconducting fast RF flux lines, $8$ pump lines, $8$ readin and $8$ readout lines, and other spare lines for any additional experiment. Standard readout is performed employing double-junction isolators and high-electron mobility transistors (HEMT) located at the $4\;K$ plate of the cryostat. Current investigations are devoted to integrating also dispersive-engineered TWPAs in the readout chain. Partenope hosts a $64$ flux-tunable transmon-based quantum processor from Quantware (Tenor), where couplings between qubits in a rectangular lattice is implemented through high-frequency bus resonators (panel (c)). Currently, the processor is under characterization.
	
	Ligea uses a Bluefors LD400 dilution refrigerator with a cooling power of $400 \, \mu W$ hosting a reduced number of connections compared to Partenope (panel (b)):  $21$ drive lines, $21$ flux-tuning lines, $4$ readin and $4$ readout lines. Also in Ligea, standard readout is performed employing double-junction isolators and high-electron mobility transistors (HEMT) at the $4\;K$ plate of the cryostat. Ligea hosts a $21$ flux-tunable transmon-based quantum processor from Quantware (Contralto-D), where couplings and circuit design follow very similar features as the Tenor chip (panel (d)). Further details on the processor can be found in references \cite{ahmad2025_magic,Ahmad26}.
	
	Every qubit features a dedicated superconducting readout resonator, distributed evenly across four readout feedlines to enable multiplexed readout. Additionally, independent drive and flux lines are assigned to each qubit to control single-qubit $X$-$Y$ and $Z$ gates. The native two-qubit gate of this sQPU is the Conditional-$Z$ (CZ) gate. This is implemented by applying sudden-net-zero (SNZ) flux pulses to the higher-frequency qubit within a given pair. Furthermore, the flux lines provide the static magnetic flux required to park the qubits at specific operational working points. Throughout the experimental procedures, the active qubits were operated at their respective flux sweet spots. Conversely, the remaining idle qubits in the matrix were parked at their anti-sweet spots (nominally at zero frequency) to minimize qubit crosstalk and frequency crowding.
	
	At room temperature, the processors connect to two independent sets of control electronics from Qblox \cite{QbloxArchitectureDocs}, connected to a host PC via LAN. Qblox clusters house FPGA-based modules for qubit control, readout, and fast flux pulsing. In Partenope, room-temperature control is achieved with Quantum Control Modules (QCM-RF), and readout through Quantum Readout Modules (QRM-RF) (figure \ref{electronics} (a)). In Ligea, we have replaced these modules with a more compact solution that allows us to perform both readout and control within the same module, Quantum Readout-Control (QRC) modules. The main difference between the two, though, is the integration of DDS mechanisms for up- and down-conversion of microwave signals in the QRC, compared to standard analogue mixing with the QCM- and QRM-RF modules, which require mixer calibration to implement high-fidelity quantum gates (figure \ref{electronics} (b)). 
	
	Control modules (QRC and QCM-RF) are used to implement native $X_{\theta}$ and $Y_{\theta}$ gates. We specifically exploit DRAG microwave pulses of $60\;ns$, where the rotation angles $\theta$ are set by changing the amplitude of such pulses, while the rotation axis is selected by changing the phase of the microwave signals. Microwave pulses are also used to excite the dedicated superconducting resonators, and readout is performed by digitizing the output signal through the readout modules. Meanwhile, Z single-qubit gates are implemented virtually via phase updates to the control signal. All other single-qubit rotations, like the Hadamard gate, are then decomposed in terms of the X, Y, and Z rotations as shown section 3.1.
	
	As for flux tunability, both Ligea and Partenope use QCM modules to generate fast RF flux pulses for the implementation of the native two-qubit CZ gate. To park the qubits at their operating flux points (sweetspots or antisweet-spots, depending on the application) static flux biases are generated by a low-noise DC electronics (SPI, Sample Processor Interface) rack in Partenope, and combined with RF fast flux pulses with the use of bias-tees at room temperature. In Ligea, offsets are generated directly through QCM. 
	
	The entire Qblox hardware stacks are orchestrated via a Python-based framework, Quantify \cite{Quantify} in Partenope and Qblox-Scheduler \cite{QbloxSchedulerDocs} in Ligea, respectively, which led us to design experimental schedules to run calibration protocols or quantum circuits directly at the pulse-level. Currently, efforts are focused on building an in-house control-stack built on top of these open-source frameworks.
	
	The experimental data reported in the review have been collected especially on the Contralto device, and acquired in two different batches, during which the Contralto processor was first hosted in Partenope, and then in Ligea. 
	
	\subsubsection{The processor}
	The Contralto-D sQPU manufactured by Quantware (panel (b), figure \ref{partenope}) consists of 21 operational, symmetric, flux-tunable transmon qubits arranged in a rectangular 2D matrix and interconnected via fixed high-frequency bus resonators. To prevent frequency crowding and crosstalk, the qubits are structurally distributed across three distinct frequency bandwidths:
	\begin{itemize}
		\item \textbf{High-frequency qubits:} Operating at approximately $6\text{ GHz}$ (indicated in red).
		\item \textbf{Medium-frequency qubits:} Operating at approximately $5\text{ GHz}$ (indicated in cyan).
		\item \textbf{Low-frequency qubits:} Operating in a range around $4\text{ GHz}$ (indicated in green).
	\end{itemize}
	Every qubit features a dedicated superconducting readout resonator. These resonators are distributed evenly across four readout feedlines to enable multiplexed readout. Additionally, independent drive and flux lines are assigned to each qubit to control single-qubit $X$-$Y$ and $Z$ gates. The native two-qubit gate of this \textit{sQPU} is the Conditional-$Z$ (CZ) gate. This is implemented by applying sudden-net-zero (SNZ) flux pulses to the higher-frequency qubit within a given pair. Furthermore, the flux lines provide the static magnetic flux required to park the qubits at specific operational working points. Throughout the experimental procedures, the active qubits were operated at their respective flux sweetspots. Conversely, the remaining idle qubits in the matrix were parked at their anti-sweet spots (nominally at zero frequency) to minimize qubit crosstalk and frequency crowding \cite{Koch07, Negirneac21}.


\newpage
	
	
	
	\begin{thebibliography}{198}
		
		\bibitem{Deutsch1985} Deutsch D. 1985 ``Quantum theory, the Church--Turing principle and the universal quantum computer'' \textit{Proceedings of the Royal Society of London. A. Mathematical and Physical Sciences} \textbf{400} (1818) 97--117 doi:10.1098/rspa.1985.0070.
		
		\bibitem{Shor94} Shor P. W. 1994 ``Algorithms for quantum computation: discrete logarithms and factoring'' in \textit{Proceedings 35th Annual Symposium on Foundations of Computer Science} pp. 124-134 doi:10.1109/SFCS.1994.365700.
		
		\bibitem{Grover96} Grover L. K. 1996 ``A fast quantum mechanical algorithm for database search'' in \textit{Proceedings of the Twenty-Eighth Annual ACM Symposium on Theory of Computing} pp. 212--219 Association for Computing Machinery New York, NY, USA doi:10.1145/237814.237866.
		
		\bibitem{Bennett2014} Bennett C. H., Brassard G. 2014 ``Quantum Cryptography: Public Key Distribution and Coin Tossing'' \textit{Theoretical Computer Science} \textbf{560} 7--11 doi:10.1016/j.tcs.2014.05.025.
		
		\bibitem{Sanders2023} Sanders B. C. 2023 ``Quantum Computing for Data Science'' \textit{Journal of Physics: Conference Series} \textbf{2438} (1) 012007 doi:10.1088/1742-6596/2438/1/012007.
		
		\bibitem{Zheng2026} Zheng Y. et al. 2026 ``Large-scale quantum communication networks with integrated photonics'' \textit{Nature} \textbf{651} (8104) 68--75 doi:10.1038/s41586-026-10152-z.
		
		\bibitem{Feynman82} Feynman R. P. 1982 ``Simulating Physics with Computers'' \textit{International Journal of Theoretical Physics} \textbf{21} (6) 467--488 doi:10.1007/BF02650179.
		
		\bibitem{Fauseweh2024} Fauseweh B. 2024 ``Quantum many-body simulations on digital quantum computers: State-of-the-art and future challenges'' \textit{Nature Communications} \textbf{15} (1) 2123 doi:10.1038/s41467-024-46402-9.
		
		\bibitem{HeadMarsden2021} Head-Marsden K., Flick J., Ciccarino C. J., Narang P. 2021 ``Quantum Information and Algorithms for Correlated Quantum Matter'' \textit{Chemical Reviews} \textbf{121} (5) 3061--3120 doi:10.1021/acs.chemrev.0c00620.
		
		\bibitem{Ollitrault2021} Ollitrault P. J., Miessen A., Tavernelli I. 2021 ``Molecular Quantum Dynamics: A Quantum Computing Perspective'' \textit{Accounts of Chemical Research} \textbf{54} (23) 4229--4238 doi:10.1021/acs.accounts.1c00514.
		
		\bibitem{NAEIJ2024} Naeij H. R., Mahmoudi E., Yeganeh H. D., Akbari M. 2024 ``Molecular electronic structure calculation via a quantum computer'' \textit{Computational and Theoretical Chemistry} \textbf{1242} 114945 doi:10.1016/j.comptc.2024.114945.
		
		\bibitem{Zhou2026} Zhou Y. et al. 2026 ``Quantum-machine-assisted drug discovery'' \textit{npj Drug Discovery} \textbf{3} (1) 1 doi:10.1038/s44386-025-00033-2.
		
		\bibitem{Kim2026} Kim S., Suh I.-S., Humble T. S., Beck T., Lee E., Luo T. 2026 ``Harnessing Quantum Computing for Energy Materials: Opportunities and Challenges'' \textit{ACS Energy Letters} \textbf{11} (2) 1182--1192 doi:10.1021/acsenergylett.5c04009.
		
		\bibitem{Ahmad26} Ahmad H. G. et al. 2026 ``Quantum Circuit-Based Adaptation for Credit Risk Analysis'' \textit{IEEE Transactions on Quantum Engineering} \textbf{7} 3103316-3103316 doi:10.1109/TQE.2026.3691176.
		
		\bibitem{Woerner2019} Woerner S., Egger D. J. 2019 ``Quantum risk analysis'' \textit{npj Quantum Information} \textbf{5} (1) 15 doi:10.1038/s41534-019-0130-6.
		
		\bibitem{Weinberg2023} Weinberg S. J., Sanches F., Ide T., Kamiya K., Correll R. 2023 ``Supply chain logistics with quantum and classical annealing algorithms'' \textit{Scientific Reports} \textbf{13} (1) 4770 doi:10.1038/s41598-023-31765-8.
		
		\bibitem{Biamonte2017} Biamonte J., Wittek P., Pancotti N., Rebentrost P., Wiebe N., Lloyd S. 2017 ``Quantum machine learning'' \textit{Nature} \textbf{549} (7671) 195--202 doi:10.1038/nature23474.
		
		\bibitem{Alexeev2025} Alexeev Y. et al. 2025 ``Artificial intelligence for quantum computing'' \textit{Nature Communications} \textbf{16} (1) 10829 doi:10.1038/s41467-025-65836-3.
		
		\bibitem{Sivak2026} Sivak V. et al. 2026 ``Reinforcement learning control of quantum error correction'' \textit{Nature} doi:10.1038/s41586-026-10759-2.
		
		\bibitem{caldeira1983} Caldeira A. O., Leggett A. J. 1983 ``Path integral approach to quantum dissipative systems'' \textit{Annals of Physics} \textbf{149} (2) 374--456.
		
		\bibitem{Preskill2018} Preskill J. 2018 ``Quantum {C}omputing in the {NISQ} era and beyond'' \textit{{Quantum}} \textbf{2} 79 doi:10.22331/q-2018-08-06-79.
		
		\bibitem{Main2025} Main D., Drmota P., Nadlinger D. P., Ainley E. M., Agrawal A., Nichol B. C., Srinivas R., Araneda G., Lucas D. M. 2025 ``Distributed quantum computing across an optical network link'' \textit{Nature} \textbf{638} (8050) 383--388 doi:10.1038/s41586-024-08404-x.
		
		\bibitem{Acharya2025} Acharya R. et al. 2025 ``Quantum error correction below the surface code threshold'' \textit{Nature} \textbf{638} (8052) 920--926 doi:10.1038/s41586-024-08449-y.
		
		\bibitem{Takeda2022} Takeda K., Noiri A., Nakajima T., Kobayashi T., Tarucha S. 2022 ``Quantum error correction with silicon spin qubits'' \textit{Nature} \textbf{608} (7924) 682--686 doi:10.1038/s41586-022-04986-6.
		
		\bibitem{gottesman2010} Gottesman D. 2010 ``An introduction to quantum error correction and fault-tolerant quantum computation'' in \textit{Quantum information science and its contributions to mathematics, Proceedings of Symposia in Applied Mathematics} pp. 13--58.
		
		\bibitem{Cerf97} Cerf N. J., Cleve R. 1997 ``Information-theoretic interpretation of quantum error-correcting codes'' \textit{Phys. Rev. A} \textbf{56} (3) 1721--1732 doi:10.1103/PhysRevA.56.1721.
		
		\bibitem{Guillaud19} Guillaud J. e. e., Mirrahimi M. 2019 ``Repetition Cat Qubits for Fault-Tolerant Quantum Computation'' \textit{Phys. Rev. X} \textbf{9} (4) 041053 doi:10.1103/PhysRevX.9.041053.
		
		\bibitem{Takada24} Takada Y., Fujii K. 2024 ``Improving Threshold for Fault-Tolerant Color-Code Quantum Computing by Flagged Weight Optimization'' \textit{PRX Quantum} \textbf{5} (3) 030352 doi:10.1103/PRXQuantum.5.030352.
		
		\bibitem{Fowler2021} Fowler A. G., Mariantoni M., Martinis J. M., Cleland A. N. 2012 ``Surface codes: Towards practical large-scale quantum computation'' \textit{Phys. Rev. A} \textbf{86} (3) 032324 doi:10.1103/PhysRevA.86.032324.
		
		\bibitem{Egan2021} Egan L. et al. 2021 ``Fault-tolerant control of an error-corrected qubit'' \textit{Nature} \textbf{598} (7880) 281--286 doi:10.1038/s41586-021-03928-y.
		
		\bibitem{Schneider23} Schneider J., Gammelmark M., Karlsson S. 2023 ``Modeling of Errors in Quantum Computers with Generated Structural Circuits'' in \textit{2023 IEEE International Conference on Quantum Computing and Engineering (QCE)} pp. 122-126 doi:10.1109/QCE57702.2023.10195.
		
		\bibitem{campbell2017} Campbell E. T., Terhal B. M., Vuillot C. 2017 ``Roads towards fault-tolerant universal quantum computation'' \textit{Nature} \textbf{549} (7671) 172--179.
		
		\bibitem{riverlane2025qec} Riverlane 2025 \textit{Quantum error correction is crucial, but the ecosystem isn't ready} Official Blog Riverlane.
		
		\bibitem{Field2024} Field M., Chen A. Q., Scharmann B., Sete E. A., Oruc F., Vu K., Kosenko V., Mutus J. Y., Poletto S., Bestwick A. 2024 ``Modular superconducting-qubit architecture with a multichip tunable coupler'' \textit{Phys. Rev. Appl.} \textbf{21} (5) 054063 doi:10.1103/PhysRevApplied.21.054063.
		
		\bibitem{Chakraborty2024} Chakraborty S., Joshi R. V. 2024 ``Cryogenic CMOS Design for Qubit Control: Present Status, Challenges, and Future Directions [Feature]'' \textit{IEEE Circuits and Systems Magazine} \textbf{24} (2) 34-46 doi:10.1109/MCAS.2024.3383808.
		
		\bibitem{Herrmann23} Herrmann N., Arya D., Doherty M. W., Mingare A., Pillay J. C., Preis F., Prestel S. 2023 ``Quantum utility -- definition and assessment of a practical quantum advantage'' in \textit{2023 IEEE International Conference on Quantum Software (QSW)} pp. 162-174 doi:10.1109/QSW59989.2023.00028.
		
		\bibitem{Montanaro2016} Montanaro A. 2016 ``Quantum algorithms: an overview'' \textit{npj Quantum Information} \textbf{2} (1) 15023 doi:10.1038/npjqi.2015.23.
		
		\bibitem{Lao22} Lao L., van Someren H., Ashraf I., Almudever C. G. 2022 ``Timing and Resource-Aware Mapping of Quantum Circuits to Superconducting Processors'' \textit{IEEE Transactions on Computer-Aided Design of Integrated Circuits and Systems} \textbf{41} (2) 359-371 doi:10.1109/TCAD.2021.3057583.
		
		\bibitem{darpaqbisolicitation} Defense Advanced Research Projects Agency (DARPA) 2026 \textit{Quantum Benchmarking Initiative Topic (QBIT): Stage A}.
		
		\bibitem{eurohpc} European High Performance Computing Joint Undertaking (EuroHPC JU) 2026 \textit{EuroHPC JU Official Website} \url{https://www.eurohpc-ju.europa.eu/index_en}.
		
		\bibitem{Karuppasamy2025} Karuppasamy K., Puram V., Johnson S., Thomas J. P. 2025 ``A Comprehensive Review of Quantum Circuit Optimization: Current Trends and Future Directions'' \textit{Quantum Reports} \textbf{7} (1) 2 doi:10.3390/quantum7010002.
		
		\bibitem{Chong2017} Chong F. T., Franklin D., Martonosi M. 2017 ``Programming Languages and Compiler Design for Realistic Quantum Hardware'' \textit{Nature} \textbf{549} (7671) 180--187 doi:10.1038/nature23459.
		
		\bibitem{Haener_2018} H{\"a}ner T., Steiger D. S., Svore K., Troyer M. 2018 ``A software methodology for compiling quantum programs'' \textit{Quantum Science and Technology} \textbf{3} (2) 020501 doi:10.1088/2058-9565/aaa5cc.
		
		\bibitem{Fellous23} Fellous-Asiani M., Chai J. H., Thonnart Y., Ng H. K., Whitney R. S., Auff\`eves A. 2023 ``Optimizing Resource Efficiencies for Scalable Full-Stack Quantum Computers'' \textit{PRX Quantum} \textbf{4} (4) 040319 doi:10.1103/PRXQuantum.4.040319.
		
		\bibitem{Acharya2023} Acharya R. et al. 2023 ``Suppressing quantum errors by scaling a surface code logical qubit'' \textit{Nature} \textbf{614} (7949) 676--681 doi:10.1038/s41586-022-05434-1.
		
		\bibitem{Bravyi2024} Bravyi S., Cross A. W., Gambetta J. M., Maslov D., Rall P., Yoder T. J. 2024 ``High-threshold and low-overhead fault-tolerant quantum memory'' \textit{Nature} \textbf{627} (8005) 778--782 doi:10.1038/s41586-024-07107-7.
		
		\bibitem{Paetznick2026} Paetznick A. et al. 2026 ``Improved quantum processor logical error rates via correction and detection'' \textit{Nature} \textbf{654} (8118) 349--355 doi:10.1038/s41586-026-10628-y.
		
		\bibitem{DiVincenzo} DiVincenzo D. P. 2000 ``The Physical Implementation of Quantum Computation'' \textit{Fortschritte der Physik} \textbf{48} (9-11) 771-783 doi:10.1002/1521-3978(200009)48:9/11<771::AID-PROP771>3.0.CO;2-E.
		
		\bibitem{wong2026} Wong H. Y. 2026 \textit{Review of Superconducting Qubit Devices and Their Large-Scale Integration} arXiv:2602.04831.
		
		\bibitem{Siddiqi2021} Siddiqi I. 2021 ``Engineering high-coherence superconducting qubits'' \textit{Nature Reviews Materials} \textbf{6} (10) 875-891 doi:10.1038/s41578-021-00370-4.
		
		\bibitem{MURRAY2021} Murray C. E. 2021 ``Material matters in superconducting qubits'' \textit{Materials Science and Engineering: R: Reports} \textbf{146} 100646 doi:10.1016/j.mser.2021.100646.
		
		\bibitem{Bylander2017} Bylander J. 2017 \textit{Superconducting Quantum Bits of Information -- Coherence and Design Improvements} in \textit{The Oxford Handbook of Small Superconductors} Oxford University Press doi:10.1093/oxfordhb/9780198738169.013.18.
		
		\bibitem{lidar2008} Lidar D. A. 2008 ``Towards fault-tolerant quantum computation with dynamical decoupling'' \textit{Physical Review Letters} \textbf{100} (16) 160506.
		
		\bibitem{Cai2023} Cai Z., Babbush R., Benjamin S. C., Endo S., Huggins W. J., Li Y., McClean J. R., O'Brien T. E. 2023 ``Quantum error mitigation'' \textit{Rev. Mod. Phys.} \textbf{95} (4) 045005 doi:10.1103/RevModPhys.95.045005.
		
		\bibitem{Vool2017} Vool U., Devoret M. 2017 ``Introduction to quantum electromagnetic circuits'' \textit{International Journal of Circuit Theory and Applications} \textbf{45} (7) 897-934 doi:10.1002/cta.2359.
		
		\bibitem{Arute2019} Arute F. et al. 2019 ``Quantum supremacy using a programmable superconducting processor'' \textit{Nature} \textbf{574} (7779) 505--510 doi:10.1038/s41586-019-1666-5.
		
		\bibitem{Gold2021} Gold A. et al. 2021 ``Entanglement across separate silicon dies in a modular superconducting qubit device'' \textit{npj Quantum Information} \textbf{7} (1) 142 doi:10.1038/s41534-021-00484-1.
		
		\bibitem{Walter26} Walter J., Weis A. C., Tsai K.-T., Yu M.-J., Katam N., Kirichenko A. F., Mukhanov O. A., Han S.-J., Vernik I. V. 2026 ``Single Flux Quantum Circuit Operation at Millikelvin Temperatures'' \textit{IEEE Transactions on Applied Superconductivity} \textbf{36} (5) 1-6 doi:10.1109/TASC.2025.3639528.
		
		\bibitem{devoret2004} Devoret M. H., Wallraff A., Martinis J. M. 2004 ``Superconducting qubits: A short review'' \textit{arXiv preprint cond-mat/0411174}.
		
		\bibitem{JOSEPHSON1962} Josephson B. D. 1962 ``Possible new effects in superconductive tunnelling'' \textit{Physics Letters} \textbf{1} (7) 251--253 doi:10.1016/0031-9163(62)91369-0.
		
		\bibitem{tafuri2019} Tafuri, Francesco 2019 \textit{Fundamentals and Frontiers of the Josephson Effect} Springer, Cham doi:10.1007/978-3-030-20726-7.
		
		\bibitem{Tolpygo2016} Tolpygo S. K. 2016 ``Superconductor digital electronics: Scalability and energy efficiency issues (Review Article)'' \textit{Low Temperature Physics} \textbf{42} (5) 361-379 doi:10.1063/1.4948618.
		
		\bibitem{Degen2017} Degen C. L., Reinhard F., Cappellaro P. 2017 ``Quantum sensing'' \textit{Rev. Mod. Phys.} \textbf{89} (3) 035002 doi:10.1103/RevModPhys.89.035002.
		
		\bibitem{Ghosh2026} Ghosh P., Konar T. K., Rakshit D., De A. S., Sen U. 2026 \textit{Journey in quantum metrology and sensing from foundations to applications: a review} arXiv:2605.21702.
		
		\bibitem{Barone1982} 1982 \textit{Front Matter} in \textit{Physics and Applications of the Josephson Effect} John Wiley \& Sons, Ltd pp. i-xix doi:10.1002/352760278X.fmatter.
		
		\bibitem{Oliver_2013} Oliver W. D., Welander P. B. 2013 ``Materials in superconducting quantum bits'' \textit{MRS Bulletin} \textbf{38} (10) 816--825 doi:10.1557/mrs.2013.229.
		
		\bibitem{Hornibrook2015} Hornibrook J. M. et al. 2015 ``Cryogenic Control Architecture for Large-Scale Quantum Computing'' \textit{Phys. Rev. Appl.} \textbf{3} (2) 024010 doi:10.1103/PhysRevApplied.3.024010.
		
		\bibitem{kjaergaard2020superconducting} Kjaergaard M., Schwartz M. E., Braum{\"u}ller J., Krantz P., Wang J. I. J., Gustavsson S., Oliver W. D. 2020 ``Superconducting Qubits: Current State of Play'' \textit{Annual Review of Condensed Matter Physics} \textbf{11} 369--395.
		
		\bibitem{Jiang25n} Jiang Y.-Y. et al. 2025 ``Advancements in superconducting quantum computing'' \textit{National Science Review} \textbf{12} (8) nwaf246 doi:10.1093/nsr/nwaf246.
		
		\bibitem{Clarke2008} Clarke J., Wilhelm F. K. 2008 ``Superconducting quantum bits'' \textit{Nature} \textbf{453} (7198) 1031--1042 doi:10.1038/nature07128.
		
		\bibitem{Nakamura99} Nakamura Y., Pashkin Y. A., Tsai J. S. 1999 ``Coherent control of macroscopic quantum states in a single-Cooper-pair box'' \textit{Nature} \textbf{398} (6730) 786--788 doi:10.1038/19718.
		
		\bibitem{Koch07} Koch J., Yu T. M., Gambetta J., Houck A. A., Schuster D. I., Majer J., Blais A., Devoret M. H., Girvin S. M., Schoelkopf R. J. 2007 ``Charge-insensitive qubit design derived from the Cooper pair box'' \textit{Phys. Rev. A} \textbf{76} (4) 042319 doi:10.1103/PhysRevA.76.042319.
		
		\bibitem{Krantz19} Krantz P., Kjaergaard M., Yan F., Orlando T. P., Gustavsson S., Oliver W. D. 2019 ``A quantum engineer's guide to superconducting qubits'' \textit{Applied Physics Reviews} \textbf{6} (2) 021318 doi:10.1063/1.5089550.
		
		\bibitem{Hutchings2017} Hutchings M. D., Hertzberg J. B., Liu Y., Bronn N. T., Keefe G. A., Brink M., Chow J. M., Plourde B. L. T. 2017 ``Tunable Superconducting Qubits with Flux-Independent Coherence'' \textit{Phys. Rev. Appl.} \textbf{8} (4) 044003 doi:10.1103/PhysRevApplied.8.044003.
		
		\bibitem{Barends13} Barends R. et al. 2013 ``Coherent Josephson Qubit Suitable for Scalable Quantum Integrated Circuits'' \textit{Phys. Rev. Lett.} \textbf{111} (8) 080502 doi:10.1103/PhysRevLett.111.080502.
		
		\bibitem{Rahamim17} Rahamim J., Behrle T., Peterer M. J., Patterson A., Spring P. A., Tsunoda T., Manenti R., Tancredi G., Leek P. J. 2017 ``Double-sided coaxial circuit QED with out-of-plane wiring'' \textit{Applied Physics Letters} \textbf{110} (22) 222602 doi:10.1063/1.4984299.
		
		\bibitem{Manucharyan2009} Manucharyan V. E., Koch J., Glazman L. I., Devoret M. H. 2009 ``Fluxonium: Single Cooper-Pair Circuit Free of Charge Offsets'' \textit{Science} \textbf{326} (5949) 113--116 doi:10.1126/science.1175552.
		
		\bibitem{Somoroff2023} Somoroff A., Ficheux Q., Mencia R. A., Xiong H., Kuzmin R., Manucharyan V. E. 2023 ``Millisecond Coherence in a Superconducting Qubit'' \textit{Phys. Rev. Lett.} \textbf{130} (26) 267001 doi:10.1103/PhysRevLett.130.267001.
		
		\bibitem{Brooks13} Brooks P., Kitaev A., Preskill J. 2013 ``Protected gates for superconducting qubits'' \textit{Phys. Rev. A} \textbf{87} (5) 052306 doi:10.1103/PhysRevA.87.052306.
		
		\bibitem{Gyenis21} Gyenis A. a., Mundada P. S., Di Paolo A., Hazard T. M., You X., Schuster D. I., Koch J., Blais A., Houck A. A. 2021 ``Experimental Realization of a Protected Superconducting Circuit Derived from the $0$--$\ensuremath{\pi}$ Qubit'' \textit{PRX Quantum} \textbf{2} (1) 010339 doi:10.1103/PRXQuantum.2.010339.
		
		\bibitem{wang2020qudits} Wang Y., Hu Z., Sanders B. C., Kais S. 2020 ``Qudits and High-Dimensional Quantum Computing'' \textit{Frontiers in Physics} \textbf{8} doi:10.3389/fphy.2020.589504.
		
		\bibitem{MIRRAHIMI2016} Mirrahimi M. 2016 ``Cat-qubits for quantum computation'' \textit{Comptes Rendus Physique} \textbf{17} (7) 778-787 doi:10.1016/j.crhy.2016.07.011.
		
		\bibitem{Teoh2023} Teoh J. D. et al. 2023 ``Dual-rail encoding with superconducting cavities'' \textit{Proceedings of the National Academy of Sciences} \textbf{120} (41) e2221736120 doi:10.1073/pnas.2221736120.
		
		\bibitem{Bal2024} Bal M. et al. 2024 ``Systematic improvements in transmon qubit coherence enabled by niobium surface encapsulation'' \textit{npj Quantum Information} \textbf{10} (1) 43 doi:10.1038/s41534-024-00840-x.
		
		\bibitem{Brosco2024} Brosco V., Serpico G., Vinokur V., Poccia N., Vool U. 2024 ``Superconducting Qubit Based on Twisted Cuprate Van der Waals Heterostructures'' \textit{Phys. Rev. Lett.} \textbf{132} (1) 017003 doi:10.1103/PhysRevLett.132.017003.
		
		\bibitem{Bland2025} Bland M. P. et al. 2025 ``Millisecond lifetimes and coherence times in 2D transmon qubits'' \textit{Nature} \textbf{647} (8089) 343-348 doi:10.1038/s41586-025-09687-4.
		
		\bibitem{Kim2024} Kim S., Abdurakhimov L. V., Pham D., Qiu W., Terai H., Ashhab S., Saito S., Yamashita T., Semba K. 2024 ``Superconducting flux qubit with ferromagnetic Josephson $\pi$-junction operating at zero magnetic field'' \textit{Communications Materials} \textbf{5} (1) 216 doi:10.1038/s43246-024-00659-1.
		
		\bibitem{Larsen2015} Larsen T. W., Petersson K. D., Kuemmeth F., Jespersen T. S., Krogstrup P., Nyg\aa{}rd J., Marcus C. M. 2015 ``Semiconductor-Nanowire-Based Superconducting Qubit'' \textit{Phys. Rev. Lett.} \textbf{115} (12) 127001 doi:10.1103/PhysRevLett.115.127001.
		
		\bibitem{Riechert2025} Riechert H. et al. 2025 ``The carbon nanotube gatemon qubit'' \textit{Nature Communications} \textbf{16} (1) 7197 doi:10.1038/s41586-019-1666-5.
		
		\bibitem{Ahmad22} Ahmad H. G. et al. 2022 ``Hybrid ferromagnetic transmon qubit: Circuit design, feasibility, and detection protocols for magnetic fluctuations'' \textit{Phys. Rev. B} \textbf{105} (21) 214522 doi:10.1103/PhysRevB.105.214522.
		
		\bibitem{Satariano2025} Satariano R., Ferraiuolo R., Calloni F., Ahmad H. G., Gatta D., Tafuri F., Bruno A., Massarotti D. 2025 ``Submicrometer tunnel ferromagnetic Josephson junctions with transmon energy scale'' \textit{Applied Physics Letters} \textbf{127} (25) 252601 doi:10.1063/5.0303194.
		
		\bibitem{calloni2026} Calloni F. et al. 2026 \textit{Phase dynamics and dissipation in tunnel ferromagnetic Josephson junctions} arXiv:2605.18586.
		
		\bibitem{Ahmad2025_transmon} Ahmad H. G. et al. 2025 ``Towards Novel Tunability Schemes for Hybrid Ferromagnetic Transmon Qubits'' \textit{IEEE Transactions on Applied Superconductivity} \textbf{35} (5) 1-7 doi:10.1109/TASC.2025.3535674.
		
		\bibitem{Kim16} Kim M. D., Kim J. 2016 ``Coupling qubits in circuit-QED cavities connected by a bridge qubit'' \textit{Phys. Rev. A} \textbf{93} (1) 012321 doi:10.1103/PhysRevA.93.012321.
		
		\bibitem{Wallraff2004} Wallraff A., Schuster D. I., Blais A., Frunzio L., Huang R. S., Majer J., Kumar S., Girvin S. M., Schoelkopf R. J. 2004 ``Strong coupling of a single photon to a superconducting qubit using circuit quantum electrodynamics'' \textit{Nature} \textbf{431} (7005) 162--167 doi:10.1038/nature02851.
		
		\bibitem{forati2025} Forati E., Langley B. W., Nersisyan A., Molavi R. 2025 \textit{On the electromagnetic couplings in superconducting qubit circuits} arXiv:2405.11426.
		
		\bibitem{DiCarlo09} DiCarlo L. et al. 2009 ``Demonstration of two-qubit algorithms with a superconducting quantum processor'' \textit{Nature} \textbf{460} (7252) 240--244 doi:10.1038/nature08121.
		
		\bibitem{Moskalenko2022} Moskalenko I. N., Simakov I. A., Abramov N. N., Grigorev A. A., Moskalev D. O., Pishchimova A. A., Smirnov N. S., Zikiy E. V., Rodionov I. A., Besedin I. S. 2022 ``High fidelity two-qubit gates on fluxoniums using a tunable coupler'' \textit{npj Quantum Information} \textbf{8} (1) 130 doi:10.1038/s41534-022-00644-x.
		
		\bibitem{yan2018} Yan F., Krantz P., Sung Y., Kjaergaard M., Campbell D. L., Orlando T. P., Gustavsson S., Oliver W. D. 2018 ``Tunable Coupling Scheme for Implementing High-Fidelity Two-Qubit Gates'' \textit{Phys. Rev. Appl.} \textbf{10} (5) 054062 doi:10.1103/PhysRevApplied.10.054062.
		
		\bibitem{sete2021} Sete E. A., Chen A. Q., Manenti R., Kulshreshtha S., Poletto S. 2021 ``Floating Tunable Coupler for Scalable Quantum Computing Architectures'' \textit{Phys. Rev. Appl.} \textbf{15} (6) 064063 doi:10.1103/PhysRevApplied.15.064063.
		
		\bibitem{Sanclemente26} Vall\'es-Sanclemente S. et al. 2026 ``Optimizing the Frequency Positioning of Tunable Couplers in a Circuit QED Processor to Mitigate Spectator Effects on Quantum Operations'' \textit{Phys. Rev. Lett.} \textbf{136} (20) 200801 doi:10.1103/wdfj-4nbt.
		
		\bibitem{Jiang2025} Jiang L., Xu P., Wu S., Sun J.-A., Dou F.-Q. 2025 ``Microwave-activated two-qubit gates for fixed-coupling and fixed-frequency transmon qubits'' \textit{Phys. Rev. A} \textbf{111} (3) 032609 doi:10.1103/PhysRevA.111.032609.
		
		\bibitem{Roth2017} Roth M., Ganzhorn M., Moll N., Filipp S., Salis G., Schmidt S. 2017 ``Analysis of a parametrically driven exchange-type gate and a two-photon excitation gate between superconducting qubits'' \textit{Phys. Rev. A} \textbf{96} (6) 062323 doi:10.1103/PhysRevA.96.062323.
		
		\bibitem{Ding23} Ding L. et al. 2023 ``High-Fidelity, Frequency-Flexible Two-Qubit Fluxonium Gates with a Transmon Coupler'' \textit{Phys. Rev. X} \textbf{13} (3) 031035 doi:10.1103/PhysRevX.13.031035.
		
		\bibitem{patent_ferro} Herrmann, J., J{\"u}nger, C. H., Ahmad, H. G., Massarotti, D., Satariano, R., Tafuri, F., and Cosenza, C.. European Patent No. EP4769240A1, published July 1, 2026. 	\textit{Assignees}: Quantrolox Finland Oy and University of Naples Federico II.
		
		\bibitem{Muthusubramanian_2024} Muthusubramanian N., Finkel M., Duivestein P., Zachariadis C., van der Meer S. L. M., Veen H. M., Beekman M. W., Stavenga T., Bruno A., DiCarlo L. 2024 ``Wafer-scale uniformity of Dolan-bridge and bridgeless Manhattan-style Josephson junctions for superconducting quantum processors'' \textit{Quantum Science and Technology} \textbf{9} (2) 025006 doi:10.1088/2058-9565/ad199c.
		
		\bibitem{patent_rigetti} J. R. Renzas K. M. J. 2022 \textit{Superconducting vias for routing electrical signals through substrates and their methods of manufacture}.
		
		\bibitem{patent_quantware} J. B. A. Marten B. C. 2026 \textit{Quantum computing assembly comprising at least one quantum chip module and an input-output structure}.
		
		\bibitem{Bravyi22} Bravyi S., Dial O., Gambetta J. M., Gil D., Nazario Z. 2022 ``The future of quantum computing with superconducting qubits'' \textit{Journal of Applied Physics} \textbf{132} (16) 160902 doi:10.1063/5.0082975.
		
		\bibitem{Ruiz2025} Ruiz D., Guillaud J., Leverrier A., Mirrahimi M., Vuillot C. 2025 ``LDPC-cat codes for low-overhead quantum computing in 2D'' \textit{Nature Communications} \textbf{16} (1) 1040 doi:10.1038/s41467-025-56298-8.
		
		\bibitem{Fu25} Fu H., Zhu M., Chen F., Zhang C., Wu J., Xie W., Li X.-Y. 2025 ``Effective and Efficient Parallel Qubit Mapper'' \textit{IEEE Transactions on Computer-Aided Design of Integrated Circuits and Systems} \textbf{44} (5) 1774-1787 doi:10.1109/TCAD.2024.3500784.
		
		\bibitem{Almudevar20} Almudever C. G., Lao L., Wille R., Guerreschi G. G. 2020 ``Realizing Quantum Algorithms on Real Quantum Computing Devices'' in \textit{2020 Design, Automation \& Test in Europe Conference \& Exhibition (DATE)} pp. 864-872 doi:10.23919/DATE48585.2020.9116240.
		
		\bibitem{Cerezo2021} Cerezo M. et al. 2021 ``Variational quantum algorithms'' \textit{Nature Reviews Physics} \textbf{3} (9) 625-644 doi:10.1038/s42254-021-00348-9.
		
		\bibitem{endo2021} Endo S., Cai Z., Benjamin S. C., Yuan X. 2021 ``Hybrid Quantum-Classical Algorithms and Quantum Error Mitigation'' \textit{Journal of the Physical Society of Japan} \textbf{90} (3) 032001 doi:10.7566/JPSJ.90.032001.
		
		\bibitem{Bharti2022} Bharti K. et al. 2022 ``Noisy intermediate-scale quantum algorithms'' \textit{Rev. Mod. Phys.} \textbf{94} (1) 015004 doi:10.1103/RevModPhys.94.015004.
		
		\bibitem{Dahlhauser2021} Dahlhauser M. L., Humble T. S. 2021 ``Modeling noisy quantum circuits using experimental characterization'' \textit{Phys. Rev. A} \textbf{103} (4) 042603 doi:10.1103/PhysRevA.103.042603.
		
		\bibitem{Loglisci2026} Loglisci C., Losavio V. N., Pascazio S., Malerba D. 2026 ``QURIOSO: Quantum gate parameter predictIon through quantum-enhanced long-short term memory'' \textit{Quantum Machine Intelligence} \textbf{8} (1) 35 doi:10.1007/s42484-026-00374-9.
		
		\bibitem{Maronese2022} Maronese M., Moro L., Rocutto L., Prati E. 2022 \textit{Quantum Compiling} in \textit{Quantum Computing Environments} Springer International Publishing Cham pp. 39-74 doi:10.1007/978-3-030-89746-8\_2.
		
		\bibitem{Rattacaso2025} Rattacaso D., Jaschke D., Ballarin M., Siloi I., Montangero S. 2025 ``Quantum circuit compilation with quantum computers'' \textit{Phys. Rev. Res.} \textbf{7} (3) 033268 doi:10.1103/hcz4-nv2y.
		
		\bibitem{Barenco95} Barenco A., Bennett C. H., Cleve R., DiVincenzo D. P., Margolus N., Shor P., Sleator T., Smolin J. A., Weinfurter H. 1995 ``Elementary gates for quantum computation'' \textit{Phys. Rev. A} \textbf{52} (5) 3457--3467 doi:10.1103/PhysRevA.52.3457.
		
		\bibitem{Pereira23} Pereira da Silva R. 2023 ``Gate-based Quantum Computing: An Overview'' \textit{SSRN Electronic Journal} doi:10.2139/ssrn.4347584.
		
		\bibitem{albayaty2024} Al-Bayaty A., Perkowski M. 2024 \textit{Cost-Effective Realization of n-Bit Toffoli Gates for IBM Quantum Computers Using the Bloch Sphere Approach and IBM Native Gates} arXiv:2410.13104.
		
		\bibitem{Shi02} Shi Y. 2003 ``{Both Toffoli and Controlled-NOT need little help to universal quantum computing}'' \textit{Quant. Inf. Comput.} \textbf{3} (1) 084--092 doi:10.26421/QIC3.1-7.
		
		\bibitem{Gottesman98} Gottesman D. 1998 ``{The Heisenberg representation of quantum computers}'' in \textit{{22nd International Colloquium on Group Theoretical Methods in Physics}} pp. 32--43.
		
		\bibitem{Alioscia2020} Zhou S., Yang Z.-C., Hamma A., Chamon C. 2020 ``{Single T gate in a Clifford circuit drives transition to universal entanglement spectrum statistics}'' \textit{SciPost Phys.} \textbf{9} 087 doi:10.21468/SciPostPhys.9.6.087.
		
		\bibitem{Oliviero_Leone_Hamma_Lloyd_2022} Oliviero S. F. E., Leone L., Hamma A., Lloyd S. 2022 ``Measuring magic on a quantum processor'' \textit{npj Quantum Information} \textbf{8} (1) 1--8 doi:10.1038/s41534-022-00666-5.
		
		\bibitem{Alioscia2025} Li G.-C. et al. 2025 ``Invested and Potential Magic Resources in Measurement-Based Quantum Computation'' \textit{Phys. Rev. Lett.} \textbf{135} (16) 160203 doi:10.1103/4yyv-hggz.
		
		\bibitem{ahmad2025_magic} Ahmad H. G. et al. 2025 \textit{Experimental demonstration of non-local magic in a superconducting quantum processor} arXiv:2511.15576.
		
		\bibitem{Kitaev97} Kitaev A. Y. 1997 ``Quantum computations: algorithms and error correction'' \textit{Russian Mathematical Surveys} \textbf{52} (6) 1191 doi:10.1070/RM1997v052n06ABEH002155.
		
		\bibitem{BOYKIN00} Boykin P. O., Mor T., Pulver M., Roychowdhury V., Vatan F. 2000 ``A new universal and fault-tolerant quantum basis'' \textit{Information Processing Letters} \textbf{75} (3) 101-107 doi:10.1016/S0020-0190(00)00084-3.
		
		\bibitem{Brassard_2002} Brassard G., Høyer P., Mosca M., Tapp A. 2002 \textit{Quantum amplitude amplification and estimation} doi:10.1090/conm/305/05215.
		
		\bibitem{Stamatopoulos2020} Stamatopoulos N., Egger D. J., Sun Y., Zoufal C., Iten R., Shen N., Woerner S. 2020 ``Option {P}ricing using {Q}uantum {C}omputers'' \textit{{Quantum}} \textbf{4} 291 doi:10.22331/q-2020-07-06-291.
		
		\bibitem{nielsen2010quantum} Nielsen M. A., Chuang I. L. 2010 \textit{Quantum Computation and Quantum Information} Cambridge University Press.
		
		\bibitem{Cory98} Cory D. G., Price M. D., Maas W., Knill E., Laflamme R., Zurek W. H., Havel T. F., Somaroo S. S. 1998 ``Experimental Quantum Error Correction'' \textit{Phys. Rev. Lett.} \textbf{81} (10) 2152--2155 doi:10.1103/PhysRevLett.81.2152.
		
		\bibitem{Georgescu14} Georgescu I. M., Ashhab S., Nori F. 2014 ``Quantum simulation'' \textit{Rev. Mod. Phys.} \textbf{86} (1) 153--185 doi:10.1103/RevModPhys.86.153.
		
		\bibitem{shende08} Shende V. V., Markov I. L. 2008 \textit{On the CNOT-cost of TOFFOLI gates} arXiv:0803.2316.
		
		\bibitem{Toffoli80} Toffoli T. 1980 ``Reversible computing'' in \textit{Automata, Languages and Programming} pp. 632--644 Springer Berlin Heidelberg Berlin, Heidelberg.
		
		\bibitem{Kim22} Kim Y., Morvan A., Nguyen L. B., Naik R. K., Jünger C., Chen L., Kreikebaum J. M., Santiago D. I., Siddiqi I. 2022 ``High-fidelity three-qubit iToffoli gate for fixed-frequency superconducting qubits'' \textit{Nature Physics} \textbf{18} (7) 783--788 doi:10.1038/s41567-022-01590-3.
		
		\bibitem{Fedorov2012} Fedorov A., Steffen L., Baur M., da Silva M. P., Wallraff A. 2012 ``Implementation of a Toffoli gate with superconducting circuits'' \textit{Nature} \textbf{481} (7380) 170-172 doi:10.1038/nature10713.
		
		\bibitem{Monz2009} Monz T., Kim K., H\"ansel W., Riebe M., Villar A. S., Schindler P., Chwalla M., Hennrich M., Blatt R. 2009 ``Realization of the Quantum Toffoli Gate with Trapped Ions'' \textit{Phys. Rev. Lett.} \textbf{102} (4) 040501 doi:10.1103/PhysRevLett.102.040501.
		
		\bibitem{Li2022} Li Y., Wan L., Zhang H., Zhu H., Shi Y., Chin L. K., Zhou X., Kwek L. C., Liu A. Q. 2022 ``Quantum Fredkin and Toffoli gates on a versatile programmable silicon photonic chip'' \textit{npj Quantum Information} \textbf{8} (1) 112 doi:10.1038/s41534-022-00627-y.
		
		\bibitem{Beterov2018} Beterov I. I., Ashkarin I. N., Yakshina E. A., Tretyakov D. B., Entin V. M., Ryabtsev I. I., Cheinet P., Pillet P., Saffman M. 2018 ``Fast three-qubit Toffoli quantum gate based on three-body F\"orster resonances in Rydberg atoms'' \textit{Phys. Rev. A} \textbf{98} (4) 042704 doi:10.1103/PhysRevA.98.042704.
		
		\bibitem{Yu13} Yu N., Duan R., Ying M. 2013 ``Five two-qubit gates are necessary for implementing the Toffoli gate'' \textit{Phys. Rev. A} \textbf{88} (1) 010304 doi:10.1103/PhysRevA.88.010304.
		
		\bibitem{Sleator95} Sleator T., Weinfurter H. 1995 ``{Realizable Universal Quantum Logic Gates}'' \textit{Phys. Rev. Lett.} \textbf{74} 4087--4090 doi:10.1103/PhysRevLett.74.4087.
		
		\bibitem{Shi20} Shi Y. et al. 2020 ``Resource-Efficient Quantum Computing by Breaking Abstractions'' \textit{Proceedings of the IEEE} \textbf{108} (8) 1353-1370 doi:10.1109/JPROC.2020.2994765.
		
		\bibitem{Siyuan2020} Niu S., Suau A., Staffelbach G., Todri-Sanial A. 2020 ``A Hardware-Aware Heuristic for the Qubit Mapping Problem in the NISQ Era'' \textit{IEEE Transactions on Quantum Engineering} \textbf{1} 1-14 doi:10.1109/TQE.2020.3026544.
		
		\bibitem{Davis2020} Davis M. G., Smith E., Tudor A., Sen K., Siddiqi I., Iancu C. 2020 ``Towards Optimal Topology Aware Quantum Circuit Synthesis'' in \textit{2020 IEEE International Conference on Quantum Computing and Engineering (QCE)} pp. 223-234 doi:10.1109/QCE49297.2020.00036.
		
		\bibitem{Li2019} Li G., Ding Y., Xie Y. 2019 \textit{Tackling the Qubit Mapping Problem for NISQ-Era Quantum Devices} arXiv:1809.02573.
		
		\bibitem{IBMQuantumTranspile} IBM Quantum 2026 \textit{Introduction to Transpilation} \url{https://quantum.cloud.ibm.com/docs/en/guides/transpile}.
		
		\bibitem{GoogleCirq} Google Quantum AI 2026 \textit{Cirq: An Open Source Framework for Programming Quantum Computers} \url{https://quantumai.google/cirq}.
		
		\bibitem{RigettiPyQuilDocs} Rigetti Computing 2026 \textit{pyQuil Documentation} \url{https://pyquil-docs.rigetti.com/en/stable/}.
		
		\bibitem{QuantinuumTKET} Quantinuum 2026 \textit{TKET} \url{https://www.quantinuum.com/glossary-item/tket}.
		
		\bibitem{JavadiAbhari2015} JavadiAbhari A., Patil S., Kudrow D., Heckey J., Lvov A., Chong F. T., Martonosi M. 2015 ``ScaffCC: Scalable compilation and analysis of quantum programs'' \textit{Parallel Computing} \textbf{45} 2--17 doi:10.1016/j.parco.2014.12.001.
		
		\bibitem{Reed2013} Reed M. 2013 \textit{Entanglement and Quantum Error Correction with Superconducting Qubits} arXiv:1311.6759.
		
		\bibitem{Kalfus20} Kalfus W. D., Lee D. F., Ribeill G. J., Fallek S. D., Wagner A., Donovan B., Ristè D., Ohki T. A. 2020 ``High-Fidelity Control of Superconducting Qubits Using Direct Microwave Synthesis in Higher Nyquist Zones'' \textit{IEEE Transactions on Quantum Engineering} \textbf{1} 1-12 doi:10.1109/TQE.2020.3042895.
		
		\bibitem{sung2021realization} 2021 ``Realization of High-Fidelity CZ and ZZ-Free iSWAP Gates with a Tunable Coupler'' \textit{Physical Review X} \textbf{11} 021058.
		
		\bibitem{Ding20} Ding Y., Gokhale P., Lin S. F., Rines R., Propson T., Chong F. T. 2020 ``Systematic Crosstalk Mitigation for Superconducting Qubits via Frequency-Aware Compilation'' in \textit{2020 53rd Annual IEEE/ACM International Symposium on Microarchitecture (MICRO)} pp. 201-214 doi:10.1109/MICRO50266.2020.00028.
		
		\bibitem{Wagner25} Wagner F., Egger D. J., Liers F. 2025 ``Optimized Noise Suppression for Quantum Circuits'' \textit{INFORMS J. on Computing} \textbf{37} (1) 22--41 doi:10.1287/ijoc.2024.0551.
		
		\bibitem{Ahmad2023} Ahmad H. G. et al. 2023 ``Investigating the Individual Performances of Coupled Superconducting Transmon Qubits'' \textit{Condensed Matter} \textbf{8} (1) 29 doi:10.3390/condmat8010029.
		
		\bibitem{Boradjiev2013} Boradjiev I. I., Vitanov N. V. 2013 ``Control of qubits by shaped pulses of finite duration'' \textit{Phys. Rev. A} \textbf{88} (1) 013402 doi:10.1103/PhysRevA.88.013402.
		
		\bibitem{Polat2026} Polat I. l., Overwater R. W. J., Rimbach-Russ M., Sebastiano F. 2026 ``Pulse shaping for ultra-fast adiabatic quantum gates'' \textit{npj Quantum Information} doi:10.1038/s41534-026-01245-8.
		
		\bibitem{Motzoi09} Motzoi F., Gambetta J. M., Rebentrost P., Wilhelm F. K. 2009 ``Simple Pulses for Elimination of Leakage in Weakly Nonlinear Qubits'' \textit{Phys. Rev. Lett.} \textbf{103} (11) 110501 doi:10.1103/PhysRevLett.103.110501.
		
		\bibitem{Rower2024} Rower D. A. et al. 2024 ``Suppressing Counter-Rotating Errors for Fast Single-Qubit Gates with Fluxonium'' \textit{PRX Quantum} \textbf{5} (4) 040342 doi:10.1103/PRXQuantum.5.040342.
		
		\bibitem{Wei2024} Wei K. X., Lauer I., Pritchett E., Shanks W., McKay D. C., Javadi-Abhari A. 2024 ``Native Two-Qubit Gates in Fixed-Coupling, Fixed-Frequency Transmons Beyond Cross-Resonance Interaction'' \textit{PRX Quantum} \textbf{5} (2) 020338 doi:10.1103/PRXQuantum.5.020338.
		
		\bibitem{Rigetti10} Rigetti C., Devoret M. 2010 ``Fully microwave-tunable universal gates in superconducting qubits with linear couplings and fixed transition frequencies'' \textit{Phys. Rev. B} \textbf{81} (13) 134507 doi:10.1103/PhysRevB.81.134507.
		
		\bibitem{ward2026} Ward T. et al. 2026 \textit{Echo Cross Resonance gate error budgeting on a superconducting quantum processor} arXiv:2601.20458.
		
		\bibitem{Sheldon16} Sheldon S., Magesan E., Chow J. M., Gambetta J. M. 2016 ``Procedure for systematically tuning up cross-talk in the cross-resonance gate'' \textit{Phys. Rev. A} \textbf{93} (6) 060302(R) doi:10.1103/PhysRevA.93.060302.
		
		\bibitem{Majer07} Majer J. et al. 2007 ``Coupling superconducting qubits via a cavity bus'' \textit{Nature} \textbf{449} (7161) 443--447 doi:10.1038/nature06184.
		
		\bibitem{Schuch03} Schuch N., Siewert J. 2003 ``Natural two-qubit gate for quantum computation using the $\mathrm{XY}$ interaction'' \textit{Phys. Rev. A} \textbf{67} (3) 032301 doi:10.1103/PhysRevA.67.032301.
		
		\bibitem{Ahmad2024_mit} Ahmad H. G. et al. 2024 ``Mitigating Errors on Superconducting Quantum Processors Through Fuzzy Clustering'' \textit{Advanced Quantum Technologies} \textbf{7} (7) 2300400 doi:10.1002/qute.202300400.
		
		\bibitem{Negirneac21} Neg\^{\i}rneac V. et al. 2021 ``High-Fidelity Controlled-$Z$ Gate with Maximal Intermediate Leakage Operating at the Speed Limit in a Superconducting Quantum Processor'' \textit{Phys. Rev. Lett.} \textbf{126} (22) 220502 doi:10.1103/PhysRevLett.126.220502.
		
		\bibitem{Caldwell18} Caldwell S. A. et al. 2018 ``Parametrically Activated Entangling Gates Using Transmon Qubits'' \textit{Phys. Rev. Appl.} \textbf{10} (3) 034050 doi:10.1103/PhysRevApplied.10.034050.
		
		\bibitem{didier2018analytical} Didier N., Bourassa J. M., Blais A. 2018 ``Analytical Modeling of Parametrically Modulated Transmon Qubits'' \textit{Physical Review A} \textbf{97} 022330.
		
		\bibitem{Hong20} Hong S. S., Papageorge A. T., Sivarajah P., Crossman G., Didier N., Polloreno A. M., Sete E. A., Turkowski S. W., da Silva M. P., Johnson B. R. 2020 ``Demonstration of a parametrically activated entangling gate protected from flux noise'' \textit{Phys. Rev. A} \textbf{101} (1) 012302 doi:10.1103/PhysRevA.101.012302.
		
		\bibitem{Martinis14} Martinis J. M., Geller M. R. 2014 ``Fast adiabatic qubit gates using only ${\ensuremath{\sigma}}_{z}$ control'' \textit{Phys. Rev. A} \textbf{90} (2) 022307 doi:10.1103/PhysRevA.90.022307.
		
		\bibitem{Rol19} Rol M. A. et al. 2019 ``Fast, High-Fidelity Conditional-Phase Gate Exploiting Leakage Interference in Weakly Anharmonic Superconducting Qubits'' \textit{Phys. Rev. Lett.} \textbf{123} (12) 120502 doi:10.1103/PhysRevLett.123.120502.
		
		\bibitem{johnson2011} Johnson B. R. 2011 \textit{Controlling photons in superconducting electrical circuits} Yale University.
		
		\bibitem{Rol2020} Rol M. A., Ciorciaro L., Malinowski F. K., Tarasinski B. M., Sagastizabal R. E., Bultink C. C., Salathe Y., Haandbaek N., Sedivy J., DiCarlo L. 2020 ``Time-domain characterization and correction of on-chip distortion of control pulses in a quantum processor'' \textit{Applied Physics Letters} \textbf{116} (5) 054001 doi:10.1063/1.5133894.
		
		\bibitem{Chow2012} Chow J. M. et al. 2012 ``Universal Quantum Gate Set Approaching Fault-Tolerant Thresholds with Superconducting Qubits'' \textit{Phys. Rev. Lett.} \textbf{109} (6) 060501 doi:10.1103/PhysRevLett.109.060501.
		
		\bibitem{Foxen20} Foxen B. et al. 2020 ``Demonstrating a Continuous Set of Two-Qubit Gates for Near-Term Quantum Algorithms'' \textit{Phys. Rev. Lett.} \textbf{125} (12) 120504 doi:10.1103/PhysRevLett.125.120504.
		
		\bibitem{Chen14} Chen Y. et al. 2014 ``Qubit Architecture with High Coherence and Fast Tunable Coupling'' \textit{Phys. Rev. Lett.} \textbf{113} (22) 220502 doi:10.1103/PhysRevLett.113.220502.
		
		\bibitem{Arute2020HartreeFock} Arute F. et al. 2020 ``Hartree-Fock on a Superconducting Quantum Computer'' \textit{Science} \textbf{369} (6507) 1084--1089 doi:10.1126/science.abb9811.
		
		\bibitem{Koch2026} Koch T. et al. 2026 ``The Quantum Optimization Benchmarking Library'' \textit{Nature Computational Science} \textbf{6} (6) 653--671 doi:10.1038/s43588-026-00991-1.
		
		\bibitem{Gambetta2012} Gambetta J. M., Córcoles A. D., Merkel S. T., Smolin J. A., Chow J. M., Poletto S., Rigetti C., Johnson B. R., Plourde B. L. T., Steffen M. 2012 ``Characterization of Addressability by Simultaneous Randomized Benchmarking'' \textit{Physical Review Letters} \textbf{109} (24) 240504 doi:10.1103/PhysRevLett.109.240504.
		
		\bibitem{Majumder2020} Majumder S., Andreta de Castro L., Brown K. R. 2020 ``Real-time calibration with spectator qubits'' \textit{npj Quantum Information} \textbf{6} (1) 19 doi:10.1038/s41534-020-0251-y.
		
		\bibitem{Proctor2022} Proctor T., Rudinger K., Young K., Nielsen E., Blume-Kohout R. 2022 ``Measuring the capabilities of quantum computers'' \textit{Nature Physics} \textbf{18} (1) 75--79 doi:10.1038/s41567-021-01409-7.
		
		\bibitem{Hashim2025} Hashim A. et al. 2025 ``Practical Introduction to Benchmarking and Characterization of Quantum Computers'' \textit{PRX Quantum} \textbf{6} (3) 030202 doi:10.1103/PRXQuantum.6.030202.
		
		\bibitem{Nielsen2021} Nielsen E., Gamble J. K., Rudinger K., Scholten T., Young K., Blume-Kohout R. 2021 ``Gate {S}et {T}omography'' \textit{{Quantum}} \textbf{5} 557 doi:10.22331/q-2021-10-05-557.
		
		\bibitem{Merkel13} Merkel S. T., Gambetta J. M., Smolin J. A., Poletto S., C\'orcoles A. D., Johnson B. R., Ryan C. A., Steffen M. 2013 ``Self-consistent quantum process tomography'' \textit{Phys. Rev. A} \textbf{87} (6) 062119 doi:10.1103/PhysRevA.87.062119.
		
		\bibitem{Gu21} Gu Y., Mishra R., Englert B.-G., Ng H. K. 2021 ``Randomized Linear Gate-Set Tomography'' \textit{PRX Quantum} \textbf{2} (3) 030328 doi:10.1103/PRXQuantum.2.030328.
		
		\bibitem{Chen21} Chen S., Yu W., Zeng P., Flammia S. T. 2021 ``Robust Shadow Estimation'' \textit{PRX Quantum} \textbf{2} (3) 030348 doi:10.1103/PRXQuantum.2.030348.
		
		\bibitem{Vinas2025} Vi{\~n}as P., Bermudez A. 2025 ``Microscopic parametrizations for gate set tomography under coloured noise'' \textit{npj Quantum Information} \textbf{11} (1) 23 doi:10.1038/s41534-025-00976-4.
		
		\bibitem{Huang2021} Huang H.-Y., Kueng R., Preskill J. 2021 ``Information-theoretic bounds on quantum advantage in machine learning'' \textit{Physical Review Letters} \textbf{126} (19) 190505.
		
		\bibitem{Stasino2025} Stasino V. et al. 2025 ``Implementation and Readout of Maximally Entangled Two-Qubit Gates Quantum Circuits in a Superconducting Quantum Processor'' \textit{Journal of Superconductivity and Novel Magnetism} \textbf{38} (2) 129 doi:10.1007/s10948-025-06968-x.
		
		\bibitem{Aktar22} Aktar S., Bärtschi A., Badawy A.-H. A., Eidenbenz S. 2022 ``A Divide-and-Conquer Approach to Dicke State Preparation'' \textit{IEEE Transactions on Quantum Engineering} \textbf{3} 1-16 doi:10.1109/TQE.2022.3174547.
		
		\bibitem{Yamamoto10} Yamamoto T. et al. 2010 ``Quantum process tomography of two-qubit controlled-Z and controlled-NOT gates using superconducting phase qubits'' \textit{Phys. Rev. B} \textbf{82} (18) 184515 doi:10.1103/PhysRevB.82.184515.
		
		\bibitem{Esposito2021} Esposito M., Ranadive A., Planat L., Roch N. 2021 ``Perspective on traveling wave microwave parametric amplifiers'' \textit{Applied Physics Letters} \textbf{119} (12) 120501 doi:10.1063/5.0064892.
		
		\bibitem{Neeley10} Neeley M. et al. 2010 ``Generation of three-qubit entangled states using superconducting phase qubits'' \textit{Nature} \textbf{467} (7315) 570--573 doi:10.1038/nature09418.
		
	\end{thebibliography}

\begin{thebibliography}{20}
	\bibitem{ahmad2025_magic} Ahmad H. G. et al. 2025 \textit{Experimental demonstration of non-local magic in a superconducting quantum processor} arXiv:2511.15576.
	
	\bibitem{Ahmad26} Ahmad H. G. et al. 2026 ``Quantum Circuit-Based Adaptation for Credit Risk Analysis'' \textit{IEEE Transactions on Quantum Engineering} \textbf{7} 3103316-3103316 doi:10.1109/TQE.2026.3691176.
	
	\bibitem{QbloxArchitectureDocs} Qblox, “Qblox electronics.”
	\url{https://docs.qblox.com/en/main/products/architecture/index.html}, 2026. Accessed:
	2026-06-25.
	
	\bibitem{Quantify} Quantify, “Quantify python framework.” \url{https://quantify-os.org/}, 2025. Quantify
	Documentation. Accessed: 2025-12-13.
	
	\bibitem{QbloxSchedulerDocs} Qblox, “Qblox scheduler documentation.”
	\url{https://docs.qblox.com/en/main/products/qblox_scheduler/index.html}, 2026.
	Accessed: 2026-06-25.
	
	\bibitem{Koch07} Koch J., Yu T. M., Gambetta J., Houck A. A., Schuster D. I., Majer J., Blais A., Devoret M. H., Girvin S. M., Schoelkopf R. J. 2007 ``Charge-insensitive qubit design derived from the Cooper pair box'' \textit{Phys. Rev. A} \textbf{76} (4) 042319 doi:10.1103/PhysRevA.76.042319.
	
	\bibitem{Negirneac21} Neg\^{\i}rneac V. et al. 2021 ``High-Fidelity Controlled-$Z$ Gate with Maximal Intermediate Leakage Operating at the Speed Limit in a Superconducting Quantum Processor'' \textit{Phys. Rev. Lett.} \textbf{126} (22) 220502 doi:10.1103/PhysRevLett.126.220502
	
	\end{thebibliography}
\end{document}